\documentclass[twocolumn, showpacs, aps, prd, superscriptaddress, showkeys, floatfix, a4paper, nofootinbib]{revtex4-1} 

\usepackage[english]{babel}

\usepackage{amsmath} 
\usepackage{amssymb}

\usepackage{url}
\usepackage{textcase} 
\usepackage{bm}
\usepackage{bbm}

\usepackage{graphics}
\usepackage{graphicx}

\usepackage{booktabs}

\usepackage[ansinew]{inputenc}
\usepackage[T1]{fontenc}
\usepackage{lmodern}

\usepackage{textcomp}

\usepackage{color}
\usepackage{xcolor}

\usepackage[caption=false]{subfig}

\usepackage{hyperref}
\hypersetup{colorlinks=true,citecolor=blue, urlcolor=blue, linkcolor=blue}

\usepackage{multirow}

\definecolor{MyBlue}{RGB}{50,50,130}
\definecolor{MyBlue2}{RGB}{0,128,255}
\definecolor{MyGreen}{RGB}{0,128,0}
\definecolor{MyOrange}{RGB}{255,149,14}
\definecolor{MyOrange2}{RGB}{255,66,14}
\definecolor{MyOrange3}{RGB}{240,113,0}
\definecolor{MyRed}{RGB}{128,0,0}
\definecolor{MyGrey}{RGB}{69,69,69}
\definecolor{MyYellow}{RGB}{250,220,0}
\definecolor{MyGrey2}{RGB}{100,100,100}
\definecolor{colorgraylight}{gray}{.8}
\definecolor{MyLila}{RGB}{255,0,200}

\begin{document}

\title{Neutrino Flavor Evolution in Binary Neutron Star Merger Remnants}

\author{Maik Frensel}
\email{Maik.Frensel@unibas.ch}
\affiliation{Department of Physics, University of Basel, Klingelbergstrasse 82, 4056 Basel, Switzerland}

\author{Meng-Ru Wu}
\email{mwu@theorie.ikp.physik.tu-darmstadt.de}
\affiliation{Institut f\"ur Kernphysik (Theoriezentrum), Technische Universit\"at Darmstadt,
Schlossgartenstra{\ss}e 2, 64289 Darmstadt, Germany}

\author{Cristina Volpe}
\email{volpe@apc.in2p3.fr}
\affiliation{Astro-Particule et Cosmologie (APC), CNRS UMR 7164, Universit\'{e} Denis Diderot,
10, rue Alice Domon et L\'{e}onie Duquet, 75205 Paris Cedex 13, France}

\author{Albino Perego}
\email{albino@theorie.ikp.physik.tu-darmstadt.de}
\affiliation{Institut f\"ur Kernphysik (Theoriezentrum), Technische Universit\"at Darmstadt,
Schlossgartenstra{\ss}e 2, 64289 Darmstadt, Germany}

\keywords{Neutrino oscillations, Accretion disk, Neutron star merger, Matter-neutrino resonance}

\date{\today}

\begin{abstract}
We study the neutrino flavor evolution in the neutrino-driven
wind from a binary neutron star merger remnant consisting
of a massive neutron star surrounded by an accretion disk. 
With the neutrino emission characteristics and
the hydrodynamical profile of the remnant consistently 
extracted from a three-dimensional simulation, we
compute the flavor evolution by taking into account 
neutrino coherent forward scattering off ordinary matter
and neutrinos themselves. 
We employ a ``single-trajectory'' approach to investigate the dependence of the flavor evolution on the neutrino emission location and angle. We also show
that the flavor conversion in the merger remnant can
affect the (anti-)neutrino absorption rates on free nucleons and may thus impact the $r$-process nucleosynthesis in the wind. 
We discuss the sensitivity of such results on the 
change of neutrino emission characteristics, also
from different neutron star merger simulations.
\end{abstract}

\maketitle

\begin{section}{Introduction}\label{sec:intro}

The first gravitational wave signal detection from a binary black hole
merger observed by the Virgo-LIGO collaboration has opened the
era of gravitational wave astronomy \cite{LIGO:2016}.
Since binary neutron star (BNS) mergers \cite{Rosswog:2015, Faber:2012} 
are one of the major sources of
gravitational waves, a measurement of such a signal is anticipated.
Moreover, BNS mergers are considered as the likely
production site for rapid neutron capture 
(\textit{r}-process) nucleosynthesis 
\cite{Lattimer:1974, Eichler:1989} 
and as a potential source of short gamma-ray bursts
\cite{Paczynski:1986, Eichler:1989, Nakar:2007}.

Similar to core-collapse supernovae, the dynamics of such
astrophysical environments is expected to be affected by
neutrinos. 
A significant amount of energy is carried by them and their
interaction with matter affects the neutron-to-proton ratio
(or equivalently, the electron fraction $Y_{e}$), which is
a crucial elements for nucleosynthesis. 
Since the main processes of transporting energy and
altering the composition are neutrino flavor-dependent,
any mechanism that changes the flavor content
of neutrinos should be studied in order to fully
access their role in these environments.

Since the first proposals of neutrino oscillations by the
pioneering works of Pontecorvo 
\cite{Pontecorvo:1957, Pontecorvo:1958, Pontecorvo:1967}, it took
almost half a century before neutrino flavor oscillations were finally discovered by the Super-Kamiokande collaboration \cite{Fukuda:1998}
and the Sudbury Neutrino Observatory \cite{Ahmad:2002}. 
It was early recognized that if neutrinos are on their way 
through a dense background medium, they acquire a
refractive index due to coherent forward scattering with
the background particles \cite{Wolfenstein:1978}. 
This can possibly lead to flavor conversions, like in the
Sun, where the Mikheyev-Smirnov-Wolfenstein (MSW) effect
\cite{Wolfenstein:1978, Mikheyev_Smirnov:1986} takes place. 
Furthermore, neutrinos themselves can constitute a 
significant background. This occurs in the early universe 
and in astrophysical environments, such as 
core-collapse supernovae, BNS mergers, and collapsars,
where large neutrino fluxes are present so that their
number density are comparable to or larger than that of 
matter.
In these environments, the neutrino coherent forward
scattering off neutrinos produces flavor-diagonal
\cite{Fuller:1987, Noetzold:1988} as well as off-diagonal
contributions to the neutrino refractive index matrix, 
as realized by Pantaleone \cite{Pantaleone-1, Pantaleone-2}.
The neutrino self-interaction contribution couples
their flavor evolution non-linearly and causes collective oscillations 
where different types of collective phenomena 
(synchronized and bipolar oscillations, spectral splits/swaps) 
can arise (see \cite{Chakraborty:2016, DuanReview2015, DuanReview2010} and references therein).

In environments with a disk geometry (e.g., in collapsars or BNS mergers)
another effect associated with neutrino self-interactions 
was revealed through numerical calculations
\cite{Malkus:2012, Malkus:2014, Malkus:2016}: 
If the matter and neutrino self-interaction potentials
almost cancel each other, matter-neutrino resonances (MNR) can occur and cause
flavor conversion in regions above the emitting disk.
Different from the case of a deleptonizing proto-neutron
star, the material in a binary neutron star merger 
starts with a huge neutron excess. 
The prevailing temperatures of the remnant 
(several $\mathrm{MeV}$ \cite{Rosswog:2015}) 
allow positron captures on neutrons 
($n + e^{+} \to p + \bar{\nu}_{e}$) 
to increase the electron fraction and to release more 
electron antineutrinos than electron neutrinos.
Initially, this larger number flux of electron
antineutrinos causes a different sign in the neutrino 
self-interaction potential compared to the neutrino-matter
potential. 
Depending on the matter profile, this can allow an almost
cancellation of the two potentials at some point.
As neutrinos leave the emission surface, the role of
geometry becomes more important \cite{Malkus:2016}: 
Since the electron antineutrinos decouple deeper inside
the remnant than electron 
neutrinos, the latter have a larger emission surface. 
In the neutrino self-interaction potential this difference
in geometry can induce a flip of sign at some point and can allow for symmetric MNR \cite{Malkus:2016},
as first found in the context of
collapsar-type disks \cite{Malkus:2012}. In \cite{Malkus:2014} another type
of MNR, later called standard MNR \cite{Malkus:2016}, 
where the neutrino self-interaction potential does not
change its sign, was found.
In \cite{Malkus:2016}, both the standard and the symmetric
MNR were investigated within models with equal and
different disk sizes for each neutrino species.
In addition, for the symmetric MNR, the possible
impact on disk wind nucleosynthesis was investigated and
it was found that it could potentially favor the formation
of \textit{r}-process elements \cite{Malkus:2012, Malkus:2016}.

The investigation of this phenomenon in schematic models
shows that the underlying mechanism can be understood in terms of adiabatic solutions similar to the
MSW flavor transformation \cite{Wu-MNR:2016, 
Vaananen-MNR:2016}.
It should be mentioned that the MNR shares common features
with the non-linear feedback in conjunction with helicity
transformations \cite{Vlasenko-2}.
Furthermore, we note that the occurence of the MNR is not
restricted to disk scenarios. Since different signs in the
matter and neutrino potentials are necessary, this effect
could potentially occur in other environments, too. 
For example in core collapse supernovae by incorporating 
active-sterile neutrino mixing \cite{Wu:2016} or non-standard
neutrino interactions \cite{Stapleford:2016}.

Due to the non-linear nature of the problem and the
anisotropic astrophysical environments, 
numerically solving flavor evolution problems including
neutrino self-interactions requires some assumptions.
As will be discussed in detail, 
one typically assumes that the initial symmetry of the
system is maintained. Within this assumption, the flavor evolution of neutrinos in a spherically symmetric environment becomes
solvable in the so-called ``bulb-model'' 
\cite{bulb-model}. This approach is usually called 
``multi-angle approximation'' when the radial coordinate and the angular variable are retained to specify the neutrino propagation.
In contrast, ``single-angle approximation'', which was
also often used in studying such problems, further assumes
that the flavor evolution of neutrinos only depends on the
radial coordinate \cite{Qian-1995}. 
It was found that in the context of supernovae, the
solutions of single-angle and multi-angle approximations
can be similar 
(e.g., \cite{bulb-model,Fogli:2007}) but sometimes
different (e.g., \cite{Dasgupta:2009, Duan:2011}).
Note that based on the ``bulb-model'', 
it was shown that, the matter potential can induce kinematical decoherence,  
suppress flavor conversion, or the flavor instability could be shifted compared 
to the single-angle case when performing a multi-angle treatment \cite{bulb-model}.

In a system with a disk-like geometry, however, the
problem is intrinsically different from a spherically 
symmetric one as the disk itself defines a particular
direction with the disk center. In this case, one naturally
expects that the flavor evolution history of neutrinos
emitted from different parts of the disk with different
emission angles should be different.
In the first flavor evolution works 
with a disk geometry \cite{Malkus:2012, Malkus:2014, Malkus:2016} neutrinos
were followed on $45^{\circ}$-trajectories from accretion
disks around black holes. 
The disk model parameters were chosen to be consistent with studies 
of the collapse of rotating massive stars \cite{Malkus:2012} 
or the mergers of a black hole and a neutron star \cite{Malkus:2014, Malkus:2016}.

In this work, we study the trajectory
dependence of the neutrino flavor evolution
in the neutrino-driven wind from a binary neutron star merger 
remnant before black hole formation. To explore 
the dependence of flavor evolution on the neutrino emission
location and angles, we use a ``single-trajectory''
approximation which assumes that at every point of a given
neutrino trajectory, the flavor states of all neutrinos with 
the same energy identically contribute to the self-interaction.
To this aim we use results from the detailed simulations of
\cite{Perego:2014}, in particular, matter profiles (density,
electron fraction and temperature), neutrino luminosities,
and mean neutrino energies. We present numerical results on the 
flux-averaged neutrino and antineutrino probabilities for
several trajectories where neutrino self-interaction and
matter potentials differ. We then discuss the potential
impact on nucleosynthesis by showing the change in the
(anti-)neutrino capture rates on free nucleons, relevant for
$r$-process nucleosynthesis, due to flavor evolution 
along these trajectories.
We also investigate the sensitivity of the flavor evolution 
to different emission characteristics within the same model, 
or considering uncertainties from available simulations.

The paper is organized as follows. In
Sec.~\ref{sec:optical_depth}, we explain 
the procedure to determine the neutrino emission surfaces. 
In Sec.~\ref{sec:method}, we discuss the equations of
motion governing the neutrino flavor evolution and the 
method we adopted in this work to investigate
the trajectory dependence.
In Sec.~\ref{sec:Trajectory_dependence}, 
we describe the unoscillated potentials along chosen trajectories.
In Sec.~\ref{sec:results}, we present our numerical
results of the trajectory dependence and the impact on the
capture rates. We comment on the dependence of the results on the initial emission parameters. 
We discuss the implications and conclude in Sec.~\ref{sec:conclusion}.
If not otherwise stated, we employ natural units: $\hbar \equiv c \equiv k_{\mathrm{B}} \equiv 1$.
\end{section}

\begin{section}{Disk structure and neutrino surfaces in binary neutron star merger remnants} \label{sec:optical_depth}
\begin{subsection}{BNS merger remnant}\label{sec:remnant}
Our discussion is based on a long-term 
three-dimensional Newtonian hydrodynamics simulation of
the neutrino-driven wind that emerges from the remnant of
the merger of two non-spinning $1.4 \, M_{\odot}$ neutron
stars \cite{Perego:2014}. As a result of the merging 
process, a massive neutron star (MNS) forms in the
central region, surrounded by an accretion disk.
The MNS has a rest
mass possibly larger then the maximally allowed rest mass of
a non-rotating neutron star
\cite{Hotokezaka:2013}.
In the case of a gravitational unstable object, its
temporary stability against gravitational collapse is
expected to be provided primarily by differential rotation
\cite{Baumgarte:2000}, but also other mechanisms, like
thermal pressure, could give additional support
\cite{Baumgarte:2000, Kaplan:2014}. 
For this reason, the MNS is assumed to stay stable during
the simulation time, $\sim \mathcal{O}(100)$~ms after the merger, and is treated 
as a stationary rotating object.

Typical timescales of the disk are given by the dynamical
timescale $t_{\mathrm{dyn}}^{\mathrm{disk}} \sim
\mathcal{O}(10)$~ms and the much longer
viscous timescale $t_{\mathrm{visc}}^{\mathrm{disk}} \sim
\mathcal{O}(300)$~ms which gives an estimate of the
lifetime of the disk \cite{Perego:2014}.
The latter is characterized by a typical radius
$R_{\mathrm{disk}} \sim \mathcal{O}(100)$~km
and innermost density $\rho_{\mathrm{disk}} \sim 5 \times
10^{11} \mathrm{g} \, \mathrm{cm}^{-3}$, while the central density of the MNS
is a few $10^{14} \, \mathrm{g} \, \mathrm{cm}^{-3}$ as can be inferred from 
Fig.~\ref{Fig:density}, where we plot the density at
$100 \, \mathrm{ms}$ after the merger.
Due to the high densities of the remnant, neutrinos act as
the major cooling source and other particles are
essentially trapped on the relevant timescales.

\begin{figure}[htbp!]
 \centering
 \includegraphics[width=0.5\textwidth]{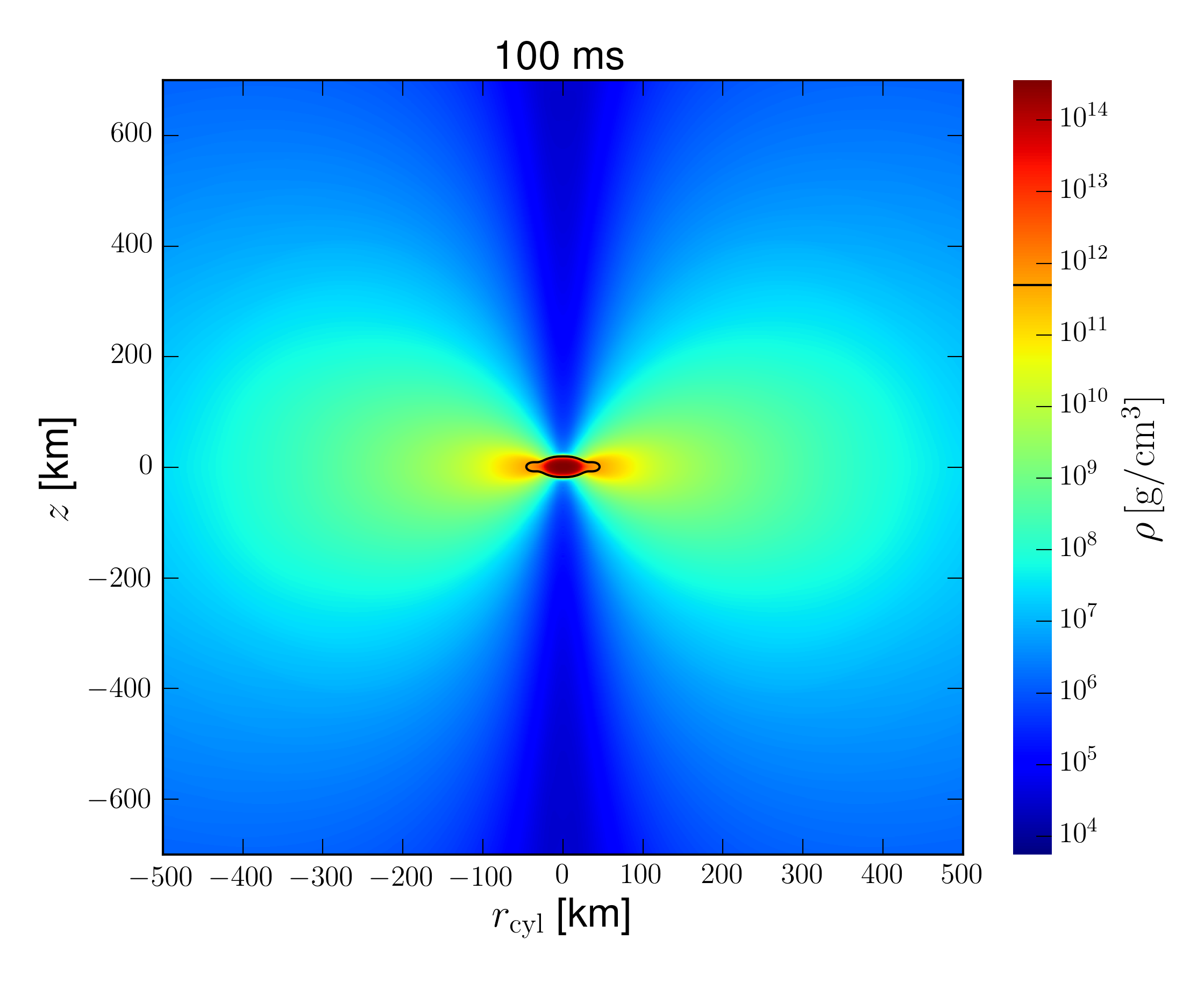}  
 \caption{Density profile as a function of cylindrical coordinates $z$ and $r_{\mathrm{cyl}}$ at $t = 100 \, \mathrm{ms}$ after the merger. The contour $\rho \approx 5 \times 10^{11} \, \mathrm{g} / \mathrm{cm}^{3}$ delimits the innermost part of the disk 
 that is comparable to the surface density of a proto-neutron star \cite{Perego:2014}.} 
 \label{Fig:density}
\end{figure}

\begin{figure}[htbp!]
 \centering
 \includegraphics[width=0.5\textwidth]{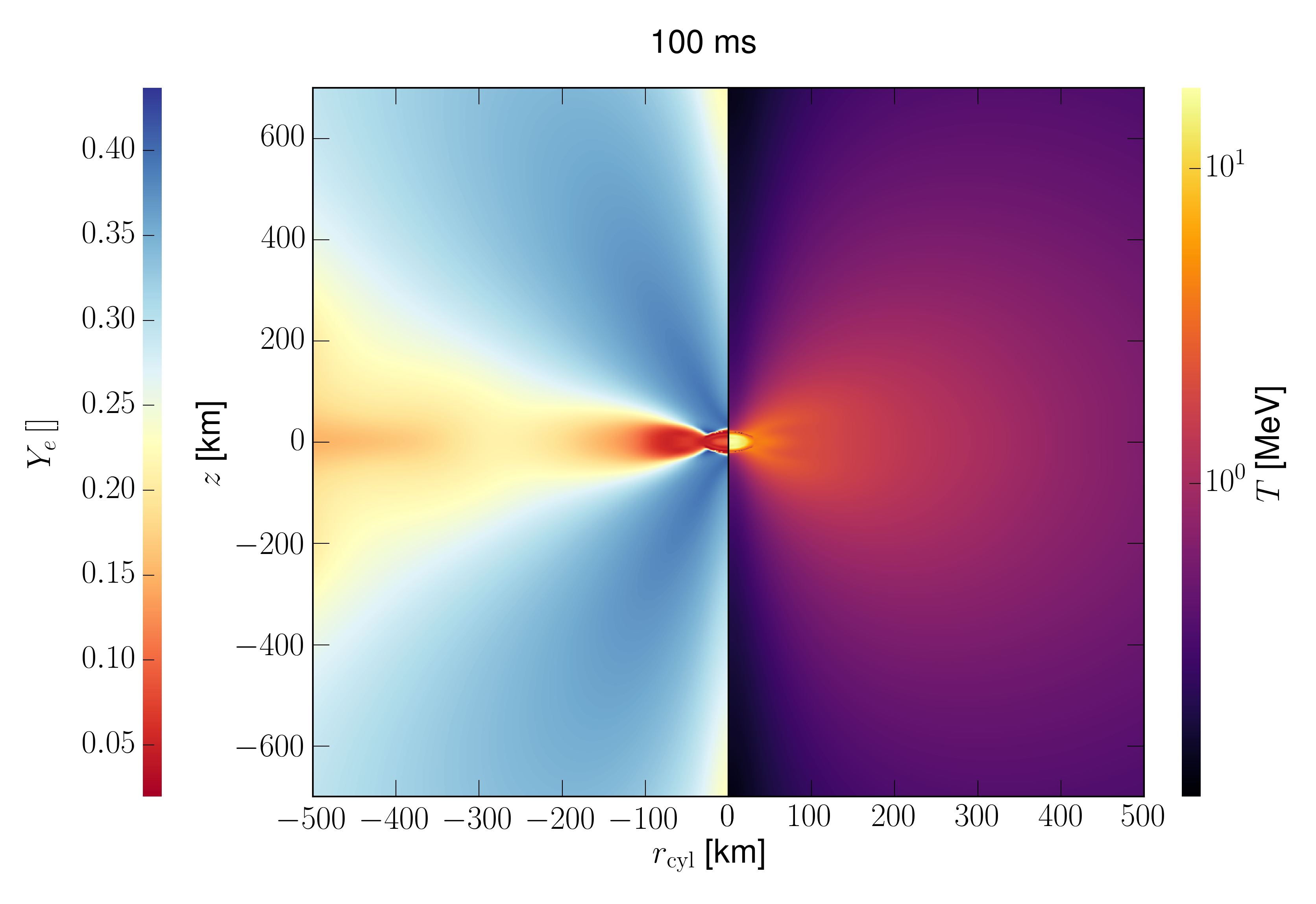} 
 \caption{Electron fraction (left panel) and matter temperature (right panel) as functions of cylindrical coordinates $z$ and $r_{\mathrm{cyl}}$ at $t = 100 \, \mathrm{ms}$ after the merger.} 
 \label{Fig:plot_ye_temp}
\end{figure}

We consider the emission and the absorption of neutrinos from the MNS 
and the surrounding disk \cite{Perego.etal:2016}. Similar to the case of a proto-neutron star, those neutrinos can cause a mass outflow, called neutrino-driven
wind, by energy deposition via absorption and scattering
processes \cite{Duncan:1986, Qian:1996}. 
This wind, together with viscously-driven ejecta, is blown away mainly from the disk \cite{Just:2015, Metzger:2014, Perego:2014, Dessart:2009, Surman:2008}.
Since the rotational period of the accretion disk and of 
the MNS is much smaller than the neutrino diffusion
timescale and the disk lifetime, after a few orbits the
remnant approaches a quasi-axisymmetric configuration.
Thus, we assume rotational symmetry around the MNS
rotational axis and use the axisymmetric averages of
hydrodynamical quantities (matter density, temperature and
electron fraction) from the simulation
\cite{Perego:2014} for our calculations below. 
These quantities are shown in
Figs.~\ref{Fig:density} and~\ref{Fig:plot_ye_temp}.
Local deviations of the three-dimensional quantities with 
respect to the cylindrically averaged values are usually
$\lesssim \, 15\%$ inside the densest part
of the remnant.
\end{subsection}

\begin{subsection}{Neutrino surfaces}
\label{sec:nu_emis}

In this Sec., we construct a neutrino emission disk from the simulation
result described in Sec.~\ref{sec:remnant}.
We first determine the neutrino emission surface 
by calculating the neutrino opacity in the remnant.
The reactions giving the most relevant contributions to
the optical depth are listed in 
Table~\ref{table:neutrino_reactions}.
For their corresponding cross sections $\sigma$ we use the
expressions described in \cite{Burrows:2006} without weak
magnetism corrections 
(see Appendix~\ref{sec:Cross_sections}).

\begin{table}
\center
\caption{Neutrino reactions considered in our model, where $N \in \lbrace n, p \rbrace$. 
For reaction (i), $\nu$ refers to all neutrino species.
In the second column we denote the associated mean free
paths, where (sc) refers to scattering while (ab) to
absorption.
The corresponding cross sections are taken from
\cite{Burrows:2006}.}
\label{table:neutrino_reactions}
\begin{ruledtabular}
\begin{tabular}{cccc}
 & Reaction & Inverse mean free path \hfill \\ \hline
(i) & $\nu + N \rightarrow \nu + N$ & $\lambda_{\nu N, \mathrm{sc}}^{-1}$ \\
(ii) & $\nu_{e} + n \rightarrow e^{-} + p$ & $\lambda_{\nu_{e}, \mathrm{ab}}^{-1}$ \\
(iii) & $\bar{\nu}_{e} + p \rightarrow e^{+} + n$ & $\lambda_{\bar{\nu}_{e}, \mathrm{ab}}^{-1}$ \\
\end{tabular}
\end{ruledtabular}
\end{table}

One main contribution to the opacity for all neutrino 
species is due to elastic neutrino scattering off free nucleons 
($N = n, p$).
Due to the presence of neutron-rich matter, the absorption 
of $\nu_{e}$s by free neutrons becomes the dominant
(though comparable to neutrino-nucleon scattering) opacity
source, while absorption of $\bar{\nu}_{e}$s by free
protons is less effective. 
The $\nu_{x}$s (short for $\nu_{\mu}$, $\bar{\nu}_{\mu}$,
$\nu_{\tau}$, $\bar{\nu}_{\tau}$) only scatter off
nucleons. As a consequence, matter is most opaque for
$\nu_{e}$s and most transparent for $\nu_{x}$s.

The region where those reactions freeze out and
(anti)neutrinos start to stream off freely is called
neutrino surface. Since neutrino opacities have a
significant dependence on the neutrino energy, this
surface is energy dependent and is usually defined in
terms of the neutrino optical depth $\tau_{\nu}$:
\begin{equation}
S_{\nu} : \left \lbrace (r_{\mathrm{cyl}}, z) \mid \tau_{\nu}(E, r_{\mathrm{cyl}}, z) = 2/3 \right \rbrace.
\end{equation}

The spectral optical depth is computed via the line 
integral 
\begin{equation} \label{optical_depth_def_1}
\tau_{\nu}^{\mathrm{d}}(E, r_{\mathrm{cyl}}, z) = \int_{\mathcal{C}_{\mathrm{d}}} \, \mathrm{d}s \, \lambda_{\nu}^{-1} (E, r_{\mathrm{cyl}}', z'),
\end{equation}
where $\mathcal{C}_{\mathrm{d}}$ corresponds to the path 
of integration,
\begin{equation}
\lambda_{\nu}^{-1}(E) = \sum_{i} \lambda_{i}^{-1} = \sum_{i} n_{i} \, \sigma_{i}(E) 
\end{equation}
denotes the inverse mean-free-path and $n_{i}$ the target 
number density corresponding to the reaction with cross
section $\sigma_{i}$. 
The index $i$ runs over all reactions in
Table~\ref{table:neutrino_reactions} relevant for the
neutrino species under consideration. 

For the optical depth $\tau_{\nu}^{\mathrm{d}}$ 
calculation, we employ a local ray-by-ray approach:
At each point $(r_{\mathrm{cyl}}, z)$ on the cylindrical 
domain, we follow a straight line path in one of the seven
directions ($d = 1, \ldots, 7$) described in
\cite{Perego:2014} until the edge of the computational
domain is reached.
Finally, we take the minimum values among all
$\tau_{\nu}^{\mathrm{d}}$ to specify the actual optical
depth at one point \cite{Perego-MODA:2014}:
\begin{equation} \label{optical_depth_def_2}
\tau_{\nu}(E, r_{\mathrm{cyl}}, z) = \min_{1 \leq \mathrm{d} \leq 7} \left \lbrace \tau_{\nu}^{\mathrm{d}}(E, r_{\mathrm{cyl}}, z) \right \rbrace.
\end{equation}

Since we are interested in obtaining an estimate of the
size of the surface where neutrinos last scatter, we focus
on the transport surfaces and perform spectral averages
using a (normalized) distribution function of Fermi-Dirac 
shape with vanishing degeneracy parameter
\begin{equation} \label{Eq:Fermi-Dirac-energy-distribution}
f_{\nu}(E, T) = \dfrac{1}{F_{2}(0)} \dfrac{1}{T^{3}} \dfrac{E^{2}}{\exp(E / T) + 1},
\end{equation}
which is completely determined by the local matter 
temperature $T = T(r_{\mathrm{cyl}}, z)$. In this
expression, we have $F_{2}(0) = \frac{3}{2} \zeta(3)
\approx 1.80$, and $F_{k}(0)$ corresponds to the 
Fermi-Dirac integral of order $k$ with zero degeneracy
parameter,
\begin{equation}
F_{k}(0) \equiv \int_{0}^{\infty} \mathrm{d}x \, \dfrac{x^{k}}{\exp(x) + 1}.
\end{equation}
The results are shown in Fig.~\ref{Fig:optical_depth} and 
the opacities reflect the density structure of the 
remnant.

The mean energies are taken from the simulations performed
in \cite{Perego:2014} and are essentially determined at
the energy surface.  
There, we assume thermal equilibrium such that the 
neutrino temperature can be obtained from the mean
energies via the Fermi relation 
$\langle E_{\nu} \rangle = (F_{3}(0) / F_{2}(0)) \,
T_{\nu}  \approx 3.15 \, T_{\nu}$, 
where $F_{3}(0) = \frac{7 \pi^{4}}{120}$.

\begin{figure*}[!htbp]
\centering
\includegraphics[scale=0.27]{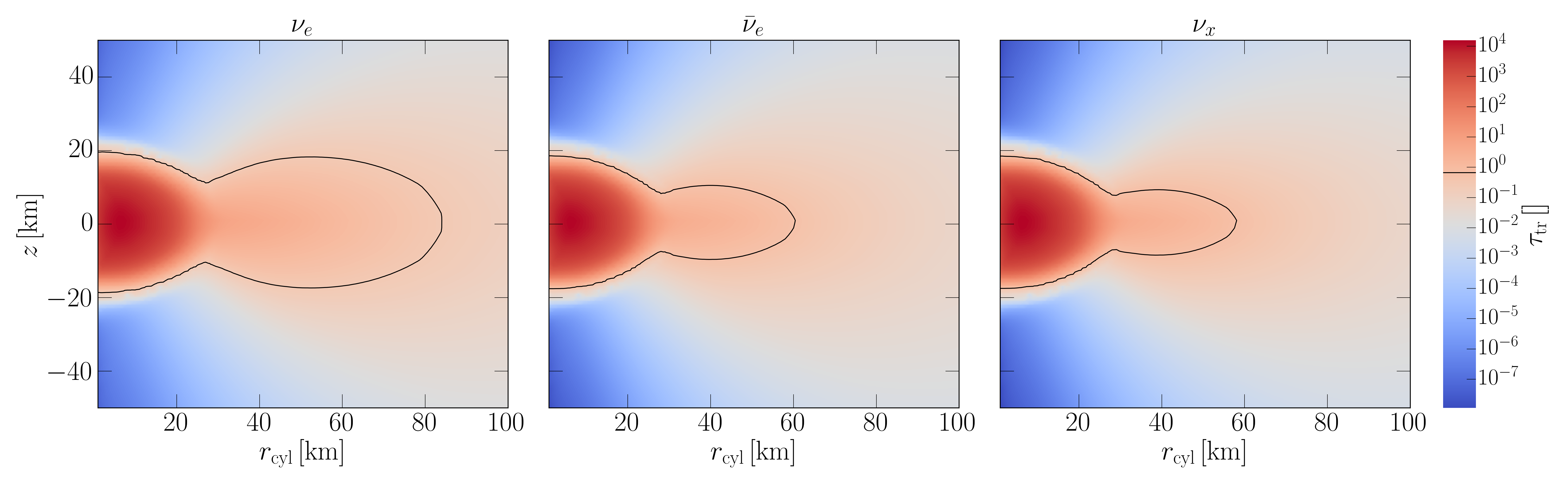} 
\caption{Transport optical depths $\tau$ (color coded) as functions of cylindrical coordinates $z$ and $r_{\mathrm{cyl}}$
at $t = 100 \, \mathrm{ms}$ after the merger. The contours (where $\tau \approx 2/3$) correspond 
to the neutrino transport surfaces associated with $\nu_{e}$ (left), $\bar{\nu}_{e}$ (middle) and $\nu_{x}$ (right).}
\label{Fig:optical_depth}
\end{figure*}

In the following we construct an infinitely thin disk, i.e., 
we turn the neutrino surface into a flat disk, assume a
constant temperature, and define the neutrino disk radius
$R_{\nu}$ as the outermost radius of the neutrino surface,
\begin{equation}
R_{\nu} \equiv \max_{(r_{\mathrm{cyl}}, z) \in S_{\nu}} \left \lbrace \sqrt{r_{\mathrm{cyl}}^2 + z^2} \right \rbrace.
\end{equation}
As can be seen from the results shown in
Table~\ref{table:neutrino_surface_radii}, the differences 
in the neutrino surface radii for the two time snapshots of 60
ms and 100 ms, that we have used in our calculations, are
only minor.
The neutrino mean energies and luminosities are
approximately stationary during the time of simulation
\cite{Perego:2014}. 
The values, used in our calculations, are listed in 
Table~\ref{table:emission_parameters}. 

\begin{table}
\center
\caption{Outermost radii of the neutrino surfaces (at 60 ms and 100 ms) (see text).}
\label{table:neutrino_surface_radii}
\begin{ruledtabular}
 \begin{tabular}{cccc}
t [ms] & $R_{\nu_{e}}$ [km] & $R_{\bar{\nu}_{e}}$ [km] & $R_{\nu_{x}}$ [km] \hfill \\ \hline
60 & $90$ & $64$ & $61$ \\
100 & $84$ & $60$ & $58$ \\
\end{tabular}
\end{ruledtabular}
\end{table}

\begin{table} 
\center
\caption{Emission parameters with $\nu_{x} \in \lbrace \nu_{\mu}, \nu_{\tau}, \bar{\nu}_{\mu}, \bar{\nu}_{\tau} \rbrace$.}
\label{table:emission_parameters}
\begin{ruledtabular}
 \begin{tabular}{ccccc}
Neutrino species & $\langle E_{\nu} \rangle$ [MeV] & $(L_{\nu} / 10^{51})$ [erg/s] \hfill \\ \hline
$\nu_{e}$ & $10.6$ & $15$  \\
$\bar{\nu}_{e}$ & $15.3$ & $30$  \\
$\nu_{x}$ & $17.3$ & $8$ \\
\end{tabular}
\end{ruledtabular}
\end{table}

\end{subsection}
\end{section}

\begin{section}{Neutrino flavor transformation: Method}\label{sec:method}

\begin{subsection}{Equations of motion} \label{sec:eom}
To follow the flavor evolution of neutrinos 
emitted from the disk, we describe a mixed neutrino 
ensemble by Wigner distributions $\rho(\mathbf{p}', \mathbf{x})$ with 
momentum $\mathbf{p}'$ at location $\mathbf{x}$
\cite{Sigl:1993}.
In flavor space\footnote{If not otherwise stated, we will
work in flavor space.}, these represent generalized
occupation number matrices.
In the neutrino free-streaming limit and assuming the 
system in a stationary state, the spatial evolution 
of $\rho(\mathbf{p}^\prime,\mathbf{x})$ obeys the
equation of motion at the lowest order \cite{Sigl:1993}:
\begin{equation} \label{eq:eom_general}
\mathbf{v}_{\mathbf{p}'} \cdot \bm{\nabla}_{\mathbf{x}} 
\rho(\mathbf{p}',\mathbf{x}) = -i \left[ \mathsf{H}
(\mathbf{p}', \mathbf{x}), \rho(\mathbf{p}', \mathbf{x})
\right],
\end{equation}
where 
\begin{equation} \label{eq:Hamiltonian_general}
\mathsf{H} = \mathsf{H}_{\mathrm{vac}} +
\mathsf{H}_{\mathrm{matt}} + \mathsf{H}_{\nu \nu}
\end{equation}
denotes the Hamiltonian including the vacuum, matter and
neutrino potentials. Notice that we do not include
external forces (such as gravity) acting on neutrinos
$(\nabla_{\mathbf{p}}\rho=0)$ and follow neutrinos on
straight-line paths.
For antineutrinos we use $\bar{\rho}(\mathbf{p}',
\mathbf{x})$ with the same definition\footnote{This ensures that both neutrinos 
and antineutrinos transform in the same way under $\mathrm{SU}(3)$.} 
as in \cite{Sigl:1993} so that an analogous equation holds with
the replacement $\mathsf{H}_{\mathrm{vac}} \mapsto 
-\mathsf{H}_{\mathrm{vac}}$. 
On the left hand side of Eq.~\eqref{eq:eom_general} one recognizes the drift term of the Liouville-Vlasov operator, 
that is caused by the free streaming of neutrinos propagating with velocity $\mathbf{v}_{\mathbf{p}'}$. 
In the following we use the ultra-relativistic approximation $\vert\mathbf{v}_{\mathbf{p}'} \vert \approx c$.

Before giving the explicit form of terms in 
Eq.~(\ref{eq:Hamiltonian_general}), we make a few remarks regarding Eq.~(\ref{eq:eom_general}). 
First, this equation of motion is valid at the mean-field
level. However, the most general mean-field approximation 
includes extra contributions, in particular neutrino-antineutrino pairing correlations 
and mass corrections \cite{Volpe:2013, Vlasenko, Serreau:2014, Cirigliano:2015, Kartavtsev:2015, Volpe:2015}. 
The role of these terms still needs to be fully assessed.
Second, we only study the limit where
coherent forward scattering applies and sharply separate
the dense region inside the neutrino surface, where neutrinos
are trapped by collisions and the free streaming region. However, it was shown that
in the context of core-collapse supernovae,
the inclusion of a small backward scattered neutrino flux can 
affect the flavor evolution significantly \cite{Cherry:2012}.
Including these effects in
numerical simulations is challenging and beyond the scope
of this work. Future efforts along these lines may be
necessary.

Now, let us discuss the different terms contributing to the
Hamiltonian in Eq.~\eqref{eq:Hamiltonian_general}. 
The first one describes the vacuum term in the flavor
basis, i.e., 
\begin{equation}\label{eq:Hamiltonian_vacuum}
\mathsf{H}_{\mathrm{vac}} = \dfrac{1}{2 E} \mathsf{U} \tilde{\mathsf{M}}^{2} \mathsf{U}^{\dagger},
\end{equation}
where $E \approx \vert \mathbf{p}' \vert$ corresponds to
the energy of the neutrino and $\mathsf{U}$ denotes the
Pontecorvo-Maki-Nakagawa-Sakata unitary mixing matrix \cite{Maki:1962} which
links weak flavor and vacuum mass eigenstates. The quantity 
$\tilde{\mathsf{M}}^{2} \equiv \mathrm{diag}[ 0, \Delta
m_{21}^{2}, \Delta m_{31}^{2} ]$ essentially corresponds to
the neutrino mass-squared matrix\footnote{Note that we
already subtracted a multiple of the identity matrix which
is not relevant for flavor oscillations.}, where $\Delta
m_{21}^{2}$ and $\Delta m_{31}^{2}$ denote the mass squared
differences.
The second term of the Hamiltonian 
Eq.~\eqref{eq:Hamiltonian_general} takes into account
neutrino coherent forward scattering off electrons. 
Explicitly, we have 
\begin{equation} \label{eq:Hamiltonian_matter}
\mathsf{H}_{\mathrm{matt}}(\mathbf{x}) = \sqrt{2} \mathrm{G}_{\mathrm{F}} n_{e}(\mathbf{x}) \mathrm{diag}[1, 0, 0],
\end{equation}
where $n_{e} = \rho_{\mathrm{matt}} Y_{e} / m_{\mathrm{u}}$ is the electron
number density, determined by the matter density $\rho_{\mathrm{matt}}$ and
electron fraction $Y_{e}$. Here, $m_{\mathrm{u}}$ denotes the unified atomic mass unit.

Similarly, we consider neutrino coherent forward scattering 
off neutrinos which introduces the non-linear nature of the
problem:
\begin{equation} \label{eq:self_interaction_Hamiltonian_general}
\mathsf{H}_{\nu \nu}(\mathbf{p}', \mathbf{x}) = \sqrt{2} \mathrm{G}_{\mathrm{F}} \int \dfrac{\mathrm{d}^{3} p}{(2 \pi)^{3}} \, (1 - \mathbf{\hat{p}}' \cdot \mathbf{\hat{p}}) \left( \rho(\mathbf{p}, \mathbf{x}) {-} \bar{\rho}(\mathbf{p}, \mathbf{x}) \right),
\end{equation}
where ${\mathbf{\hat{p}'} = \mathbf{p}' / \vert \mathbf{p}' \vert}$ and ${\mathbf{\hat{p}} = \mathbf{p} / \vert \mathbf{p} \vert}$ denote unit vectors.
In Eq.~\eqref{eq:self_interaction_Hamiltonian_general} $\rho(\mathbf{p}, \mathbf{x})$
can be decomposed 
\cite{Qian-1995, Fuller-Qian-2006} according to
\begin{equation} \label{eq:density_matrix_nu}
\dfrac{\mathrm{d}^{3} p}{(2 \pi)^{3}} \, \rho(\mathbf{p}, \mathbf{x}) = \sum_{\alpha = e, \mu, \tau} \mathrm{d} n_{\nu_{\underline{\alpha}}}(\mathbf{p}, \mathbf{x}) \, \rho_{\nu_{\underline{\alpha}}}(\mathbf{p}, \mathbf{x})
\end{equation}
for neutrinos in a differential volume element
$\mathrm{d}^{3}p$ centered at momentum $\mathbf{p}$.
A similar relation holds for antineutrinos.
In Eq. \eqref{eq:density_matrix_nu} we introduced the initial (i.e., at the neutrino 
surface, denoted by an underline) differential neutrino
number density $\mathrm{d} n_{\nu_{\underline{\alpha}}}$ 
and the single density $(3 \times 3)$-matrices
$\rho_{\nu_{\underline{\alpha}}}(\mathbf{p}, \mathbf{x})$ 
for a neutrino with initial flavor $\alpha$ and momentum
$\mathbf{p}$ at position $\mathbf{x}$. 
For normalization we choose the trace equal to one.
The diagonal elements of the single density matrices 
correspond to the probabilities that a neutrino 
with initial flavor $\alpha$ can be found in a particular 
flavor $\beta$, i.e., 
$(\rho_{\nu_{\underline{\alpha}}})_{\beta \beta} = 
P(\nu_{\underline{\alpha}} \to \nu_{\beta})$, 
while the (complex-valued) off-diagonal elements describe
quantum correlations between different neutrino flavors
with the same momentum.
An analogous relation holds for antineutrinos. 

\begin{widetext}
Using Eq.~\eqref{eq:density_matrix_nu}, 
Eq.~\eqref{eq:self_interaction_Hamiltonian_general}
becomes:
\begin{equation} \label{eq:Hamiltonian-nu-nu}
\mathsf{H}_{\nu \nu}(\mathbf{p}', \mathbf{x}) = \sqrt{2} \mathrm{G}_{\mathrm{F}} \sum_{\alpha = e, \mu, \tau} \left( \int \mathrm{d} n_{\nu_{\underline{\alpha}}} \, (1 - \mathbf{\hat{p}}' \cdot \mathbf{\hat{p}}) \rho_{\nu_{\underline{\alpha}}}(\mathbf{p}, \mathbf{x}) - \int \mathrm{d} n_{\bar{\nu}_{\underline{\alpha}}} \, (1 - \mathbf{\hat{p}}' \cdot \mathbf{\hat{p}}) \bar{\rho}_{\bar{\nu}_{\underline{\alpha}}}(\mathbf{p}, \mathbf{x}) \right).
\end{equation}

If we follow the flavor evolution of neutrinos along a
specific trajectory, we can replace
$\mathbf{v}_{\mathbf{p}}\cdot\mathbf{\nabla}_{\mathbf{x}}$ 
by a differential operator $\partial / \partial r$ along
their direction of propagation so that the equations of motion \eqref{eq:eom_general}
for the single density flavor matrices become:
\begin{align} 
\label{eq:eom_1} \dfrac{\partial}{\partial r}
\rho_{\nu_{\underline{\alpha}}}(\mathbf{p}',
\mathbf{Q}_0,r) & = -i [ \mathsf{H}_{\mathrm{vac}} +
\mathsf{H}_{\mathrm{matt}} + \mathsf{H}_{\nu\nu},
\rho_{\nu_{\underline{\alpha}}}(\mathbf{p}',\mathbf{Q}_0,
r) ],
\\
\label{eq:eom_2} \dfrac{\partial}{\partial r}
\bar{\rho}_{\bar{\nu}_{\underline{\alpha}}}
(\mathbf{p}',\mathbf{Q}_0,r) & = 
-i [ -\mathsf{H}_{\mathrm{vac}} +
\mathsf{H}_{\mathrm{matt}} + \mathsf{H}_{\nu\nu},
\bar{\rho}_{\bar{\nu}_{\underline{\alpha}}}
(\mathbf{p}',\mathbf{Q}_0, r) ],
\end{align}
where $\mathbf{Q}_0$ is the emission point of the
neutrinos and $r = r(\mathbf{x})$ is the distance they have
traveled.
\end{widetext}
\end{subsection}

\begin{subsection}{Neutrino self-interaction Hamiltonian
in disk geometry}\label{sec:raybyray}

We employ the formalism introduced in \cite{Dasgupta:2008, Malkus:2012} 
and explicitly construct the self-interaction Hamiltonian for neutrinos emitted from a disk.
The coordinate system is defined in such a way, that the following relations for 
the basis vectors hold: 
$\hat{\mathbf{e}}_{x} =\hat{\mathbf{e}}_{r_{\mathrm{cyl}}}$
and $\hat{\mathbf{e}}_{y} = \hat{\mathbf{e}}_{\phi}$.
This allows us to identify the $x$-coordinate with the 
cylindrical radius $r_{\mathrm{cyl}}$ 
(Fig.~\ref{disk-model}). 

\begin{figure*}[!tp]
\centering
\includegraphics[width=0.8\textwidth]{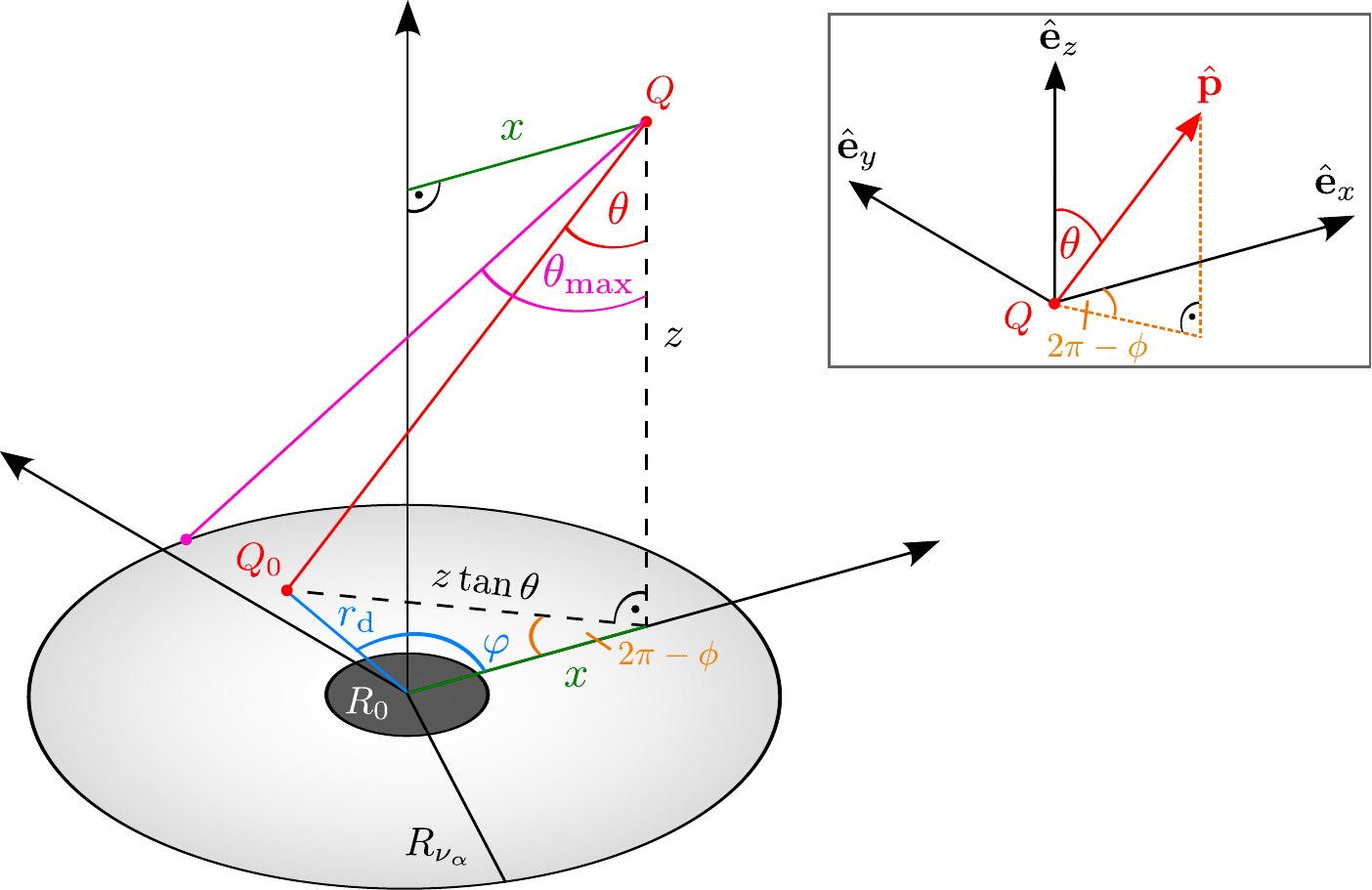}
\caption{Accretion disk with a central object (black hole or neutron star) located at the origin. 
In case of a black hole, the radius $R_{0}$ denotes the last stable orbit and defines the inner radius 
of the disk while for a neutron star, we set $R_{0} = 0$. 
The (outer) flavor dependent radii of the disk are $R_{\nu_{\alpha}}$ for neutrinos and $R_{\bar{\nu}_{\alpha}}$ for antineutrinos, respectively. 
The values used in our emission model are based on detailed simulations of a neutron star merger remnant \cite{Perego:2014} 
and are given in Table~\ref{table:neutrino_surface_radii}.
The polar angle $\theta$ and the azimuthal angle $\phi$ 
describe the direction of the neutrino momentum $\mathbf{p}$. 
Neutrinos are emitted from a point $Q_{0}$ located in the disk plane at a distance $r_{\mathrm{d}}$ 
from the origin with a relative angle $\varphi$ with respect to the 
positive $x$-axis and propagate to a point $Q$ located at a distance $x$ from the $z$-axis 
and a vertical distance $z$ from the equatorial plane.
}
\label{disk-model}
\end{figure*}

At any point $\mathbf{Q}(x,0,z)$ on the $x$-$z$ 
plane, for a neutrino which is emitted from 
a point $\mathbf{Q}_0$ on the disk and
passes through $\mathbf{Q}$, its momentum direction 
$\mathbf{\hat p}$ can be specified by a polar angle
$\theta$ and an azimuthal angle $\phi$ in the spherical 
coordinate system (see Fig.~\ref{disk-model}):
\begin{equation} \label{direction-1}
\mathbf{\hat{p}} = (\sin{\theta} \cos{\phi}, \sin{\theta} \sin{\phi}, \cos{\theta}),
\end{equation}
or by the polar coordinates $r_{\mathrm{d}}$ and $\varphi$
of the emission point
$\mathbf{Q}_0(r_{\mathrm{d}},\varphi)$ on the disk:
\begin{equation} \label{direction-2}
\mathbf{\hat{p}} = \dfrac{\mathbf{Q} - \mathbf{Q}_{0}}{\Delta^{1/2}},
\end{equation}
where
\begin{equation}
\Delta \equiv \vert \mathbf{Q} - \mathbf{Q}_{0} \vert^{2} = x^{2} + r_{\mathrm{d}}^{2} - 2 x r_{\mathrm{d}} \cos{\varphi} + z^{2}.
\end{equation}
A comparison of Eq.~\eqref{direction-1} and 
Eq.~\eqref{direction-2} yields the following coordinate
transformations:
\begin{equation}
\label{eq:coord-transformations}
\begin{split}
\cos{\theta} & = \dfrac{z}{\Delta^{1/2}}, \qquad \sin{\theta} = \dfrac{(\Delta - z^{2})^{1/2}}{\Delta^{1/2}},
\\
\cos{\phi} & = \dfrac{x - r_{\mathrm{d}} \cos \varphi}{\Delta^{1/2} \sin \theta}, \qquad \sin \phi = - \dfrac{r_{\mathrm{d}} \sin \varphi}{\Delta^{1/2} \sin \theta}
\end{split}
\end{equation}
with
$\varphi \in [0, 2 \pi]$ and $r_{\mathrm{d}} \in [R_{0}, R_{\nu}]$.
We note that 
the determinant of the Jacobian $J \equiv \partial( \cos \theta, \phi ) / \partial (r_{\mathrm{d}}, \varphi)$ turns out to be
\begin{align}
\mathrm{det} \, J & = \dfrac{\partial (\cos \theta)}{\partial r_{\mathrm{d}}} \dfrac{\partial \phi}{\partial \varphi} - \dfrac{\partial (\cos \theta) }{\partial \varphi} \dfrac{\partial \phi}{\partial r_{\mathrm{d}}}
\\
\label{eq:Jacobi_det_2} & = - \dfrac{z r_{\mathrm{d}}}{\Delta^{3/2}},
\end{align}
such that $\mathrm{d}(\cos \theta) \mathrm{d}\phi = \vert \mathrm{det} \, J \vert \, \mathrm{d}r_{\mathrm{d}} \mathrm{d}\varphi$ holds. 

Now, we consider another neutrino with momentum direction 
\begin{equation}
\mathbf{\hat{p}'} = (\sin \theta' \cos \phi', \sin \theta' \sin \phi', \cos \theta').
\end{equation}
The cosine of the scattering angle, ${\cos \Theta_{\mathbf{p} \mathbf{p'}} \equiv \mathbf{\hat{p}} \cdot \mathbf{\hat{p}}'}$, 
between the two neutrinos is then given by:
\begin{equation} \label{eq:scattering-angle}
\begin{split}
\cos \Theta_{\mathbf{p} \mathbf{p'}} & = \cos \theta \cos \theta' 
\\
& \quad + \sin \theta \sin \theta' \left( \cos \phi \cos \phi' + \sin \phi \sin \phi' \right) .
\end{split}
\end{equation}
If we make use of the above transformations we find:
\begin{equation} \label{eq:scattering-angle_transformed}
\begin{split}
\cos \Theta_{\mathbf{p} \mathbf{p'}} & = \dfrac{z \cos \theta'}{\Delta^{1/2}} + \dfrac{x \sin \theta' \cos \phi'}{\Delta^{1/2}}
\\
& - \dfrac{r_{\mathrm{d}} \sin \theta' \cos \phi' \cos \varphi}{\Delta^{1/2}} - \dfrac{r_{\mathrm{d}} \sin \theta' \sin \phi' \sin \varphi}{\Delta^{1/2}}.
\end{split}
\end{equation}

For neutrinos emitted isotropically from 
any point on the disk, the differential neutrino number
density in Eq.~\eqref{eq:Hamiltonian-nu-nu} is given by:
\begin{equation} \label{eq:diff_nu_density}
\mathrm{d} n_{\nu_{\underline{\alpha}}} \equiv \mathrm{d} n_{\nu_{\underline{\alpha}}}(\mathbf{p}) = j_{\nu_{\underline{\alpha}}}(E) \mathrm{d} E \mathrm{d} \Omega_{\nu_{\alpha}},
\end{equation}
where $E = \vert \mathbf{p} \vert \equiv p$,
$\mathrm{d} \Omega_{\nu_{\alpha}} \equiv \mathrm{d} \phi
\mathrm{d}(\cos \theta)$ and $j_{\nu_{\underline{\alpha}}}$
denotes the neutrino number flux per unit energy per solid 
angle for which we assume a Fermi-Dirac shape (see Appendix~\ref{app:fluxes})\footnote{Note that we divide by a factor of $2$, since $L_{\nu}$ corresponds to the total luminosity while we need 
the luminosity of only one hemisphere.}:
\begin{equation} \label{eq:number_flux}
j_{\nu_{\underline{\alpha}}}(E) = \dfrac{F_{\nu_{\underline{\alpha}}}}{2 \pi} f_{\nu_{\underline{\alpha}}}(E).
\end{equation}
Here, $F_{\nu_{\underline{\alpha}}} =
L_{\nu_{\underline{\alpha}}} / (\pi R_{\nu_{\alpha}}^{2}
\langle E_{\nu_{\underline{\alpha}}} \rangle)$ 
corresponds to the neutrino number flux at the neutrino
emitting surface and $f_{\nu_{\underline{\alpha}}}$ denotes 
the normalized Fermi-Dirac energy distribution function
corresponding to the right hand site of 
Eq.~\eqref{Eq:Fermi-Dirac-energy-distribution} with 
$T = T_{\nu_{\underline{\alpha}}}$.

Inserting Eqs.~\eqref{eq:diff_nu_density} 
and~\eqref{eq:number_flux} into Eq.~\eqref{eq:Hamiltonian-nu-nu}, 
we rewrite the self-interaction Hamiltonian as:
\begin{widetext}
\begin{equation} \label{eq:Hamiltonian_nu_nu_disk_general} 
\begin{split}
\mathsf{H}_{\nu \nu}(\mathbf{p}', \mathbf{Q}_{0}, r) = \dfrac{\sqrt{2} \mathrm{G}_{\mathrm{F}}}{2 \pi} \sum_{\alpha = e, \mu, \tau} \int_{0}^{\infty} \mathrm{d} E & 
\left( \int_{\Omega_{\nu_{\alpha}}} \mathrm{d} \Omega \, (1 - \cos{\Theta_{\mathbf{p} \mathbf{p}'}}) \, F_{\nu_{\underline{\alpha}}} \rho_{\nu_{\underline{\alpha}}}(\Omega, \mathbf{Q}_{0}, E, r) f_{\nu_{\underline{\alpha}}}(E) \right. 
\\
- & \left. \int_{\Omega_{\bar{\nu}_{\alpha}}} \mathrm{d} \Omega \, (1 - \cos{\Theta_{\mathbf{p} \mathbf{p}'}}) \, F_{\bar{\nu}_{\underline{\alpha}}} \bar{\rho}_{\bar{\nu}_{\underline{\alpha}}}(\Omega, \mathbf{Q}_{0}, E, r) f_{\bar{\nu}_{\underline{\alpha}}}(E) \right),
\end{split}
\end{equation}
\end{widetext}
where the angular integration is performed with the 
corresponding limits $\Omega_{\nu_{\alpha} (\bar{\nu}_{\alpha})}$ for
neutrinos and antineutrinos, respectively.
\end{subsection}

\begin{subsection}{Single-trajectory versus single- and multi-angle approximations} 
\label{sec:single_trajectory}
In order to follow the evolution,
one should solve Eqs.~\eqref{eq:eom_1}
and~\eqref{eq:eom_2} for all neutrinos with
different $\mathbf{p}$ and $\mathbf{Q}_0$ simultaneously since $\mathsf{H}_{\nu\nu}$ couples them. This is computationally extremely demanding as we will discuss shortly. Instead of solving the full problem we employ a "single-trajectory" approximation which consists in making the assumption that
in Eq.~\eqref{eq:Hamiltonian_nu_nu_disk_general} 
the density matrix is given by
\begin{equation}
\label{eq:sa}
\rho_{\nu_{\underline{\alpha}}}(\Omega, \mathbf{Q}_{0}, E, r) = \rho_{\nu_{\underline{\alpha}}}(\mathbf{p}^\prime,\mathbf{Q}_{0}, E, r),
\end{equation}
that is, it does not dependent on the angular variables. In other words, we   
suppose that at every point of a given
neutrino trajectory with momentum $\mathbf{p}^\prime$, all
neutrino states contributing to $\mathsf{H}_{\nu\nu}(\mathbf{p}^\prime, \mathbf{Q}_{0}, r)$ have the same flavor evolution as the one with
$\mathbf{p}^\prime$.
The approximation given by Eq.~\eqref{eq:sa} was already used in \cite{Malkus:2012, Malkus:2014, Malkus:2016}.
We emphasize that this approach reduces to the ``single-angle approximation'' used in the supernova context
for a spherically-symmetric system, such as the supernova bulb-model \cite{bulb-model}.
Note that the "multi-angle approximation" in the supernova bulb-model corresponds to retaining
also the $\theta$ emission angle dependence in the self-interaction Hamiltonian. 

At present, no simulations of neutrino flavor evolution in binary neutron star mergers
exist where Eqs.~\eqref{eq:eom_1} and~\eqref{eq:eom_2} are solved without making the
assumption Eq.~\eqref{eq:sa}. This is due to the fact that it may require computational capabilities beyond the current available resources. In fact, a multi-angle calculation in the supernova neutrino bulb-model, which
only evolves the flavor content in the radial coordinate with one explicit emission angle
variable, requires $\sim\mathcal{O}(10^3)$ CPU hours \cite{bulb-model}. Numerical convergence
requires a large number of angle bins, typically of the order of $10^{3}$-$10^{4}$~\cite{bulb-model, Sarikas:2012}.
In the disk case, performing a full calculation that preserves the initial symmetry of the system is much more complex than in the supernova bulb-model and requires to evolve the
flavor content in both $x$ and $z$ coordinates with three explicit variables: $r_{\mathrm{d}}$, $\theta$, $\phi$ specifying the
emission location and angles, respectively.
As for the possible effect of going from the ``single-trajectory'' approximation to 
the full flavor calculation, one can speculate that this will introduce decoherence
in the flavor evolution as in the supernova context multi-angle simulations have shown that
occurs \cite{Dasgupta:2009, Duan:2011}. Therefore the results presented here can be considered as an upper limit
for the effects of flavor evolution on the capture rates since we expect that decoherence
is likely to reduce them.

Now, under the single-trajectory approximation,
Eqs.~\eqref{eq:eom_1} and~\eqref{eq:eom_2} can be 
solved for each density matrix
$\rho_{\nu_{\underline{\alpha}}} =
\rho_{\nu_{\underline{\alpha}}}(\theta_{0}, \phi_{0}, \mathbf{Q}_0, E, r)$, and the angular integration yields a geometric factor 
\begin{equation} \label{eq:geometric_factor_general}
G_{\nu_{\alpha}}(\theta_{0}, \phi_{0}, \mathbf{Q}_0, r) =
\int_{\Omega_{\nu_{\alpha}}} \mathrm{d} \Omega \, (1 -
\cos{\Theta_{\mathbf{p} \mathbf{p}'}}),
\end{equation}
whose explicit form is described in 
Appendix~\ref{app:geometric_factor}. Here, we fixed the emission angles $\theta_{0}$ and $\phi_{0}$ 
describing the direction of momentum $\mathbf{p}'$.
In Fig.~\ref{Geometric_factor_plot} we show typical sizes
for those factors.
The ratio $G_{\nu_{e}} / G_{\bar{\nu}_{e}}$ increases as a function of distance.
\begin{figure}[!htbp]
 \centering
   \includegraphics[width=0.45\textwidth]{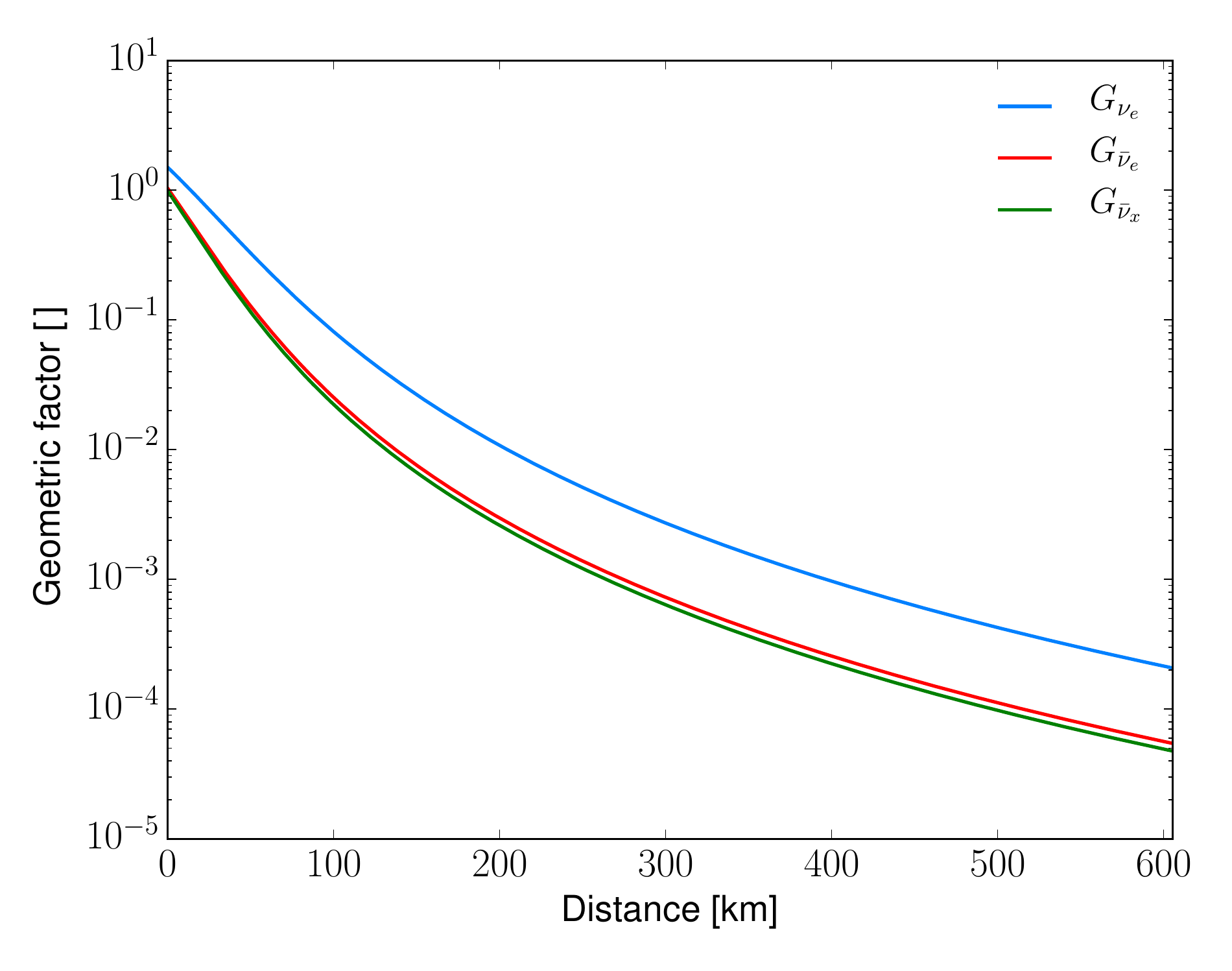}
   \caption{Geometric factors as function of the distance from the emission point $x_{0} = 10 \, \mathrm{km}$, $z_{0} = 30 \, \mathrm{km}$. 
   The emission angles correspond to $\theta = 20^{\circ}$ and $\phi = 0^{\circ}$ while the disk radii are taken from Table~\ref{table:neutrino_surface_radii}.}
   \label{Geometric_factor_plot}
\end{figure}

In the following the explicit reference to the angular labels and the emission point will be omitted and 
the density matrices will be denoted just by $\rho_{\nu_{\underline{\alpha}}}(E, r)$ 
for notational convenience. Finally, the Hamiltonian 
Eq.~\eqref{eq:Hamiltonian_nu_nu_disk_general} can be expressed in the compact form:
\begin{widetext}
\begin{equation} \label{eq:Hamiltonian-nu-nu_disk}
\begin{split}
\mathsf{H}_{\nu \nu}(\theta_{0}, \phi_{0}, \mathbf{Q}_{0}, r) = \sqrt{2} G_{\mathrm{F}} \sum_{\alpha = e, \mu, \tau} \int_{0}^{\infty} \mathrm{d}E & \, \left( \rho_{{\nu}_{\underline{\alpha}}}(E, r) j_{{\nu}_{\underline{\alpha}}}(E)
G_{\nu_{\alpha}}(\theta_{0}, \phi_{0}, \mathbf{Q}_{0}, r) \right.
\\
& \left. - \bar{\rho}_{{\bar{\nu}}_{\underline{\alpha}}}(E, r) j_{{\bar{\nu}}_{\underline{\alpha}}}(E) G_{\bar{\nu}_{\alpha}}(\theta_{0}, \phi_{0}, \mathbf{Q}_{0}, r) \right).
\end{split}
\end{equation}
\end{widetext}
\end{subsection}
\end{section}

\begin{section}{Trajectory dependence of the unoscillated potentials} \label{sec:Trajectory_dependence}

Flavor transformation through matter-neutrino resonances is 
an MSW-like phenomenon. 
Its occurrence is due to the almost cancellation of
the matter and the neutrino self-interaction
potentials, that have opposite signs. 
This condition is met for most neutrino trajectories 
in our model, since the self-interaction potential starts negative
due to the dominating electron antineutrino fluxes 
(Table~\ref{table:emission_parameters}).
However, for significant flavor conversions to occur, this nearly cancellation is not sufficient.
Similar to the MSW case, it is the adiabaticity of the evolution that determines the flavor conversion efficiency \cite{Wu-MNR:2016, Vaananen-MNR:2016, Malkus:2014} and depends, beside the mixing parameters 
and the neutrino energy, on the matter profiles and their gradients.

We choose different neutrino emission
points $(x_0,z_0)$ on the neutrino surfaces and compute their 
flavor evolution along trajectories that pass through two different reference points
$(x_{\mathrm{ref}}, z_{\mathrm{ref}})$ as given in
Tables~\ref{Table_TB100_1} and~\ref{Table_TB100_2}. 
\begin{table}[!htbp]    
\center
\caption{Parameters that specify the neutrino trajectories: emission coordinates at the neutrino surface $(x_{0}, z_{0})$
and emission angle $\theta_{0}$. The last column shows the distance between the emission point and the reference point
$x_{\mathrm{ref}} = 293 \, \mathrm{km}$, $z_{\mathrm{ref}} = 313 \, \mathrm{km}$.}
\label{Table_TB100_1}
\begin{ruledtabular}
\begin{tabular}{cccccc}
Trajectory & $x_{0}$ [km] & $z_{0}$ [km] & $\theta_{0}$ [$^{\circ}$] & Distance [km] \hfill \\ \hline
1a & $10$ & $30$ & $45.0$ & $400$
\\
1b & $-10$ & $30$ & $47.0$ & $415$
\\
1c & $-35$ & $25$ & $48.7$ & $436$
\\
1d & $50$ & $30$ & $40.7$ & $373$
\\
\end{tabular}
\end{ruledtabular}
\end{table}

\begin{table}[!htbp]    
\center
\caption{Same as Table~\ref{Table_TB100_1} for the reference point 
$x_{\mathrm{ref}} = 74 \, \mathrm{km}$, $z_{\mathrm{ref}} = 206 \, \mathrm{km}$.}
\label{Table_TB100_2}
\begin{ruledtabular}
\begin{tabular}{cccccc}
Trajectory & $x_{0}$ [km] & $z_{0}$ [km] & $\theta_{0}$ [$^{\circ}$] & Distance [km] \hfill \\ \hline
2a & $10$ & $30$ & $20.0$ & $187$
\\
2b & $-10$ & $30$ & $25.5$ & $195$
\\
2c & $-35$ & $25$ & $31.1$ & $211$
\\
2d & $50$ & $30$ & $7.8$ & $178$
\\
\end{tabular}
\end{ruledtabular}
\end{table}

To simplify the discussion, 
we implicitly assume that $\phi_{0}=0^{\circ}$, 
i.e., we do not explore the trajectory dependence on $\phi_{0}$.
The two reference points are chosen to have a
temperature $T=8 \, \mathrm{GK}$ in different regions of the wind that give rise to different nucleosynthesis outcomes \cite{Martin:2015}.
Point 1 (2) lies on $\approx 43^\circ$ $(20^{\circ})$ from the 
$z$-axis and is $\approx 429 \, (219) \, \mathrm{km}$ away from the center 
of the MNS.
Figure~\ref{Fig:trajectories} shows the chosen neutrino 
emission points
on the disk and the reference points 1 and 2 along with the
density structure of the remnant.

For the mixing parameters we take values compatible with current best-fit values \cite{Olive}:
$\Delta m_{\mathrm{21}}^{2} = 7.59 \times 10^{-5} \, \mathrm{eV}^{2}$, 
$\vert \Delta m_{\mathrm{31}}^{2} \vert = 2.43 \times 10^{-3} \, \mathrm{eV}^{2}$, 
$\sin^{2}( 2 \theta_{12} ) = 0.87$, $\theta_{13} = 0.15$, $\theta_{23} = \pi / 4$.
We use $\delta_{\mathrm{CP}} = 0$ for the CP-violating Dirac phase.
Since the neutrino mass hierarchy is still unknown \cite{Hierarchy}, 
we consider both, the normal mass hierarchy (NH), i.e., $\Delta m_{31}^{2} > 0$,
and inverted mass hierarchy (IH), i.e., $\Delta m_{31}^{2} < 0$.

Before presenting the numerical results we introduce the unoscillated potentials 
associated with the matter and the neutrino self-interaction terms of the Hamiltonian, 
as done in \cite{Malkus:2012}.
The point where the sum of these two quantities cancel already gives an idea in which spatial region MNR are expected to occur.

As a measure for the matter strength, we use the refractive energy shift between $\nu_{e}$ and $\nu_{x}$ 
and define the neutrino-matter potential as follows:
\begin{equation} \label{matter_potential}
\lambda(r) \equiv \sqrt{2} \mathrm{G}_{\mathrm{F}} n_{e}(r).
\end{equation}
For the neutrino self-interaction, the individual
contributions from $\nu_{x}$ exactly cancel at any point when
flavor transformations have not occurred yet, since we assume the same 
initial fluxes and surface sizes. Hence, it is convenient to
introduce the unoscillated neutrino self-interaction potential as follows:
\begin{widetext}
\begin{equation} \label{neutrino_potential}
\mu(r) \equiv 
\dfrac{\sqrt{2} \mathrm{G}_{\mathrm{F}}}{2 \pi^{2}} \left \lbrace \dfrac{L_{\nu_{\underline{e}}}}{\langle E_{\nu_{\underline{e}}} \rangle R_{\nu_{e}}^{2}} G_{\nu_{e}}(\theta_{0}, \phi_{0}, r) 
- \dfrac{L_{\bar{\nu}_{\underline{e}}}}{\langle E_{\bar{\nu}_{\underline{e}}} \rangle R_{\bar{\nu}_{e}}^{2}} G_{\bar{\nu}_{e}}(\theta_{0}, \phi_{0}, r) \right \rbrace.
\end{equation}
\end{widetext}
Note that the scales set by the vacuum potentials $\omega \equiv \Delta m^{2} / (2 E)$ 
($\vert \omega_{31} \vert \approx \, 0.4 \, \mathrm{km}^{-1}$ and $\omega_{21} \approx \, 0.01 \, \mathrm{km}^{-1}$ for a $15 \, \mathrm{MeV}$ neutrino)
are typically well below $\lambda(r)$ and |$\mu(r)$|.

\begin{figure}[!htbp]
 \centering
 \includegraphics[width=0.5\textwidth]{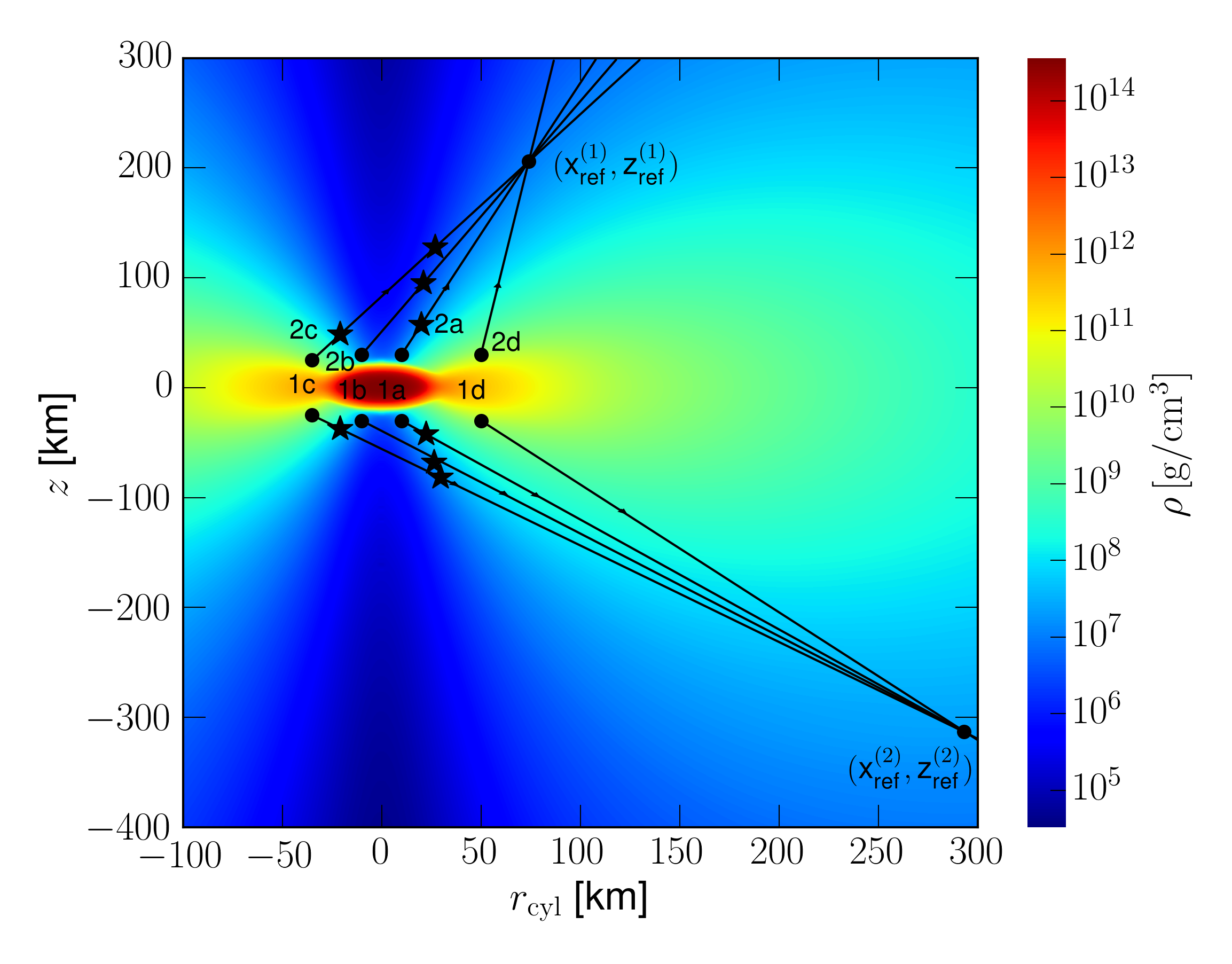} 
 \caption{Density profile at $100 \, \mathrm{ms}$ after the merger as a function of cylindrical coordinates $z$ and $r_{\mathrm{cyl}}$. 
  The neutrino trajectories, shown by the black lines, are specified in Tables~\ref{Table_TB100_1} and \ref{Table_TB100_2}. 
 We mirrored the trajectories of Table~\ref{Table_TB100_1} for clarity. The two reference points $(x_{\mathrm{ref}}^{(i)}, z_{\mathrm{ref}}^{(i)})$, $i = 1, 2$,
 are located at a temperature $T \approx 8 \, \mathrm{GK}$ and chosen as representative locations interesting for nucleosynthesis.
 The points where the matter and unoscillated neutrino self-interaction potentials cancel are marked with $\bigstar$. } 
 \label{Fig:trajectories}
\end{figure}

Let us discuss the trajectories listed in 
Tables~\ref{Table_TB100_1} and~\ref{Table_TB100_2} 
taken as representatives over the large set we explored. 
In the top panels of Figs.~\ref{Fig:results1} and~\ref{Fig:results2} we present the matter and unoscillated neutrino self-interaction potentials 
Eqs.~\eqref{matter_potential} and~\eqref{neutrino_potential} for these trajectories shown in Fig.~\ref{Fig:trajectories}. In addition we show the vacuum potentials $\omega_{21}$ and $\vert \omega_{31} \vert$ for $5 \, \mathrm{MeV}$ (anti)neutrinos.
To guide the eye we highlight the region around the location of the reference point with a color band.
The initial points of 1a (2a) and  1b (2b) are located in the low density polar region, 
where the matter potential $\lambda$ is around $3 \times 10^{3} \, \mathrm{km}^{-1}$ ($\rho_{\mathrm{matt}} \sim 2 \times 10^{7} \, \mathrm{g} \, \mathrm{cm}^{-3}$, $Y_{e} \sim 0.39$); 
while trajectory 1c (2c) starts in a low density regime of the wind, where the matter potential is much stronger $9 \times 10^{5} \, \mathrm{km}^{-1}$ 
($\rho_{\mathrm{matt}} \sim 7 \times 10^{9} \, \mathrm{g} \, \mathrm{cm}^{-3}$, $Y_{e} \sim 0.31$).
The starting point of trajectory 1d (2d) is located deeper inside the wind where $\lambda \sim 1 \times 10^{6} \, \mathrm{km}^{-1}$
($\rho_{\mathrm{matt}} \sim 2 \times 10^{10} \, \mathrm{g} \, \mathrm{cm}^{-3}$, $Y_{e} \sim 0.18$).
\begin{figure*}[tp!]
\centering
\subfloat[1a (NH): $x_{0} = 10 \, \mathrm{km}$, $z_{0} = 30 \, \mathrm{km}$, $\theta_{0} = 45.0^{\circ}$]{\includegraphics[width=0.5\textwidth]{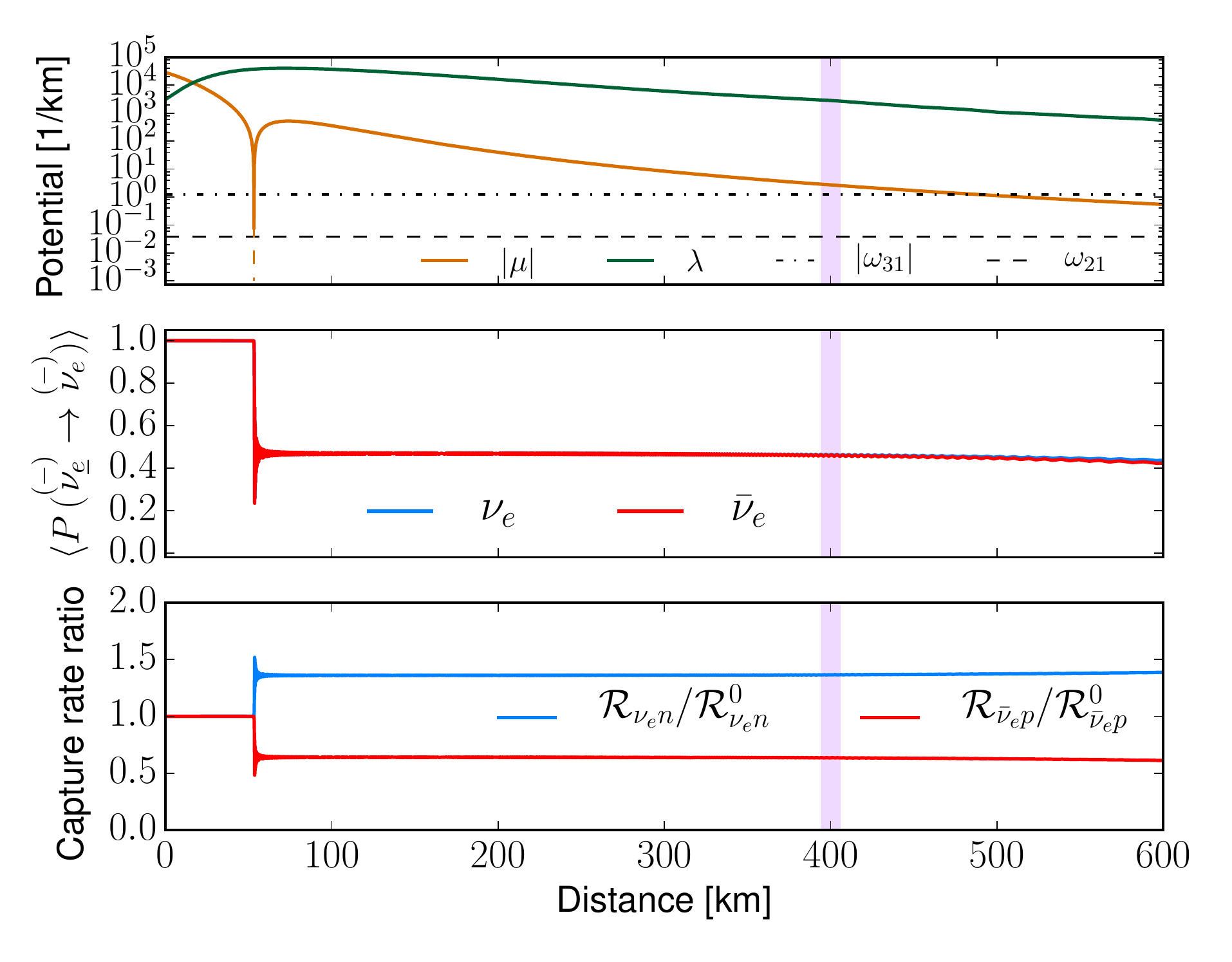}} 
\hfill
\subfloat[1b (NH): $x_{0} = -10 \, \mathrm{km}$, $z_{0} = 30 \, \mathrm{km}$, $\theta_{0} = 47.0^{\circ}$]{\includegraphics[width=0.5\textwidth]{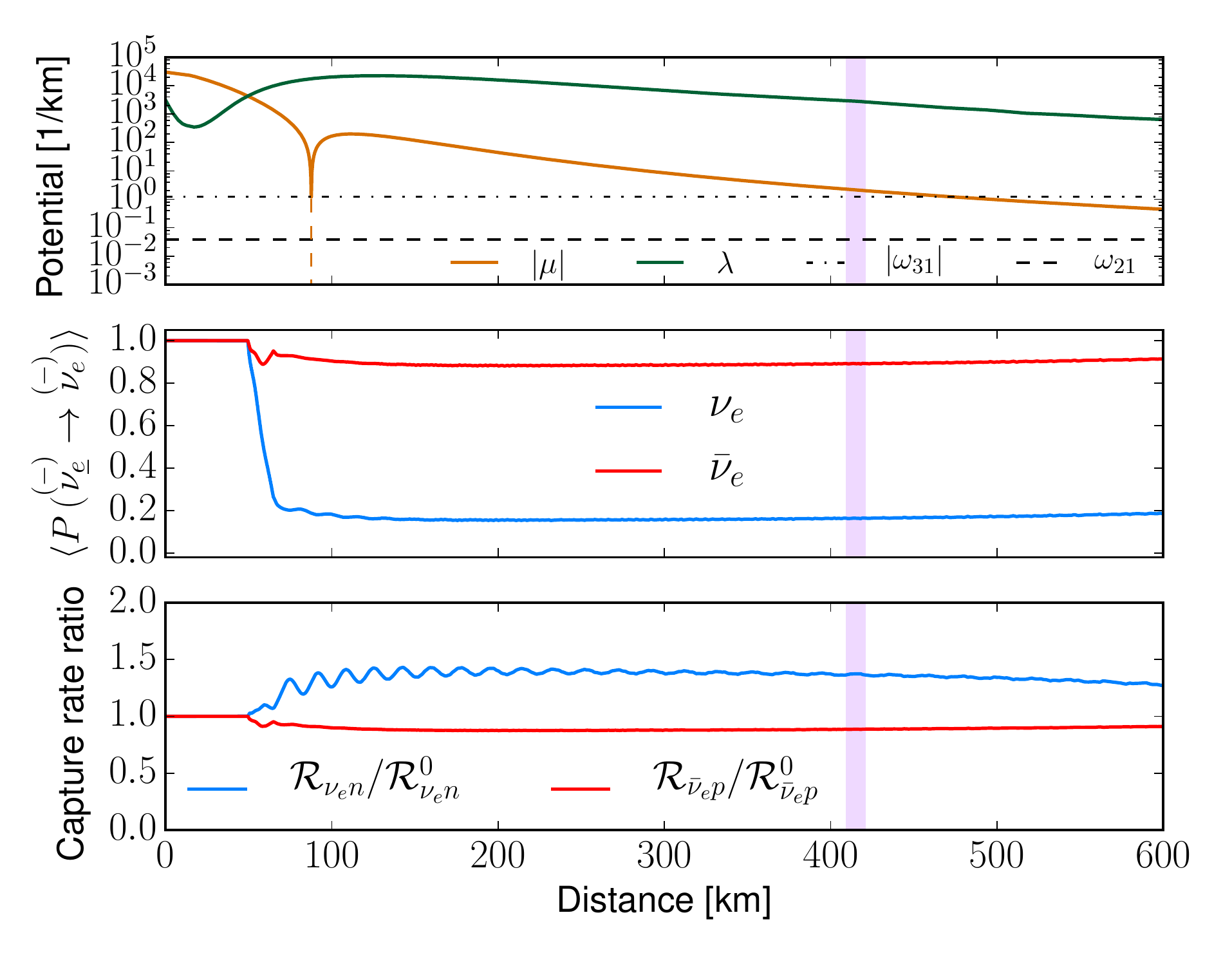}}

\subfloat[1c (NH): $x_{0} = -35 \, \mathrm{km}$, $z_{0} = 25 \, \mathrm{km}$, $\theta_{0} = 48.7^{\circ}$]{\includegraphics[width=0.5\textwidth]{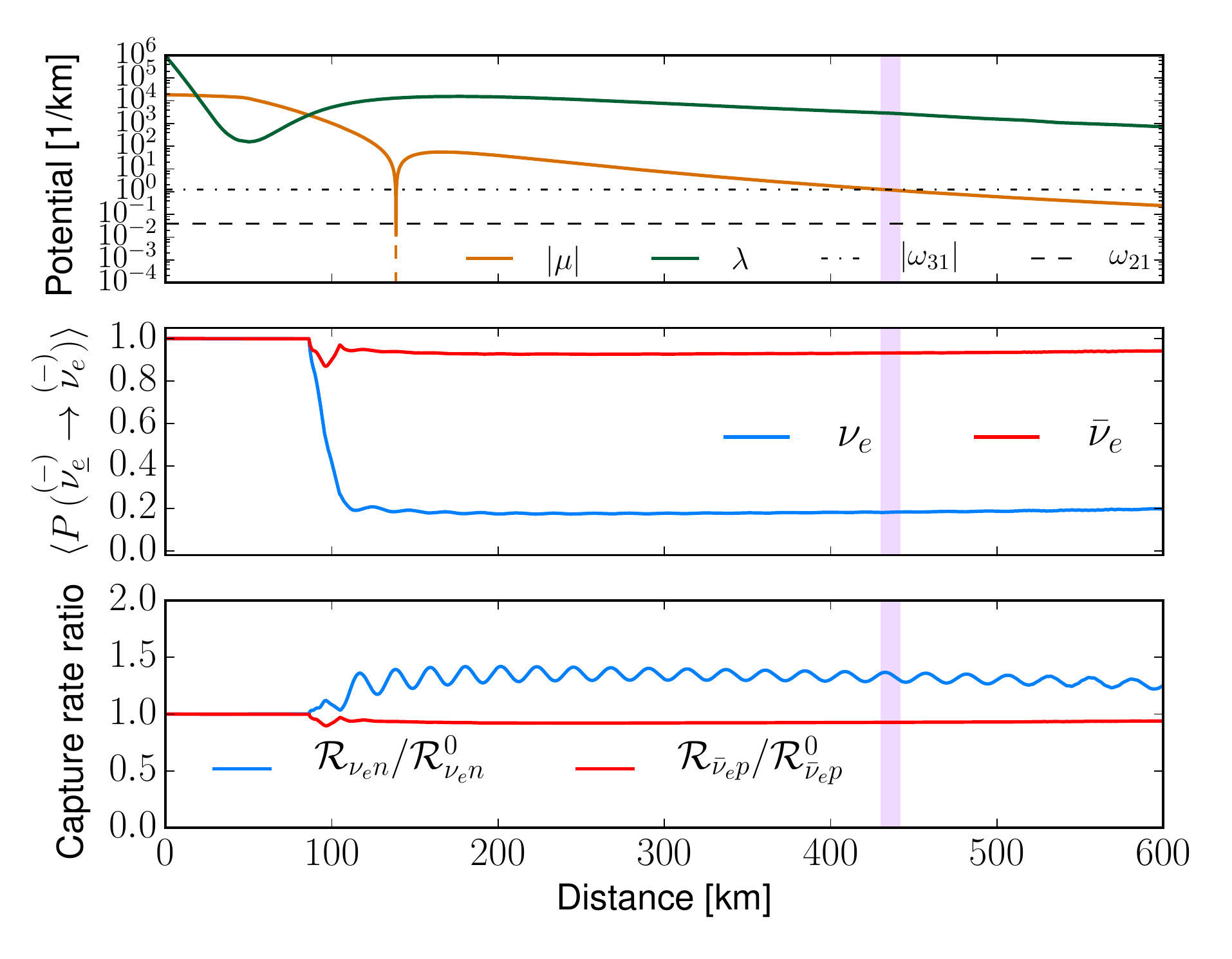}}
\hfill
\subfloat[1d (NH): $x_{0} = 50 \, \mathrm{km}$, $z_{0} = 30 \, \mathrm{km}$, $\theta_{0} = 40.7^{\circ}$]{\includegraphics[width=0.5\textwidth]{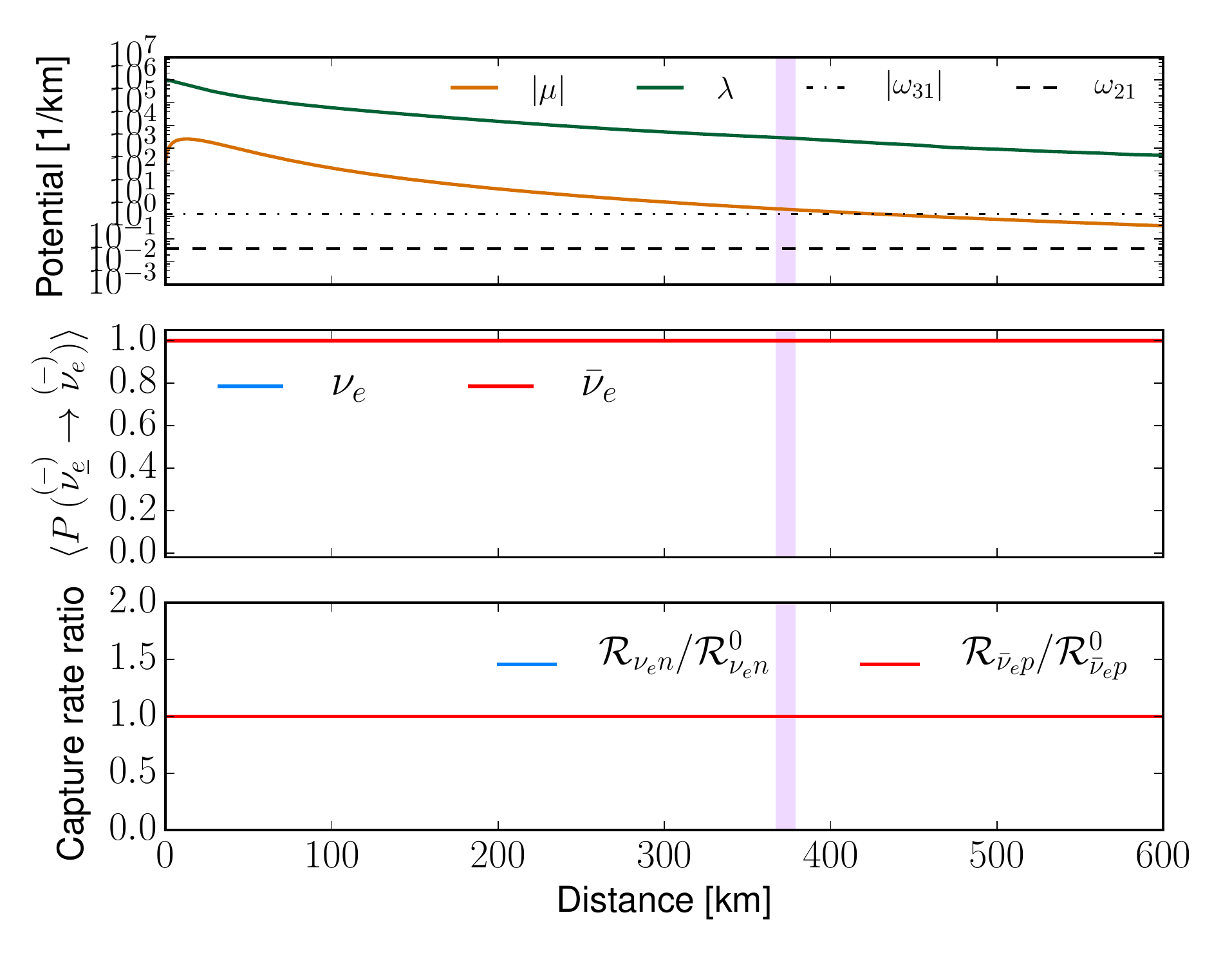}} 

\caption{Results for
selected trajectories specified in Table~\ref{Table_TB100_1}
as functions of the distance for the normal mass hierarchy. The location of the reference point is highlighted by a vertical color band.
Top panels: matter potentials ($\lambda$) and absolute values of the unoscillated neutrino potentials ($\vert \mu \vert)$ along with the vacuum potentials $\omega_{21}$ and $\vert \omega_{31} \vert$ for $5 \, \mathrm{MeV}$ (anti)neutrinos. 
The middle panels show the spectral averaged survival probabilities for the electron flavor. 
The blue curves correspond to $\nu_{e}$ and the red curves to $\bar{\nu}_{e}$, respectively. 
In the bottom panels, the ratios of capture rates per solid angle for electron neutrinos (blue) and antineutrinos (red) are presented.
}
\label{Fig:results1}
\end{figure*}
Neutrinos on their way on trajectories 1a initially experience an increasing matter potential. When the wind becomes 
more dilute, the potential decreases until they reach the reference point. 
In case of trajectories 1b and 2b, neutrinos will first pass the funnel above the MNS pole where the density is very low compared 
to the emission region. When it enters the wind region the density increases. 
Afterwards neutrinos proceed similarly like in 1a, i.e., 
they go through the dilute part of the wind (matter potential is decreasing) and arrive at the reference point. For trajectories 1c and 2c, 
neutrinos will first need to cover some distance through the dense part of the wind before entering the funnel. Afterwards, they propagate in an analogous way like in cases 1b and 2b.
The transition between wind and funnel leads to a rapid drop in the density which is clearly visible in Figs.~\ref{Fig:results1} 
and~\ref{Fig:results2}
for trajectories 1b, 1c, 2b and 2c.
A different behavior will be experienced by neutrinos following trajectories 1d, 2a, and 2d. 
They encounter a monotonically decreasing matter potential until they reach the reference point. 
\begin{figure*}[tp!]
\centering
\subfloat[2a (NH): $x_{0} = 10 \, \mathrm{km}$, $z_{0} = 30 \, \mathrm{km}$, $\theta_{0} = 20.0^{\circ}$]{\includegraphics[width=0.5\textwidth]{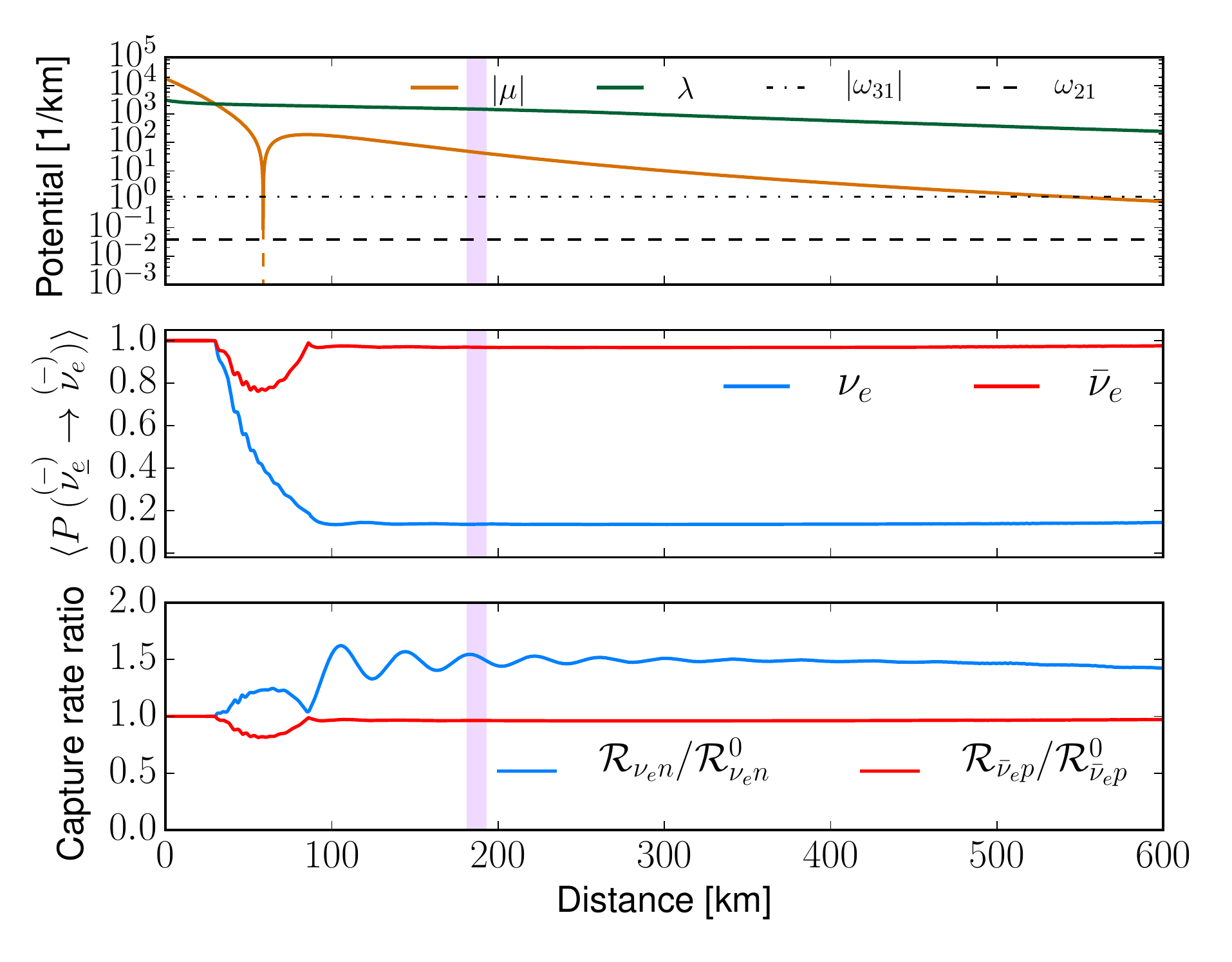}} 
\hfill
\subfloat[2b (NH): $x_{0} = -10 \, \mathrm{km}$, $z_{0} = 30 \, \mathrm{km}$, $\theta_{0} = 25.5^{\circ}$]{\includegraphics[width=0.5\textwidth]{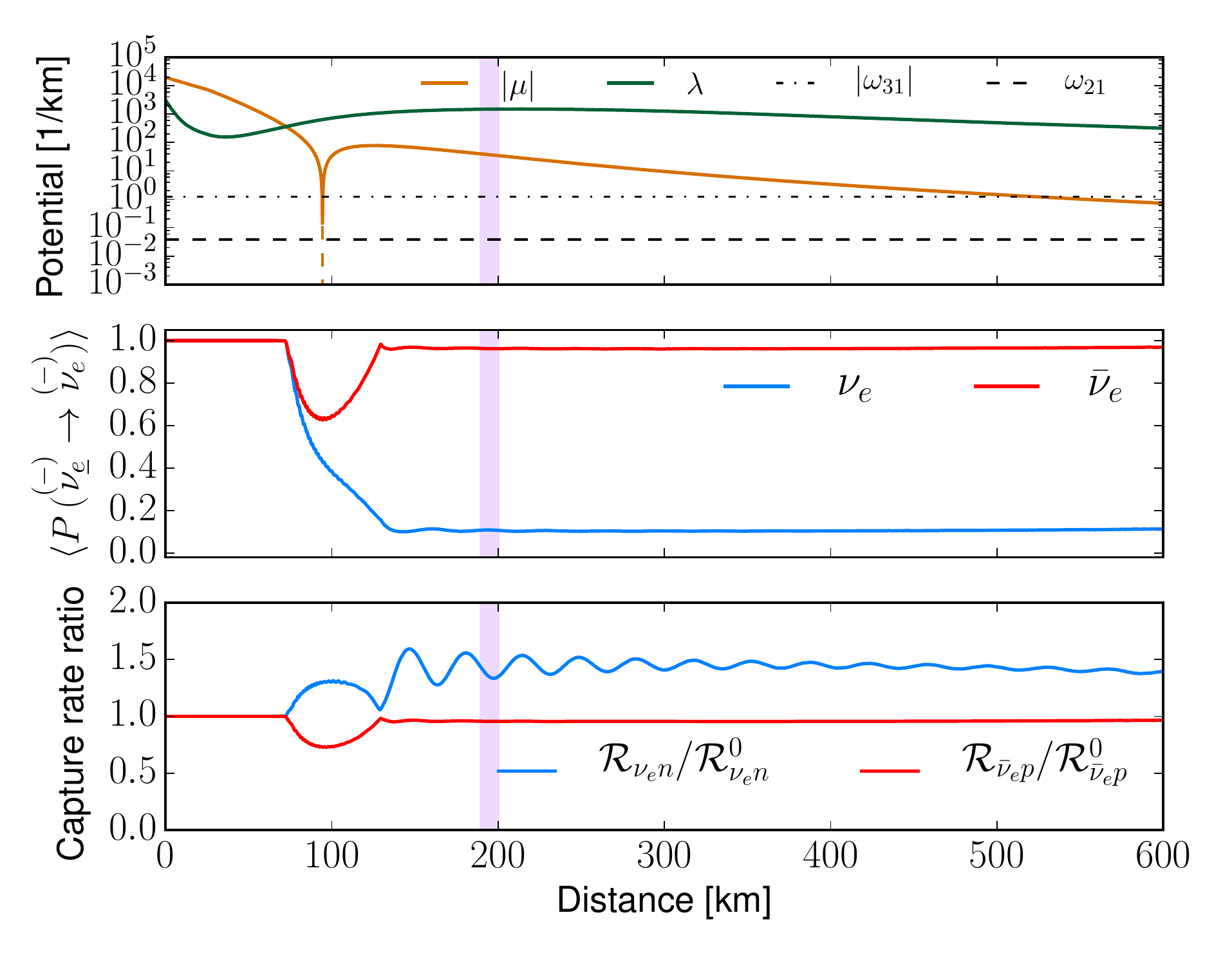}} 

\subfloat[2c (NH): $x_{0} = -35 \, \mathrm{km}$, $z_{0} = 25 \, \mathrm{km}$, $\theta_{0} = 31.1^{\circ}$]{\includegraphics[width=0.5\textwidth]{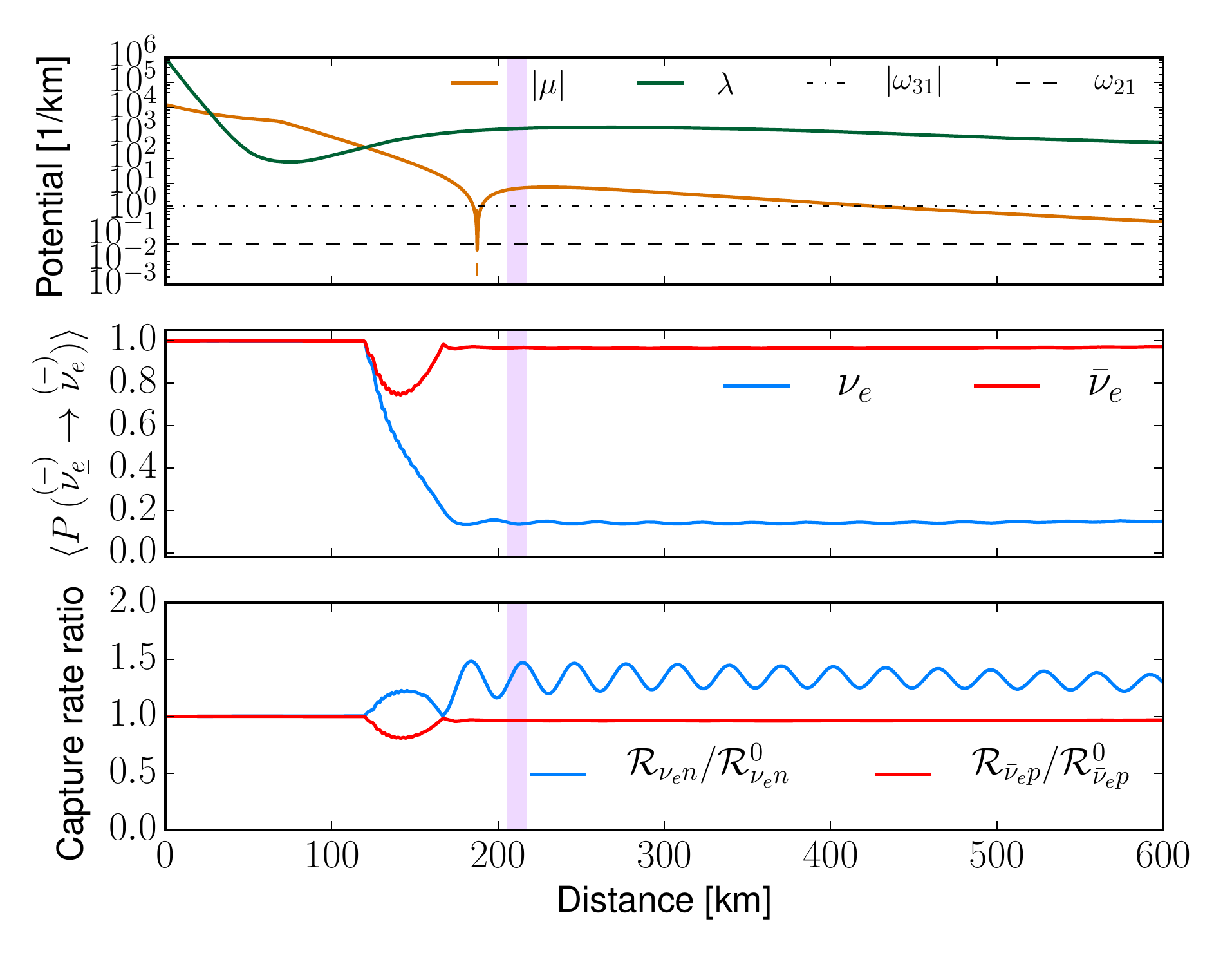}}
\hfill
\subfloat[2d (NH): $x_{0} = 50 \, \mathrm{km}$, $z_{0} = 30 \, \mathrm{km}$, $\theta_{0} = 7.8^{\circ}$]{\includegraphics[width=0.5\textwidth]{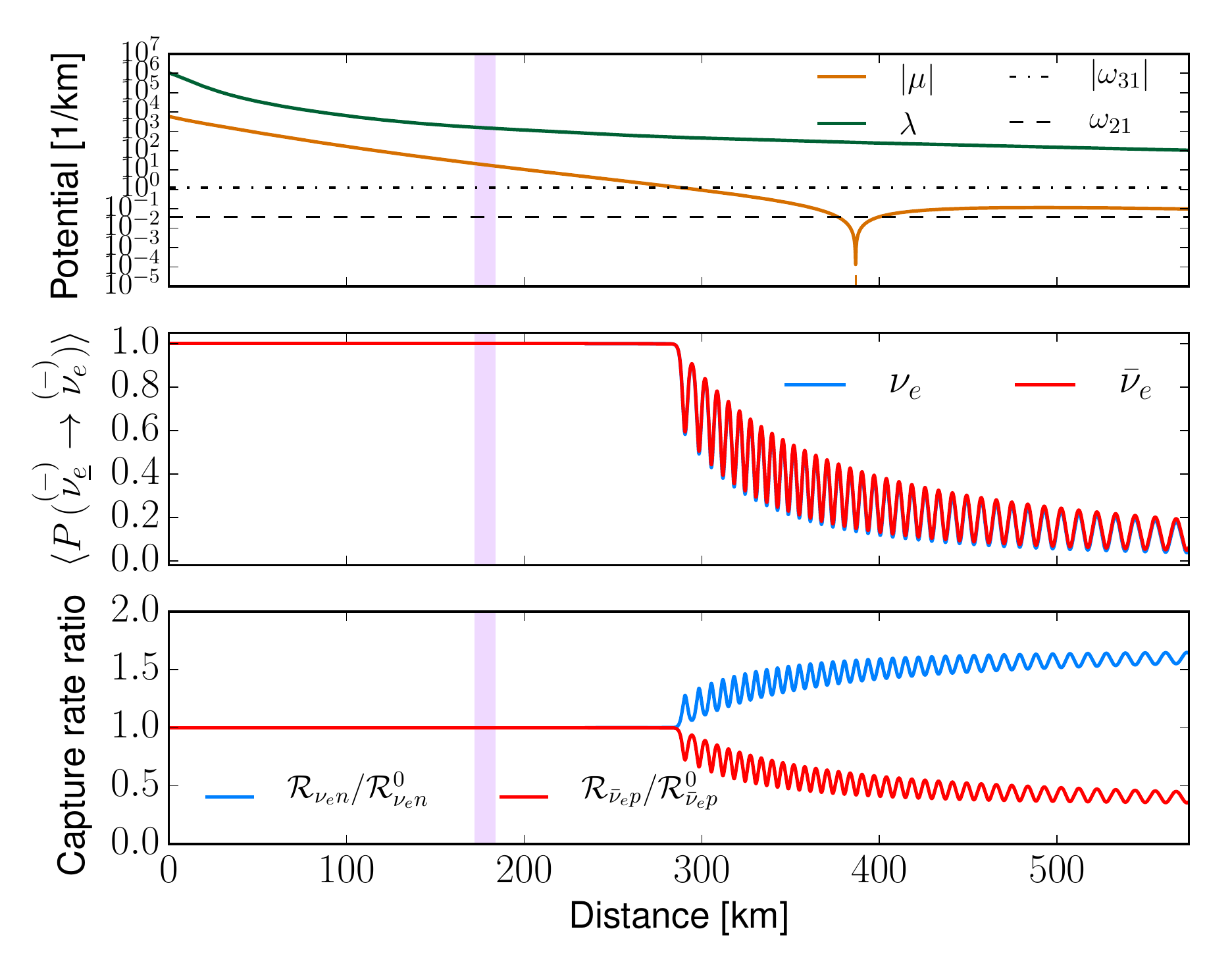}}

\caption{Same as Fig.~\ref{Fig:results1} for the trajectories specified in Table~\ref{Table_TB100_2}.}
\label{Fig:results2}
\end{figure*}

In the self-interaction potential, the relative contribution of $\nu_{e}$ and $\bar{\nu}_{e}$ changes as a function of distance,
due to the interplay of $j_{\nu_{\underline{\alpha}}}$ and $G_{\nu_{\underline{\alpha}}}$.
In particular this means that initially, the neutrino self-interaction potential is
negative, because it is dominated by the larger electron
antineutrino fluxes. 
Later, the fact that the neutrino surface for electron neutrinos is larger than that of electron antineutrinos, 
may lead to a change of sign in the self-interaction potential (as we will see in Sec.~\ref{sec:results_occurence_MNR}).
The absolute values of the unoscillated neutrino potentials $\vert \mu(r) \vert$ vary between $\sim 10^{4} \, \mathrm{km}^{-1}$ initially to $\sim 10^{-1} \, \mathrm{km}^{-1}$ at $600 \, \mathrm{km}$.
The relative magnitude of the matter and neutrino potentials and the possible presence of crossings will determine the flavor evolution as
we will see in Sec.~\ref{sec:results_occurence_MNR}. We find that the crossings happen at the edge of the funnel (see Fig. \ref{Fig:trajectories}).
This means that for neutrinos emitted around the central region and from the opposite side of the disk, their trajectories cross the funnel so that MNR may occur.

\end{section}

\begin{section}{Numerical Results} \label{sec:results}

Our goal is to show the trajectory dependence of flavor evolution for
neutrinos from the disk by presenting spectral-averaged flavor conversion probabilities (Sec.~\ref{sec:results_occurence_MNR}). 
As we will discuss we find a variety of flavor conversion behaviors.
Furthermore, we explore the potential impact on nucleosynthesis in the neutrino-driven wind by showing ratios of 
oscillated and unoscillated capture rates per solid angle (Sec.~\ref{sec:capture_rates}).
We present results based on hydrodynamical profiles obtained at 100 ms. 
We discuss possible variations with a different time snapshot (60 ms)
in Sec.~\ref{sec:timesteps}. 
Finally, we show the sensitivity of the flavor evolution when employing
different assumptions for the initial luminosities, or considering
uncertainties on the neutrino fluxes from simulations available in the literature.

In our calculations of the flavor evolution we assume that 
all (anti-)neutrinos are prepared in flavor eigenstates.
The flavor evolution of neutrinos with different energies 
is then followed by numerically solving Eqs.~(\ref{eq:eom_1}) 
and~(\ref{eq:eom_2}) with the Hamiltonian components given by 
Eqs.~(\ref{eq:Hamiltonian_vacuum}),
(\ref{eq:Hamiltonian_matter}), 
and~(\ref{eq:Hamiltonian-nu-nu_disk}) for a given 
trajectory with emission angle $\theta_{0}$.
We employed different discretizaton schemes to check for convergence of the results.

\begin{subsection}{Flavor conversion results and general behavior}  \label{sec:results_occurence_MNR} 

After obtaining the flavor evolution of neutrinos along
each trajectory, $\rho_{\nu_{\underline{\alpha}}}(E,r)$ and
$\rho_{\bar\nu_{\underline{\alpha}}}(E,r)$, we compute
spectral averages of the neutrino survival probability, i.e., 
\begin{equation} \label{eq:spectral_average}
\langle P(\nu_{\underline{e}} \to \nu_{e}) \rangle (r) = \int_{0}^{\infty} \mathrm{d} E \, f_{\nu_{\underline{e}}}(E) P(\nu_{\underline{e}} \to \nu_{e})(E, r).
\end{equation}
Notice that ${P(\nu_{\underline{\alpha}} \to \nu_\beta)=(\rho_{\nu_{\underline{\alpha}}})_{\beta\beta}}$ as 
defined in Sec.~\ref{sec:eom}.

As examples, we show the averaged survival probabilities of electron neutrinos and antineutrinos
for the trajectories 1a to 1d in the middle panels of Fig.~\ref{Fig:results1}
and 2a to 2d in Fig.~\ref{Fig:results2}.
These results are obtained in NH.

As can be seen from Figs.~\ref{Fig:results1} and~\ref{Fig:results2} (top panels), the structure of some profiles allows the unoscillated
potentials to cancel at more than one spatial location such as trajectories 1c and 2c.
We indicate these locations in Fig.~\ref{Fig:trajectories} for the trajectories defined in Tables~\ref{Table_TB100_1} and \ref{Table_TB100_2}.
If the resonance condition is fulfilled, flavor conversion only occurs
if the strength of the neutrino self-interaction\footnote{We remind that because of the dominance of the electron antineutrino 
number fluxes over the neutrino one, the neutrino self-interaction term 
starts negative.} is larger than the matter contribution ($\lambda<\vert \mu \vert$)
prior to it. If the matter term dominates the self-interaction term ($\lambda>\vert \mu \vert$),
before a cancellation point, the resonances are extremely 
non-adiabatic and nearly no flavor transformation can 
happen \cite{Wu-MNR:2016,Malkus:2014}. 
The characteristic feature of the standard MNR is that electron neutrinos can undergo significant flavor change, while
electron antineutrinos only experience little flavor conversion. This is due to the fact that 
the latter go through their resonances extremely non-adiabatically at either the beginning or the end of the MNR,
depending on the hierarchy, in a way similar to the 
results shown in \cite{Wu-MNR:2016} with 
2-flavor toy models. 
In NH, we find that survival probabilities for neutrinos propagating along trajectories 1b, 1c, 2a, 2b, and 2c exhibit the standard MNR features discussed above. 
We note that in all MNR cases, high energy $\nu_e$ are only partially converted at the end of the MNR region, resulting
in a $\sim 20\%$ averaged survival probabilities.

For trajectory 1a, despite the MNR condition is fulfilled, the flavor transformation is extremely non-adiabatic and nearly no
flavor conversion happens immediately after the MNR location. However, at $\sim 50$~km,
we see that both $\nu_e$ and $\bar\nu_e$ undergo simultaneous flavor conversions when
$\lambda(r)\gg|\mu(r)|$.
This is due to the fact that the $\nu\nu$ coupling introduces a
synchronization frequency\footnote{Note that in Eq.~\eqref{eq:synch} we suppressed the angular dependence in the geometric factors to simplify the notation.} ~\cite{Pastor:2002b}:
\begin{equation} \label{eq:synch}
\omega_{\mathrm{sync}}^{ij}(r) \approx
\dfrac{\sqrt{2} G_\mathrm{F} \int \mathrm{d}E \, \omega_{ij}[j_{\nu_{\underline{e}}}(E) G_{\nu_e}(r)
+ j_{\bar\nu_{\underline{e}}}(E) G_{\bar\nu_e}(r)]}{\mu(r)}.
\end{equation}
As $|\mu(r)|\rightarrow 0$ when it changes sign, $|\omega^{ij}_{\rm sync}(r)|\propto |1/\mu(r)|$
can be very large. 
Thus, a synchronized MSW effect (see, e.g., \cite{Wong:2002, Abazajian:2002}) 
happens when $\omega^{ij}_{\rm sync} \cos{\theta_{ij}} \sim\lambda$ so that all neutrinos and
antineutrinos with different momenta are bound together and simultaneously go through
the MSW-like flavor conversion. We note here that the flavor transformation is actually due to 
the resonance of $\omega^{21}_{\rm sync}$ with $\lambda(r)$ as the larger mixing angle
$\theta_{12}$ provides enough adiabaticity.

For 1d and 2d, the MNR condition is not met
and there is no flavor conversion in 1d. However, 2d
shows synchronized type oscillations starting at $\sim 285$~km,
resulting both $\nu_e$ and $\bar\nu_e$ flavor conversions.

For IH, the qualitative behaviors are the same as in NH when MNR
occurs (1b, 1c, 2a, 2b, 2c). The only difference is the slightly more
adiabatic flavor transformation near the end of the MNR region (see Fig.~\ref{Fig:IH} 
for the example of 2a). Regarding the other trajectories, we find for 1a (as in NH) the same synchronized MSW conversion while 1d and 2d, now show the ``bipolar'' type
of flavor transformation (see e.g.,~\cite{DuanReview2010}) so that both
$\nu_e$ and $\bar\nu_e$ are transformed, but their averaged survival probabilities are different (see Fig.~\ref{Fig:IH} for the example of 2d).

We provide a summary of the results in Table~\ref{tab:results}, where we report the type of flavor conversion mechanism.

\begin{figure}[!htbp]
 \centering
 \subfloat[2a]{\includegraphics[width=0.5\textwidth]{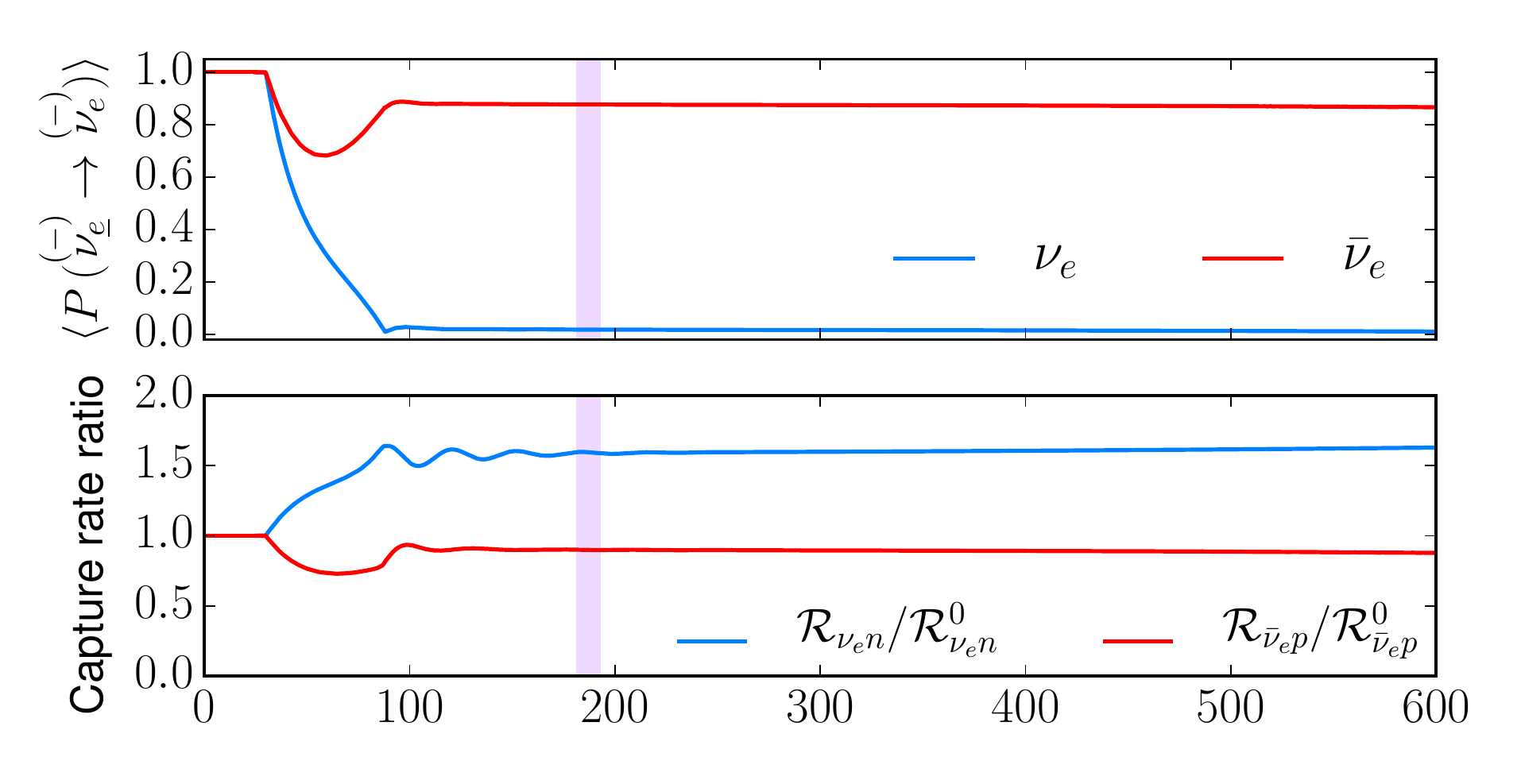}}
 \hfill
 \subfloat[2d]{\includegraphics[width=0.5\textwidth]{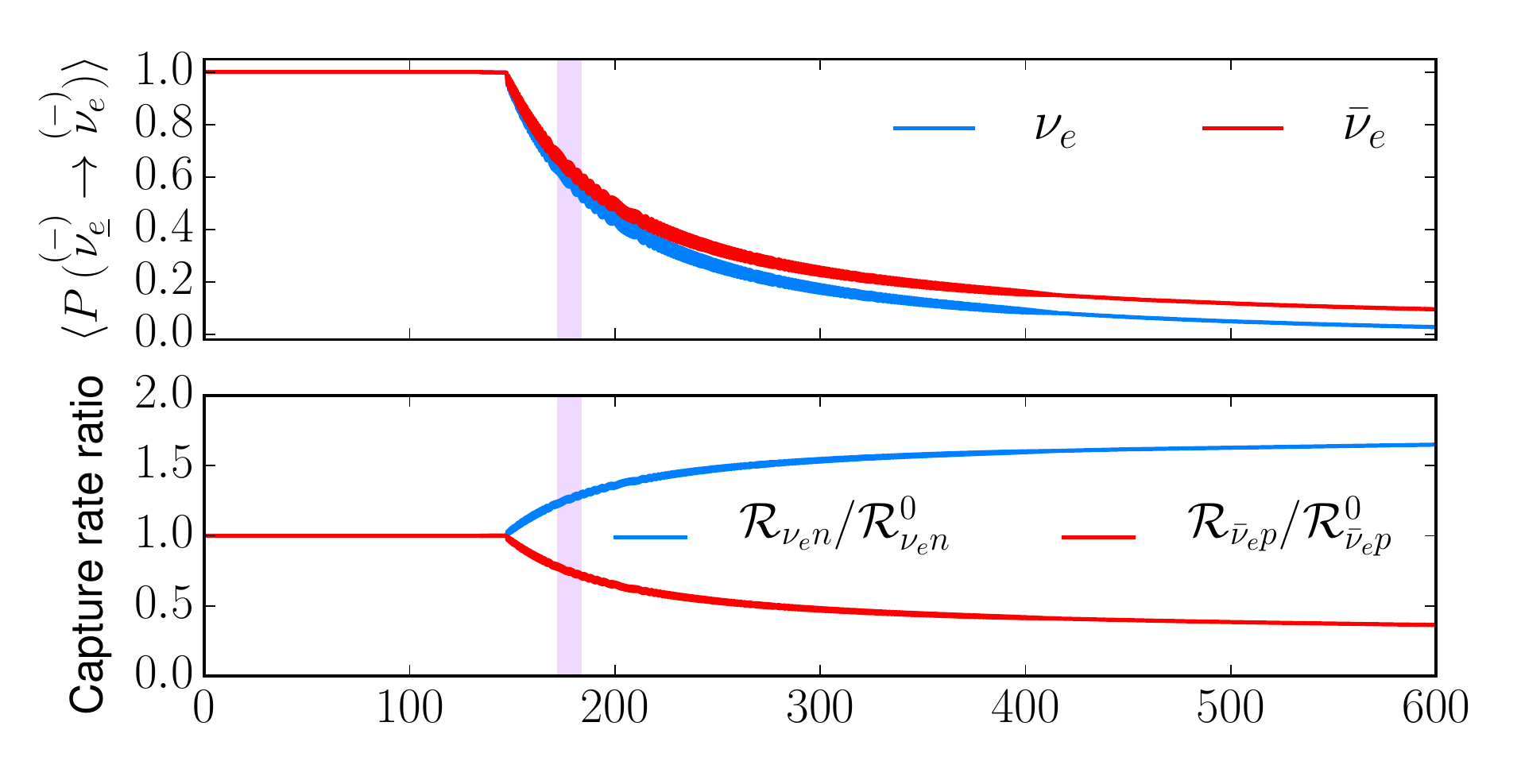}}
 \caption{Flavor evolution results obtained in IH along trajectories 2a (top) and 2d (bottom).
The top panels show the spectral averaged survival probabilities for the electron flavor. 
The blue curves correspond to $\nu_{e}$ and the red ones to $\bar{\nu}_{e}$, respectively. 
In the bottom panels the ratios of capture rates per solid angle for electron neutrinos (blue) and antineutrinos (red) are presented.} 
 \label{Fig:IH}
\end{figure}

\begin{table}
\center
\caption{For the given trajectories, we list the flavor conversion mechanism (MNR, synchronized MSW (sync.), bipolar, and no conversion (-)) 
and capture rate ratios for $\nu_{e}$ and $\bar{\nu}_{e}$ (distance-averaged around the location of the reference point ($T = 8 \, \mathrm{GK}$)) and both mass hierarchies.}
\label{tab:results}
\begin{ruledtabular}
\begin{tabular}{c*{1}{c}*{6}{c}}
Trajectory & \multicolumn{2}{c}{\, Flavor conversion} & \multicolumn{4}{c}{\qquad Capture rate ratio} \hfill \\ \hline
 & NH & IH & \multicolumn{2}{c}{NH} & \multicolumn{2}{c}{\qquad IH} \hfill \\
   & & & $\nu_{e}$ & $\bar{\nu}_{e}$ & $\nu_{e}$ & $\bar{\nu}_{e}$ \hfill \\ \hline
1a & sync. & sync. & +36\% & -36\% & +67\% & -67\% \\
1b & MNR & MNR & +37\% & -11\% & +46\% & -12\% \\
1c & MNR & MNR & +33\% & -7\% & +43\% & -7\% \\
1d & - & bipolar & - & - & +46\% & -49\% \\
2a & MNR & MNR & +52\% & -4\% & +59\% & -10\% \\
2b & MNR & MNR & +39\% & -4\% & +56\% & -8\% \\
2c & MNR & MNR & +37\% & -4\% & +53\% & -6\% \\
2d & sync. & bipolar & - & -\% & +26\% & -25\% \\
\end{tabular}
\end{ruledtabular}
\end{table}

\end{subsection}

\begin{subsection}{Differential capture rates}\label{sec:capture_rates}

We calculate the rate per solid angle for
electron neutrino captures on free neutrons with
both the oscillated and unoscillated neutrino spectra.
The oscillated rate is given by:
\begin{equation} \label{eq:capture_rate_1}
\begin{split}
\mathcal{R}_{\nu_{e} n}(r) & = 
\dfrac{1}{2 \pi} \sum_{\alpha = e, \mu, \tau}
F_{\nu_{\underline{\alpha}}} \int_{0}^{\infty} \mathrm{d} E \,
f_{\nu_{\underline{\alpha}}}(E) 
\\
& \qquad \times \sigma_{\nu_{e} n,\mathrm{abs}}(E) P(\nu_{\underline{\alpha}} \to \nu_{e})(E, r).
\end{split}
\end{equation}
For the unoscillated capture rate we use:
\begin{equation} \label{eq:capture_rate_unoscillated}
\mathcal{R}_{\nu_{e} n}^{0} = \dfrac{F_{\nu_{\underline{e}}}}{2 \pi} \int_{0}^{\infty} \mathrm{d} E \, f_{\nu_{\underline{e}}}(E) \sigma_{\nu_{e} n, \mathrm{abs}}(E).
\end{equation}

Similar expressions ($\mathcal{R}_{\bar{\nu}_{e} p}$, $\mathcal{R}_{\bar{\nu}_{e} p}^{0}$) 
hold for electron antineutrino capture on free protons with $\sigma_{\bar{\nu}_{e} p, \mathrm{abs}}$, 
where the lower bound of the integrals has to be replaced by the threshold $m_{e} + \Delta_{np}$, 
i.e., the sum of the electron mass $m_{e} \approx 0.5 \, \mathrm{MeV}$ and the neutron-proton mass difference $\Delta_{np}$.
We compute the ratio between oscillated $\mathcal{R}$ and unoscillated $\mathcal{R}^{0}$ capture rates.

From Eq.~\eqref{eq:capture_rate_1}, one can see that
when calculating $\mathcal{R}$,
the electron neutrino appearance probabilities
will be weighted by the $\nu_{\mu}$/$\nu_{\tau}$ fluxes 
(Eq.~\eqref{eq:neutrino_flux}) 
and the cross section (Eq.~\eqref{eq:sigma_abs}). 
We note that the fluxes $j_{\nu_{\underline{\alpha}}}$
are peaked around $2.2 \, T_{\nu_{\underline{\alpha}}}$
while the rates $(j_{\nu_{\underline{\alpha}}} \sigma)$
around $4.1 \, T_{\nu_{\underline{\alpha}}}$
(Fig.~\ref{Fig:fluxes}).
Therefore, for $\mathcal{R}_{\nu_{e}n}$, the contribution
of the initial non-electron flavors will enhance
it when efficient
$\nu_{e}\leftrightarrow\nu_{x}$ flavor conversion took place,
as the high energy tail of the initial $\nu_x$
dominates the capture rates indicated in
Fig.~\ref{Fig:fluxes}.
For $\mathcal{R}_{\bar\nu_ep}$, from
Fig.~\ref{Fig:fluxes} wee see that since
$(j_{\bar\nu_{\underline{e}}} \sigma_{\bar\nu_ep,{\rm
abs}})>(j_{\nu_{\underline{\mu,\tau}}}
\sigma_{\bar\nu_ep,{\mathrm{abs}}})$ for the whole energy 
spectrum, any flavor conversion of
$\bar{\nu}_{e} \leftrightarrow \bar{\nu}_{x}$ will suppress
it.

Based on the above discussions, we see from the bottom
panels of Figs.~\ref{Fig:results1}, \ref{Fig:results2},
and \ref{Fig:IH} that for $\nu_e$ which undergo
efficient flavor conversions due to MNR 
(1b, 1c, 2a, 2b, 2c), the $\nu_{e}$
capture rates are largely enhanced by up to $\sim 59\%$ after
the end of the MNR region while the $\bar{\nu}_{e}$ capture
rates are slightly decreased by up to $\sim 12\%$ in those
cases. 
For the trajectories showing the synchronized MSW
flavor transformation (1a), the 
$\nu_{e}$ ($\bar\nu_{e}$) capture
rates are increased (decreased) by up to $\sim 67\%$ $(67\%)$
as both are simultaneously transformed.
As for the cases with bipolar type of flavor
conversion (1d and 2d in IH), the capture rates for
$\nu_{e}$ ($\bar{\nu}_{e}$) are gradually changed up to
$\sim 65\%$ ($-62\%$) at $600$~km.
In most cases where flavor conversion takes place, the
capture rates are affected in regions with temperature
$T \gtrsim 8 \, \mathrm{GK}$ before all nucleons
recombine into $\alpha$-particles.
We provide a summary of the capture rate ratio,
$\mathcal{R}/\mathcal{R}^0$, in Table~\ref{tab:results}
at the reference locations ($T=8$~GK) for all
trajectories.

\begin{figure}
 \centering
 \includegraphics[width=0.5\textwidth]{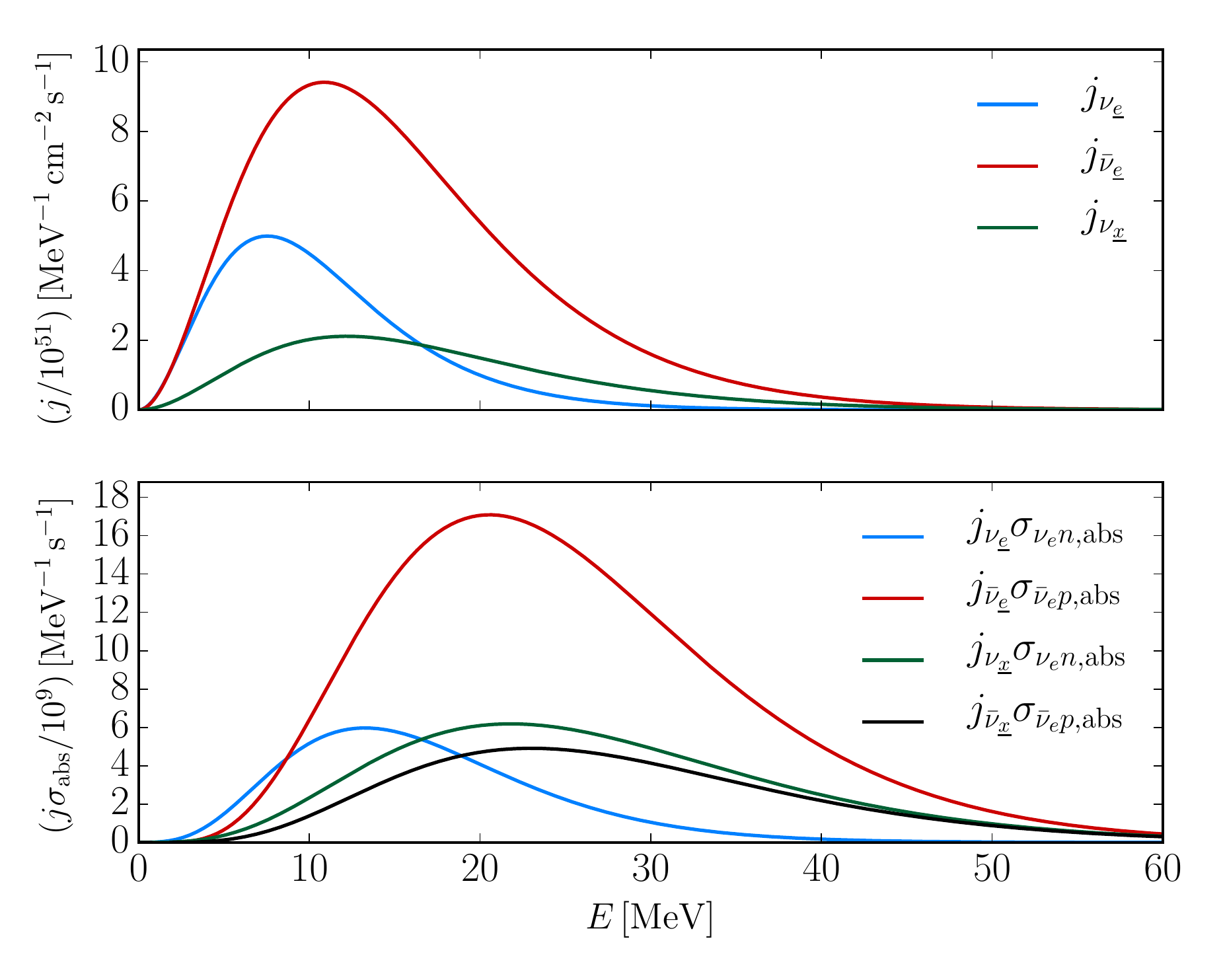} 
 \caption{Neutrino number fluxes [see \eqref{eq:number_flux}] (upper panel) 
 and number fluxes times cross sections (lower panel). All quantities are per unit energy per solid angle.} 
 \label{Fig:fluxes}
\end{figure}

\end{subsection}

\begin{subsection}{Comparison between different post-merger times} \label{sec:timesteps}

Although the neutrino luminosities and average energies remain nearly stationary over the disk evolution time, 
the wind density profiles change substantially.
In Fig.~\ref{Fig:compare_timesteps}, we compare the matter potentials for trajectories 2a (upper) and 2c (lower panel) 
at different times ($60$ and $100 \, \mathrm{ms}$).
For both cases, the matter potentials are larger at $100 \, \mathrm{ms}$ compared to the one at $60 \, \mathrm{ms}$
and show a similar overall behavior as a function of distance. 
This is because the expanding wind drives the surrounding area less neutron-rich at later times, as can be inferred from Figs.~\ref{Fig:plot_ye_temp} 
and~\ref{Fig:plot_ye_temp_60ms}. 
Consequently, the MNR locations at $100 \, \mathrm{ms}$ are shifted to smaller distances.
However, such differences do not result in qualitative changes in the overall flavor evolution behavior.

\begin{figure}[!htbp!]
 \centering
 \includegraphics[width=0.5\textwidth]{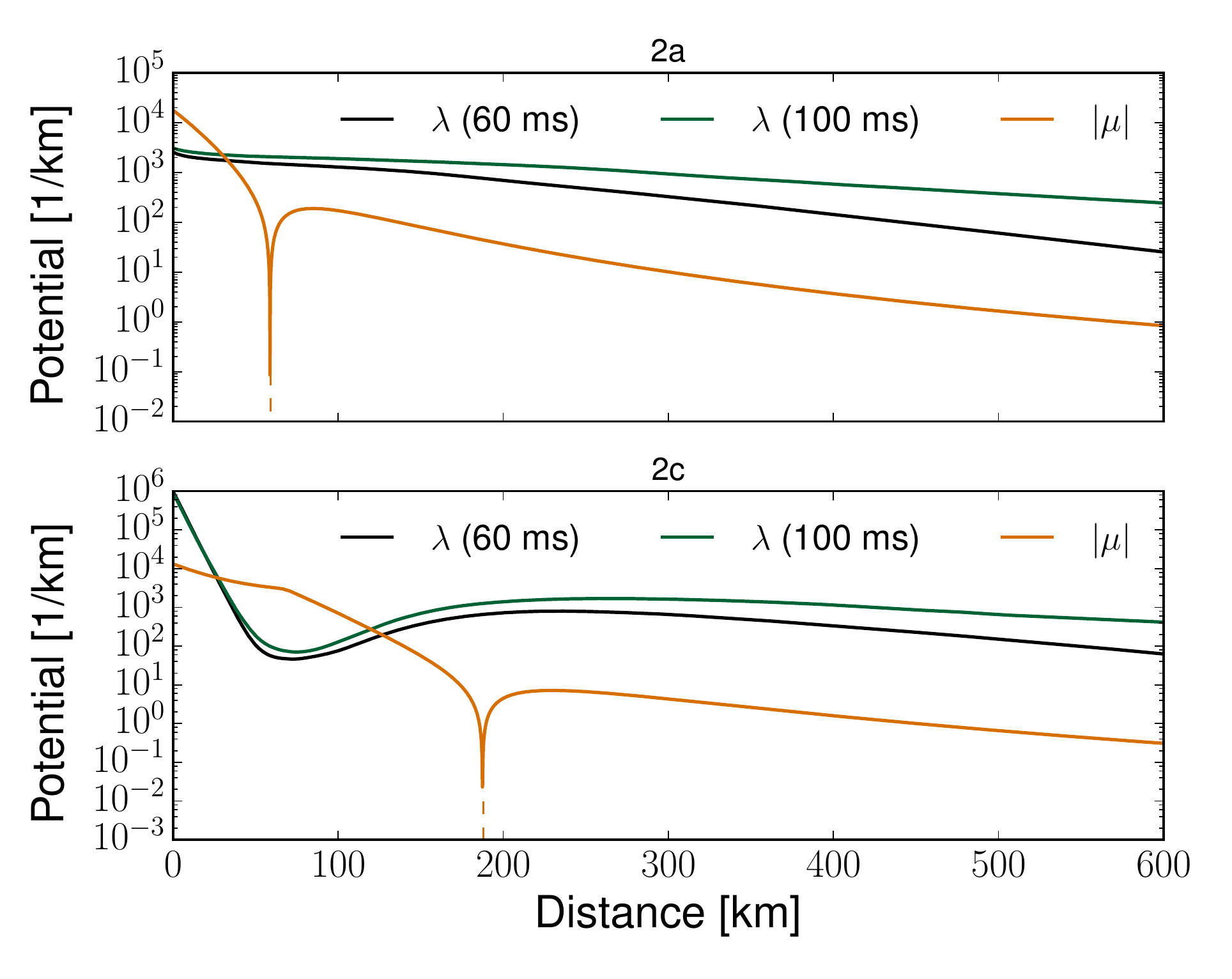} 
 \caption{Comparison of the unoscillated potentials between the time snapshots for trajectories 2a (upper) and 2c (lower panel).} 
 \label{Fig:compare_timesteps}
\end{figure}

\begin{figure}[!htbp]
 \centering
 \includegraphics[width=0.5\textwidth]{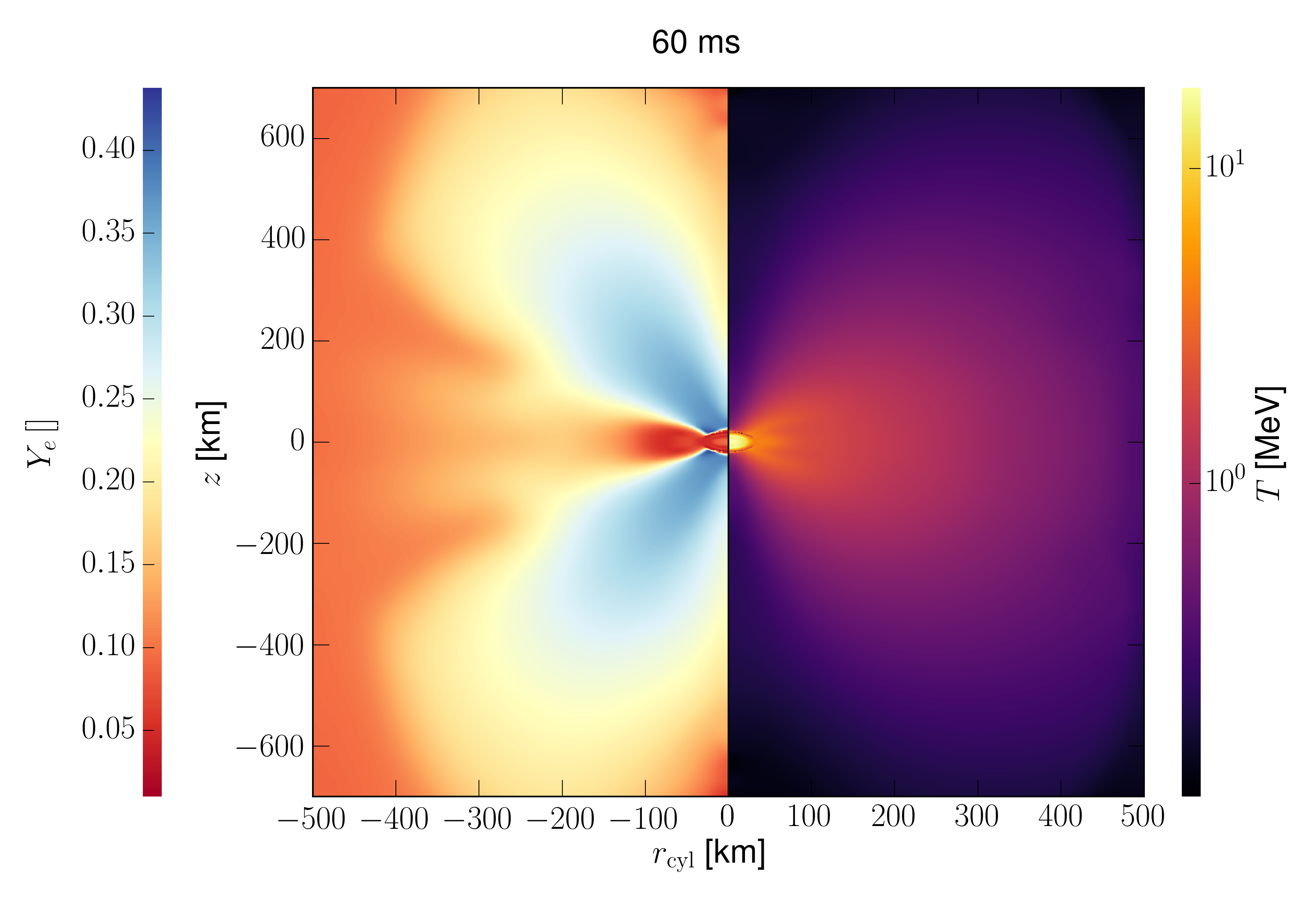} 
 \caption{Electron fraction (left panel) and matter temperature (right panel) as functions of cylindrical coordinates $z$ and 
 $r_{\mathrm{cyl}}$ at $t = 60 \, \mathrm{ms}$ after the merger.} 
 \label{Fig:plot_ye_temp_60ms}
\end{figure}

\end{subsection}

\begin{subsection}{Impact of input neutrino emission characteristics} \label{sec:uncertainties}
For the results presented so far, the calculations are
performed based on the neutrino luminosities and
mean energies given in
Table~\ref{table:emission_parameters}.
These values are obtained at distances far away
from the disk, which are called 
``net luminosities'' in \cite{Perego:2014}
(see Fig.~8 of \cite{Perego:2014}).
However, neutrinos do not completely travel
unhindered from the neutrino surfaces in a
realistic environment.
Their luminosities can be higher in regions close
to the neutrino surfaces and reduce to the net luminosities due to 
charged-current neutrino absorptions above the surfaces.
Since the environment is neutron-rich, 
the decrease of $L_{\nu_e}$ is larger than
that of $L_{\bar\nu_e}$ and $L_{\nu_x}$.
Consequently, a less negative $\mu(r)$ compared 
to the values obtained with net luminosities 
in regions close to the neutrino surfaces may be 
obtained. If for a trajectory,
$|\mu(r)|>\lambda(r)$ initially close to the 
neutrino surfaces, one expects MNR 
to occur closer to the emission point.

We explore this effect by performing additional
calculations for all trajectories 
using the ``cooling'' neutrino
luminosities from \citep{Perego:2014},
calculated by neglecting charged-current
neutrino absorptions above the neutrino
surfaces:
$L_{\nu_e,{\rm cool}}=25\times 
10^{51}$~erg~s~$^{-1}$, 
$L_{\bar\nu_e,{\rm cool}}=33\times
10^{51}$~erg~s~$^{-1}$, and
$L_{\nu_x,{\rm cool}}=L_{\nu_x}=8\times 
10^{51}$~erg~s~$^{-1}$. 
Compared to the net luminosities (see 
Table~\ref{table:emission_parameters}), 
$L_{\nu_e,{\rm cool}}=1.67 L_{\nu_e}$
and $L_{\bar\nu_e,{\rm cool}}=
1.1 L_{\bar\nu_e}$.
Figure~\ref{Fig:cooling_lum} shows the
results with cooling luminosities for 
trajectories 1c and 2a. 
In both cases, the MNR locations with a
larger $L_{\nu_e,{\rm cool}}$ are indeed closer to
their emission points when compared to
Fig.~\ref{Fig:results1} and \ref{Fig:results2}.
In 1c, MNR now occurs at $\sim 66$~km immediately
prior to the point where $\mu$ changes sign. 
This results in a complete flavor transformation for
both $\nu_e$ and $\bar\nu_e$, or symmetric MNR
\cite{Malkus:2012, Malkus:2016}. After the symmetric
MNR, another standard MNR occurs at a distance of 
$\sim 82$~km so that antineutrinos go through 
nearly complete flavor conversion as discussed in
\cite{Malkus:2016}.
For the capture rate ratio of $\nu_e$, it only change
slightly due to a much higher $L_{\nu_e,{\rm cool}}$.
For $\bar\nu_e$, the capture rate ratio is largely
suppressed in the region between the symmetric MNR
and the second standard MNR.
For case 2a, the position where MNR occurs is also
largely shifted to a smaller distance at $\sim 10$~km.
However, we find a strongly non-adiabatic behavior
resulting in no flavor transformation.

\begin{figure}[tp!]
\centering
\subfloat[1c (NH)]{\includegraphics[width=0.5\textwidth]{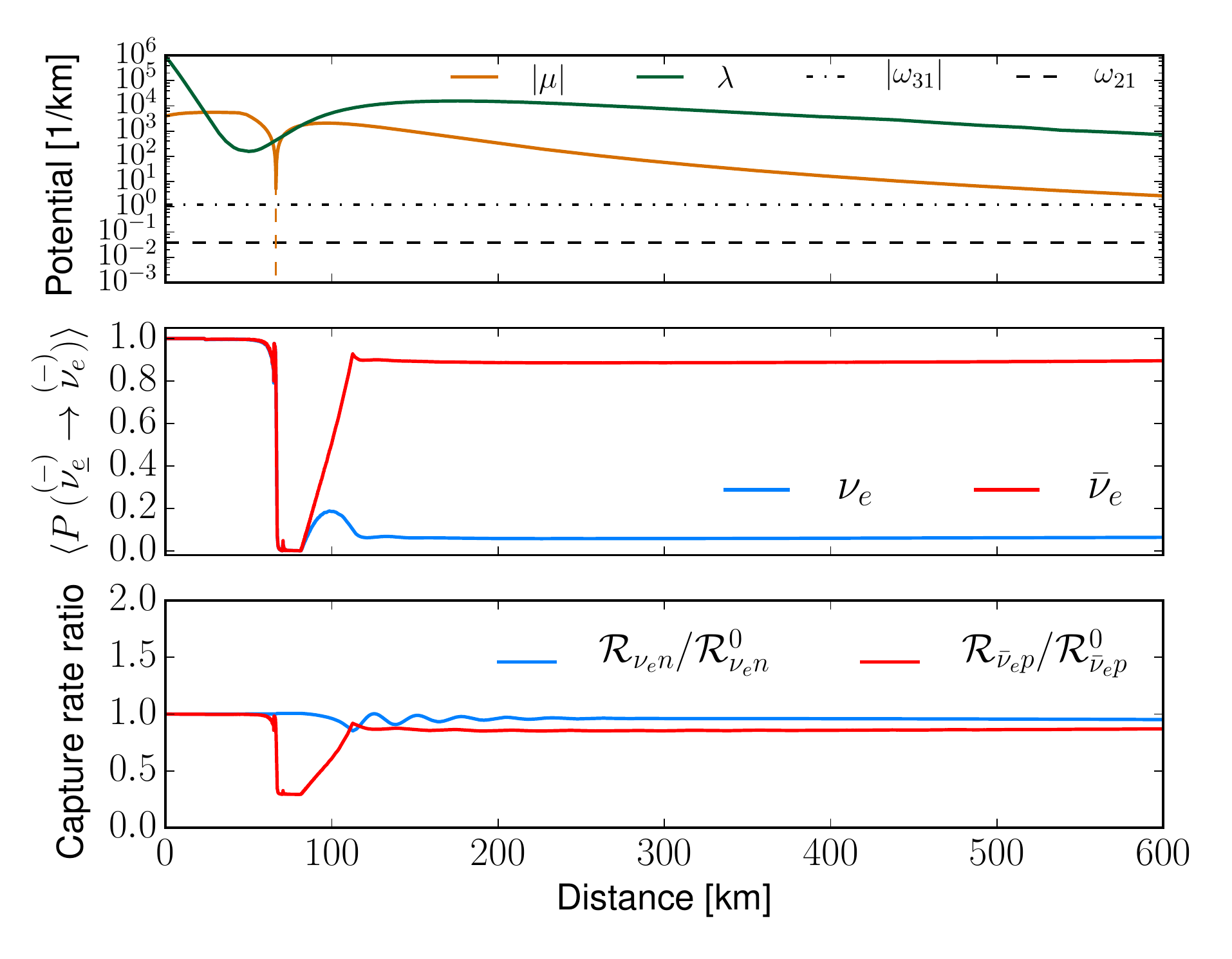}} 
\hfill
\subfloat[2a (NH)]{\includegraphics[width=0.5\textwidth]{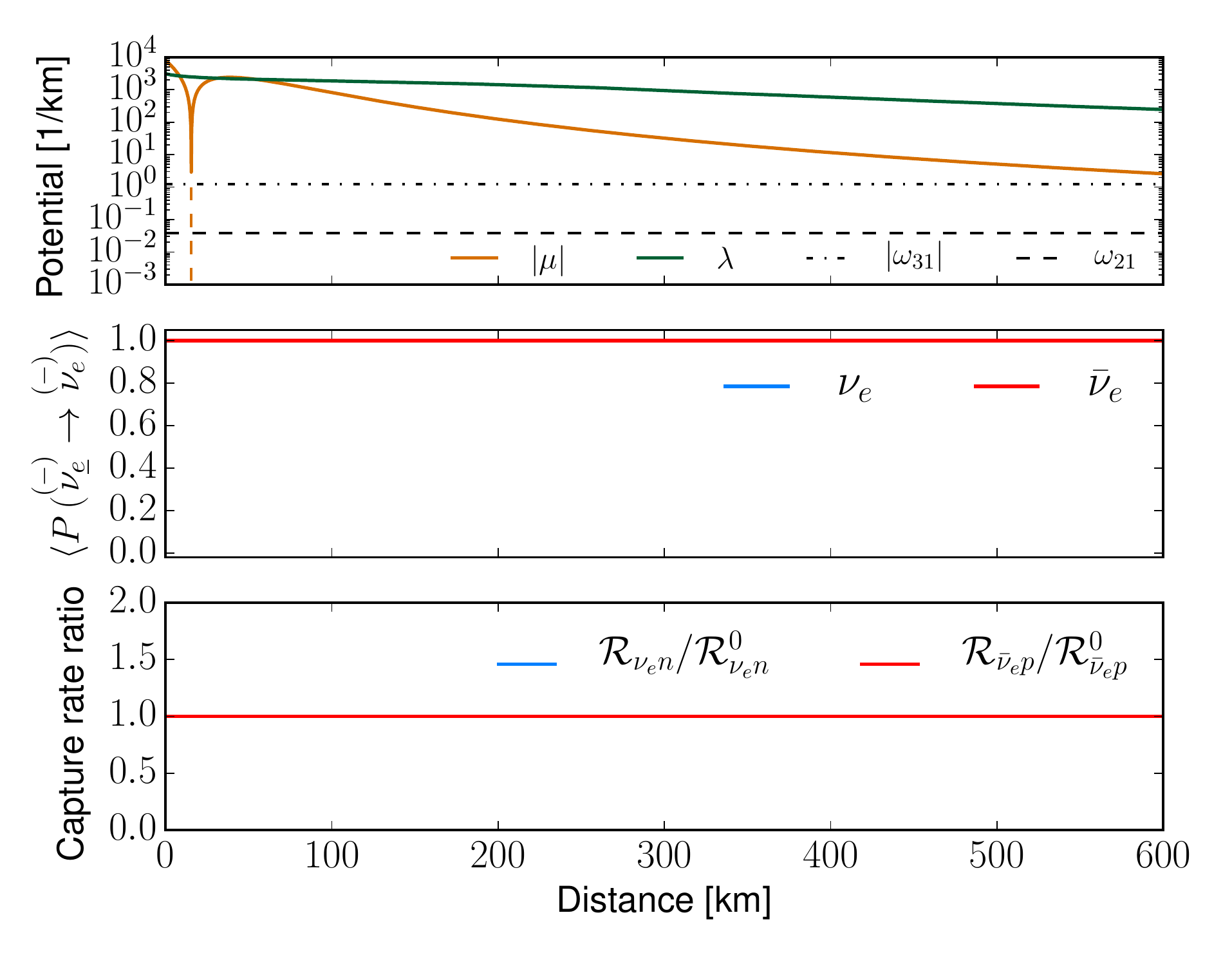}} 
 \caption{Same as Fig.~\ref{Fig:results1} for trajectories 1c (left) and 2a (right) with cooling luminosities (see text).}
 \label{Fig:cooling_lum}
\end{figure}

The model we have considered is based on one
particular simulation of the remnant from the merger
of BNS with equal mass.
The large variety of possible initial conditions of the
merger (NS masses and spins, mass ratios) is expected to  
translate into a wide range of both the neutrino
luminosities and the mean energies for different systems.
Also, the neutrino emissions can be significantly
influenced by the thermodynamical properties of the dense
nuclear matter, which is subject to the large 
uncertainties in the nuclear equation of state 
(EOS, see e.g. \cite{Sekiguchi:2015, Foucart:2016}
for the potential impact of the nuclear EOS on the neutrino
luminosities).
Lastly, the different numerical techniques and levels of
approximations that are used to model hydrodynamics,
gravity and weak interactions can all lead to
quantitatively different predictions of the relevant
neutrino quantities.
To quantify these uncertainties in our input parameters, we
have collected published values of neutrino 
luminosities and mean energies (when available) from
several different simulations of BNS merger and of 
the merger aftermath in presence of a long lived MNS. We
show these values in Table~\ref{Table_Uncertainties} in
Appendix~\ref{app:collected_data}. We see that the 
luminosities may differ by one order of magnitude while
the differences in mean energies are within a factor 
of two. We show in Figs.~\ref{Fig:ratio_luminosity} 
and~\ref{Fig:ratio_mean_energy} the ratio of 
luminosities, $L_{\bar\nu_e}/L_{\nu_e}$ and
$L_{\nu_e}/L_{\nu_x}$, and the ratio of mean energies,
$\langle E_{\bar\nu_e}\rangle/\langle E_{\nu_e}\rangle$
and $\langle E_{\nu_{e}}\rangle / \langle E_{\nu_{x}}\rangle$. 
We emphasize that our goal is not to
compare the results of the different simulations, but to
show the variety of possible ranges for these ratios.

Note that the neutrino self-interaction potential
depends on the difference between the fluxes of neutrinos
and antineutrinos, which are proportional to the neutrino
number luminosities $\sim L_{\nu_\alpha}/\langle
E_{\nu_\alpha}\rangle$. We further show the corresponding 
ratios,
$(L_{\bar\nu_e}/\langle E_{\bar\nu_e}\rangle)/
(L_{\nu_e}/\langle E_{\nu_e}\rangle)$
and $(L_{\nu_e}/\langle E_{\nu_e}\rangle)/
(L_{\nu_x}/\langle E_{\nu_x}\rangle)$ in
Fig.~\ref{Fig:ratio_flux}.

\begin{figure}[tp!]
 \centering
 \includegraphics[width=0.5\textwidth]{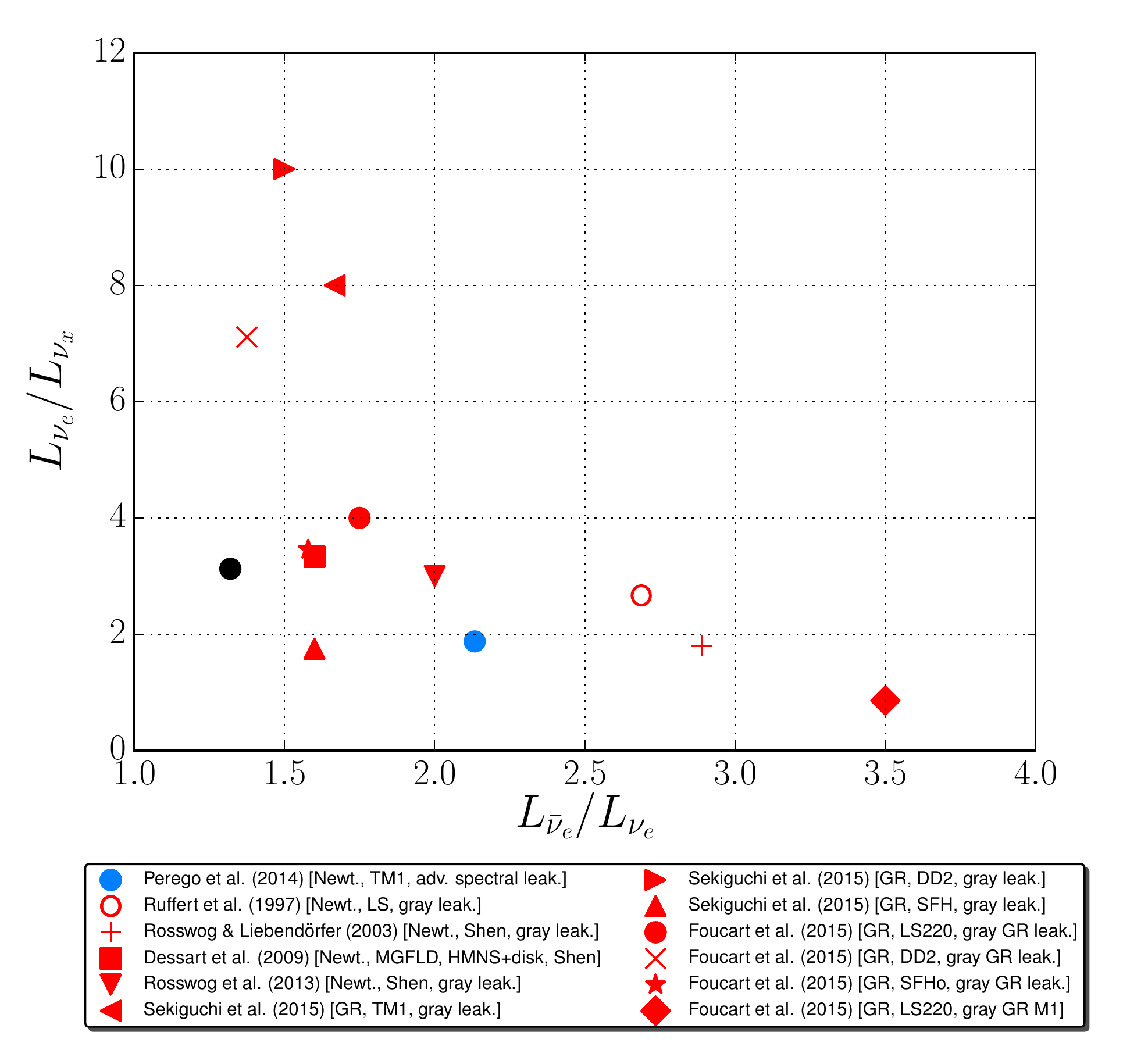} 
 \caption{Ratio of luminosities: $L_{\nu_{e}} / L_{\nu_{x}}$ vs. $L_{\bar{\nu}_{e}} / L_{\nu_{e}}$. The corresponding values are given in Table~\ref{Table_Uncertainties}.
 The black point refers to the cooling luminosities of \cite{Perego:2014} (see text).
 } 
 \label{Fig:ratio_luminosity}
\end{figure}

\begin{figure}[tp!]
 \centering
 \includegraphics[width=0.5\textwidth]{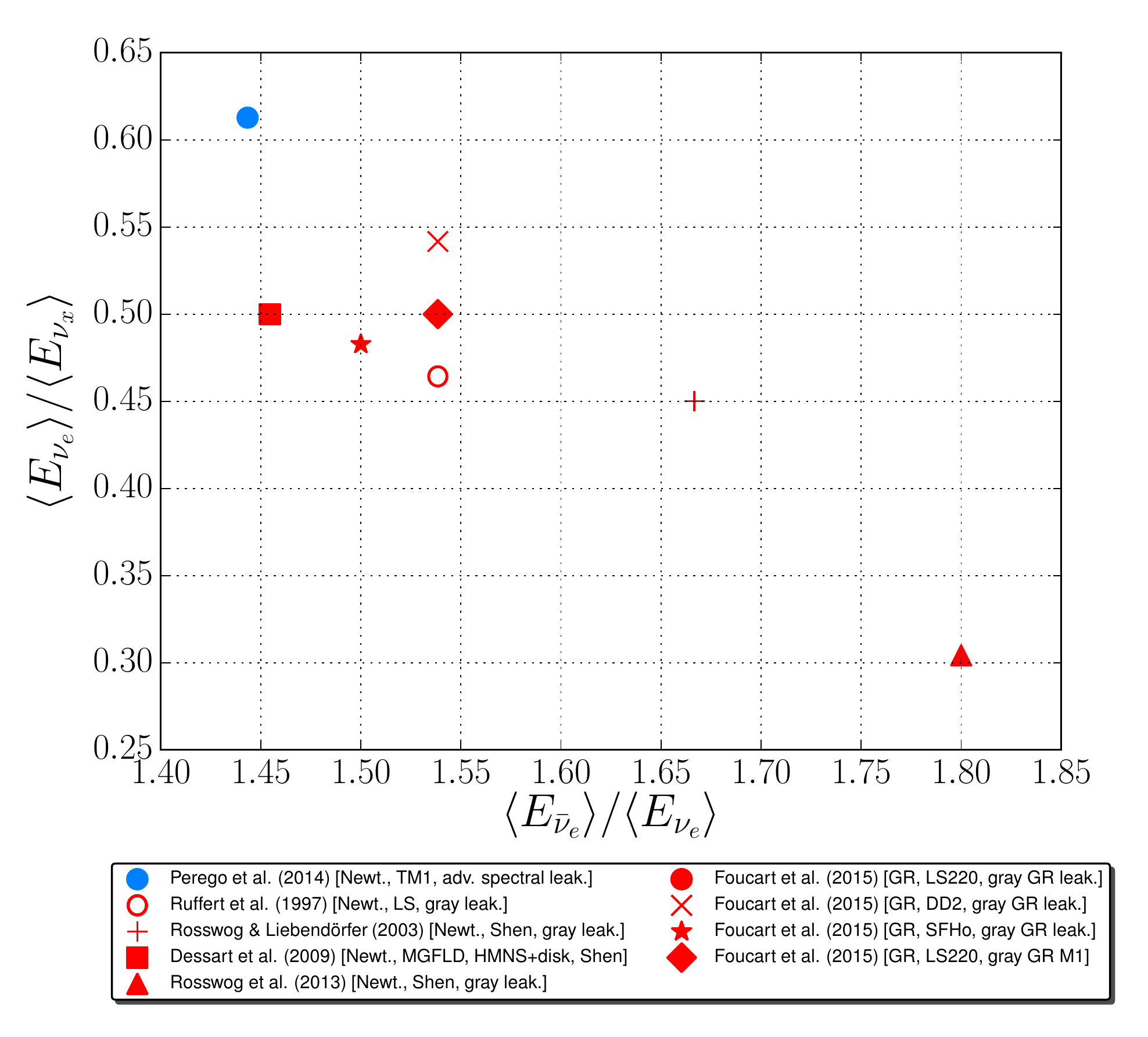} 
 \caption{Ratio of mean energies: $\langle E_{\nu_{e}} \rangle / \langle E_{\nu_{x}} \rangle$ vs. $\langle E_{\bar{\nu}_{e}} \rangle / \langle E_{\nu_{e}} \rangle$.
 The corresponding values are given in Table~\ref{Table_Uncertainties}.} 
 \label{Fig:ratio_mean_energy}
\end{figure}

\begin{figure}[tp!]
 \centering
 \includegraphics[width=0.5\textwidth]{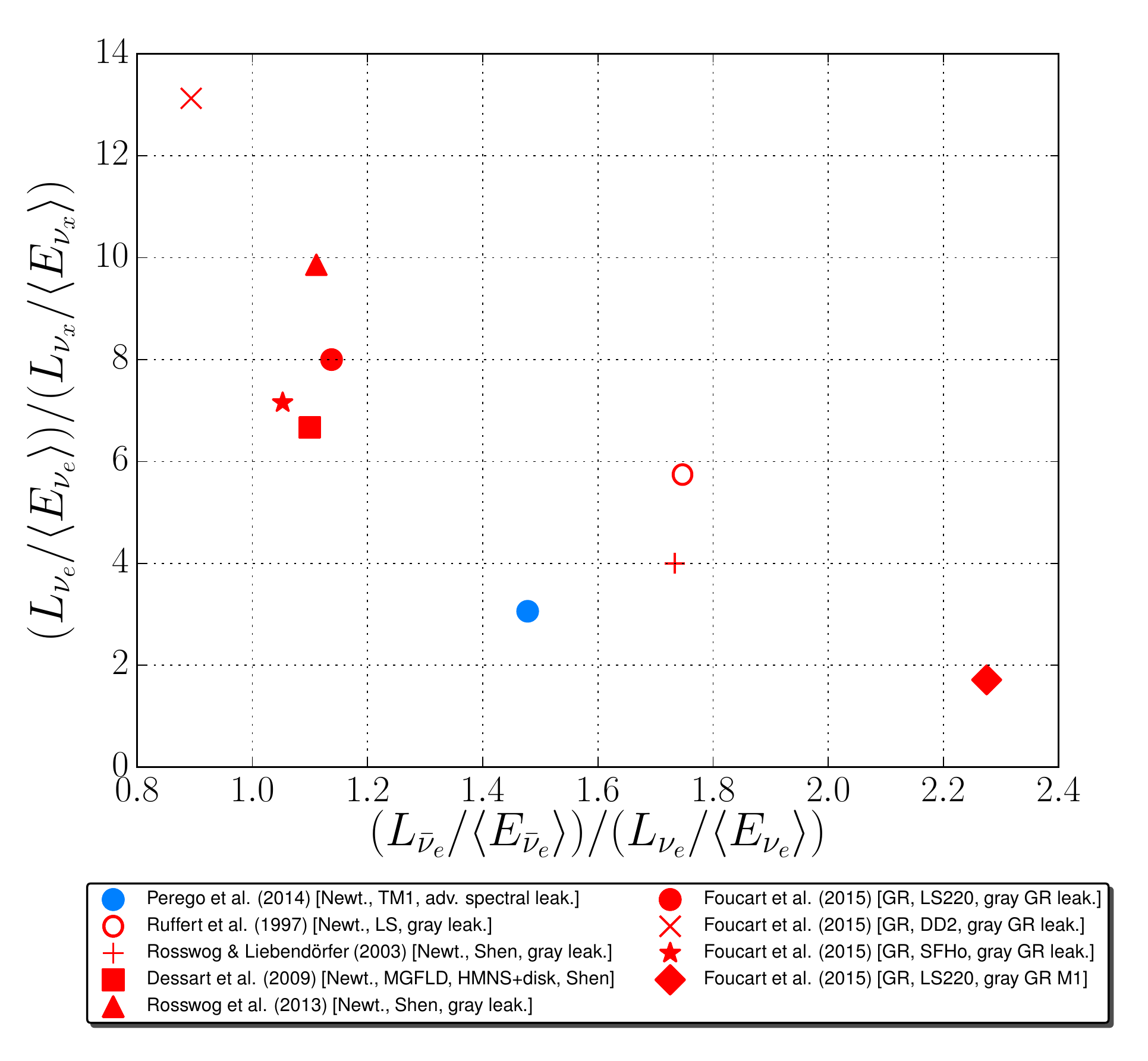} 
 \caption{Ratio of emission rates: $(L_{\nu_{e}} / E_{\nu_{e}}) / ( L_{\nu_{x}} / E_{\nu_{x}})$ vs. $(L_{\bar{\nu}_{e}} / E_{\bar{\nu}_{e}}) / (L_{\nu_{e}} / E_{\nu_{e}})$. 
 The corresponding values are given in Table~\ref{Table_Uncertainties}.
 }
 \label{Fig:ratio_flux}
\end{figure}

\begin{figure}[!htbp]
\centering
\subfloat[1c (NH): $x_{0} = -35 \, \mathrm{km}$, $z_{0} = 25 \, \mathrm{km}$, $\theta_{0} = 48.7^{\circ}$]{\includegraphics[width=0.5\textwidth]{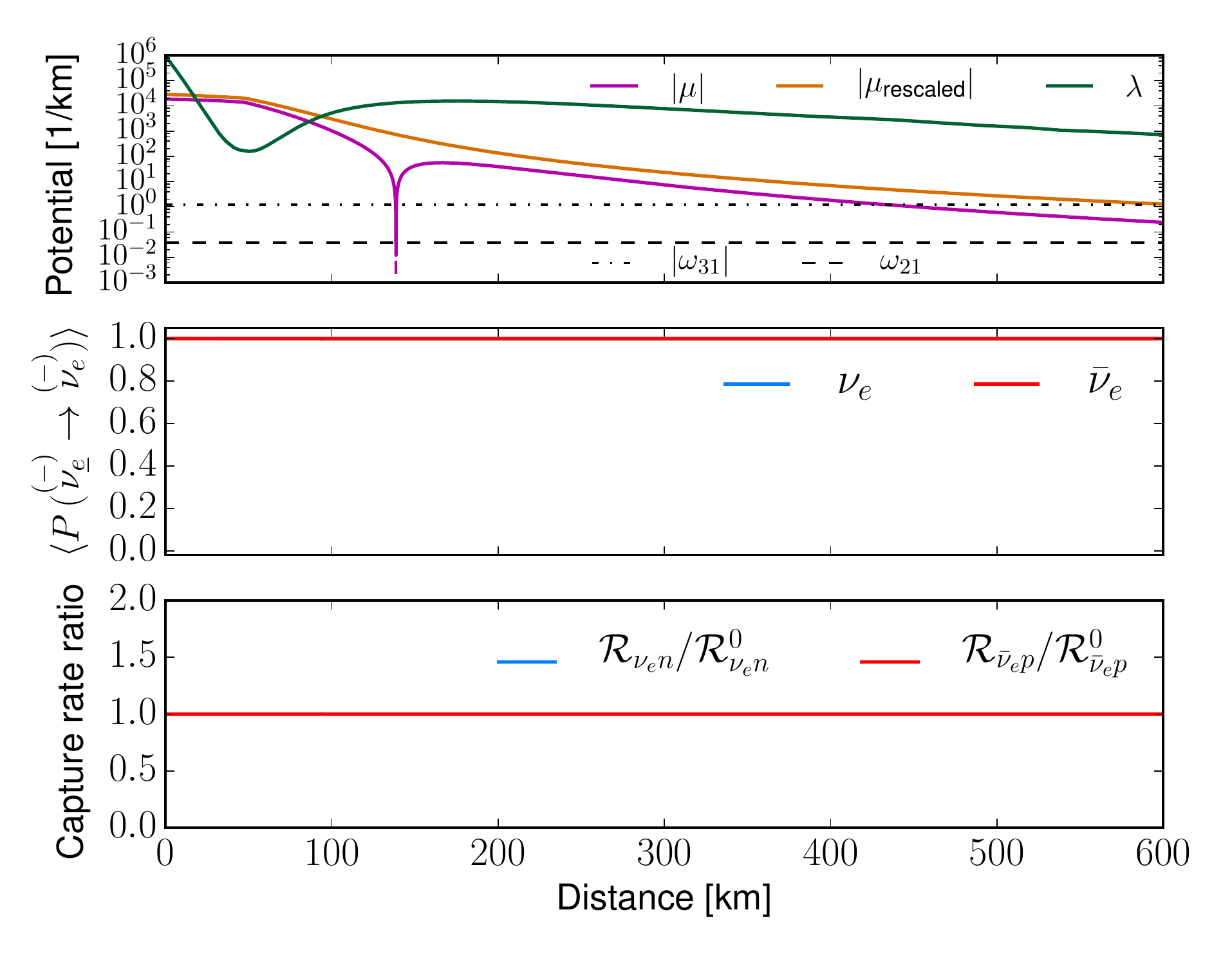}} 
\hfill
\subfloat[2a (NH): $x_{0} = 10 \, \mathrm{km}$, $z_{0} = 30 \, \mathrm{km}$, $\theta_{0} = 20.0^{\circ}$]{\includegraphics[width=0.5\textwidth]{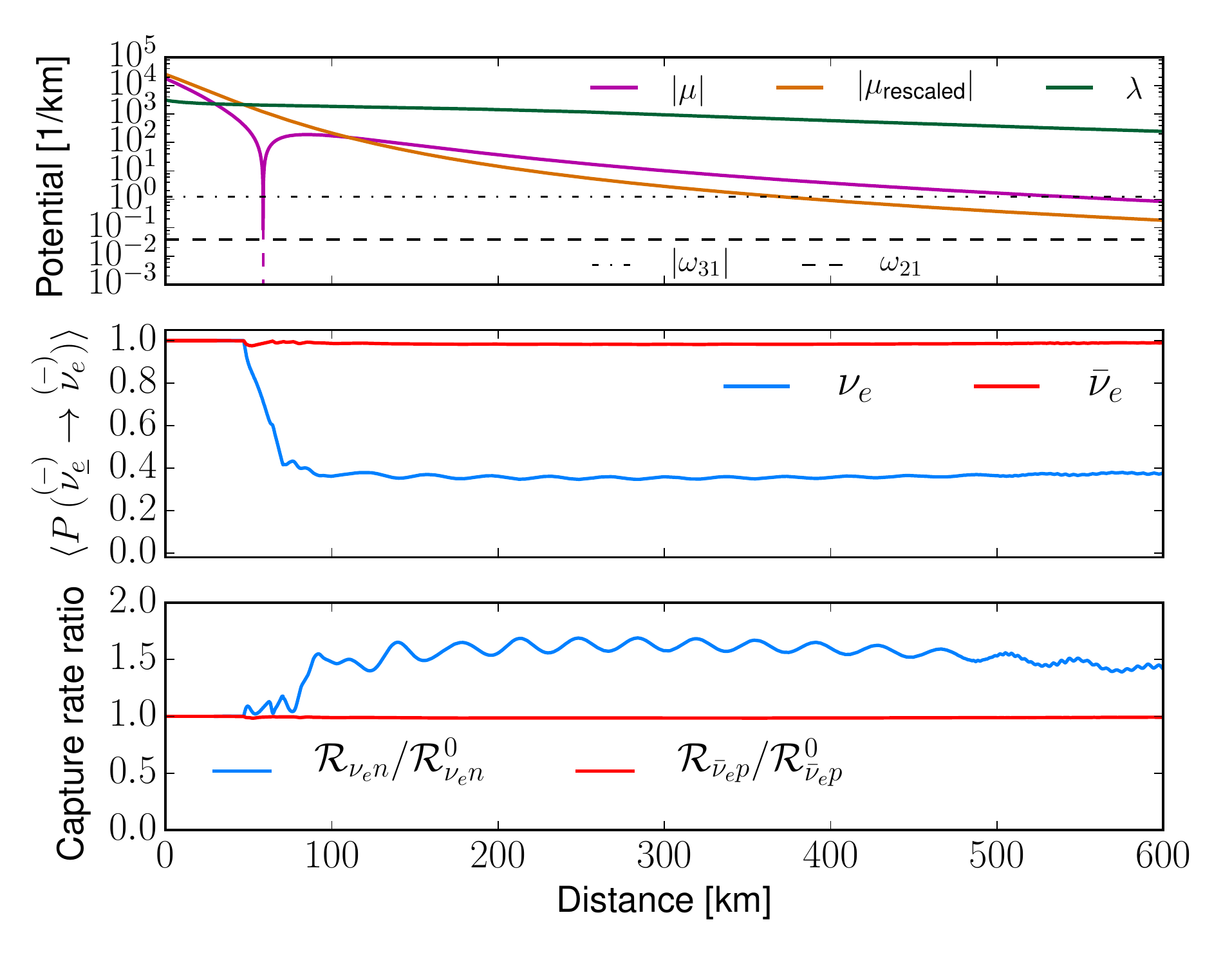}} 
\caption{
Same as Fig.~\ref{Fig:results1} for trajectories 1c and 2a with rescaled fluxes (see text).
}
\label{Fig:rescale}
\end{figure}

To explore the impact of different flux ratios on the neutrino
flavor evolution, we have varied the neutrino luminosities
of Table~\ref{table:emission_parameters} by: $L_{\nu_{e}}
\mapsto 0.65 \, L_{\nu_{e}}$ and $L_{\nu_{x}} \mapsto 1.16
\, L_{\nu_{x}}$. This change gives a similar 
$(L_{\bar\nu_e}/\langle E_{\bar\nu_e}\rangle)/
(L_{\nu_e}/\langle E_{\nu_e}\rangle)$ to the value
obtained in Foucart et al. (2016) 
[GR, gray GR M1, LS220 EOS] \cite{Foucart:2016}.
The results with such luminosities for trajectories 1c 
and 2a in NH are shown in Fig.~\ref{Fig:rescale}.
Contrary to the previously discussed exploration with
cooling luminosities, the differences between the fluxes of electron neutrinos 
and electron antineutrino becomes larger. This affects the self-interaction potential 
in such a way that no change of sign in $\mu(r)$ occurs anymore.

For trajectory 1c, the flavor evolution becomes 
strongly non-adiabatic when compared to the result shown in 
Fig.~\ref{Fig:results1}c. For trajectory 2a, 
the flavor evolution becomes less adiabatic.

\end{subsection}

\end{section}

\begin{section}{Discussion and Conclusions} \label{sec:conclusion}
In the present work, we investigated the trajectory dependence of the neutrino flavor evolution 
in a BNS merger remnant. We found that depending on which location and which polar angle neutrinos are emitted, the outcome can be different. 
In particular, we observed that flavor conversion
through MNR can occur for most neutrinos traveling
through the low density funnel. For cases without
flavor conversion across MNR due to the 
non-adiabaticity, a synchronized MSW transformation can
take place afterwards. For neutrinos that do not
encounter MNR, bipolar type of flavor oscillations may
occur for the IH.
We note that future investigations should also explore the dependence on
the azimuthal emission angle that was omitted in this work.

We found that flavor evolution can significantly affect the neutrino capture rates on protons and neutrons at the regions with $T \gtrsim 8 \, \mathrm{GK}$. 
This may change the $r$-process abundances in the neutrino wind.
In our work we presented capture rates which do not take into account the solid angle contributions.
This treatment is consistent with our single-trajectory approximation and we found that
the oscillated $\nu_e$ ($\bar\nu_e$) capture
rates can increase (decrease) up to almost 67\% (67\%).

If we apply flavor conversion probabilities derived with the 
single-trajectory approximation to all neutrinos emitted from the disk
and integrate over the solid angle (see, e.g., Fig.~\ref{Fig:solid_angle}), 
we find the $\nu_{e}$ capture rates instead always decrease.
This behavior is due to the fact that the solid angle for $\nu_{e}$ is larger so that the relative contribution from $\nu_{x}$ reduces.
We speculate that the use of the "single-trajectory" assumption is maximizing the impact of flavor conversion we find on the capture rates. 

The initial luminosities are a key input for the flavor evolution as we observed. 
In fact, a change in the luminosities can produce different flavor conversion results. 
In particular, we have shown this fact in two ways: 
either by taking the cooling luminosities as input instead of the net luminosities, or by rescaling the number luminosity ratios of different neutrino species 
according to BNS merger model predictions.
With cooling luminosities, we obtained flavor conversions for $\nu_{e}$ and, for most cases also for $\bar{\nu}_{e}$, that occur closer to the neutrino emission surface. 
Using rescaled luminosities, we also obtained standard MNR albeit the flavor transformation becomes less adiabatic.
If future realistic initial fluxes happen to produce flavor conversion very close to the neutrino surface, 
the presence of neutrino absorption in the region above the disk might require, in the long run, 
the investigation of the competition between collisions and flavor evolution in an improved treatment of the neutrino propagation.

Demanding simulations beyond the currently used approximations will tell us how 
the implementation of the full coupling of the self-interaction Hamiltonian will
modify the results presented in this work. From the studies performed in the supernova context one may speculate 
that this may introduce decoherence and possibly change the starting points of flavor conversions. 
Moreover, one should also include general relativistic effects which produce a redshift and the bending of the neutrino trajectory \cite{Caballero:2012}. 
These steps beyond the approximations employed in this work may be necessary to assess the impact of flavor conversions on the $r$-process elements produced 
in the neutrino-driven wind and the actual contribution from neutron star mergers to the observed heavy element abundances. 

\end{section}

\begin{acknowledgments}
We thank Annie Malkus, Basudeb Dasgupta and Gail McLaughlin for helping 
in clarifying differences with respect to the geometry, Kevin Ebinger, Andreas Lohs, Gabriel Mart\'{\i}nez-Pinedo, Huaiyu Duan and Yong-Zhong Qian for useful discussions. 
M. F. acknowledges support by the European Research Council (ERC; FP7) under ERC Advanced Grant Agreement No. 321263 - FISH 
and support by the Nuclear Astrophysics Virtual Institute (NAVI) of the Helmholtz Association.
M. F. acknowledges the Laboratoire Astro-Particule et Cosmologie in Paris and the Technische Universit\"at Darmstadt for their hospitality during the development of this work.
MRW acknowledges support from the Helmholtz Association through the Nuclear Astrophysics Virtual Institute (VH-VI-417).
C. V. thanks "Gravitation et physique fondamentale" (GPHYS) of the Observatoire de Paris for their support.
The work of A. P. is supported by the Helmholtz-University Investigator grant No. VH-NG-825.
Calculations were performed at \href{http://scicore.unibas.ch/}{sciCORE} scientific computing core facility at the University of Basel.
\\
\end{acknowledgments}

During the completion of this manuscript 
another work appeared \cite{Zhu:2016} that investigated flavor evolution in BNS mergers using results from the astrophysical simulations~\cite{Perego:2014}. 
In Appendix~\ref{app:MNR_locations} we briefly compare with this work.

\appendix

\begin{section}{Cross sections} \label{sec:Cross_sections}

In the following we list explicit expressions from \cite{Burrows:2006} for the cross sections without weak magnetism corrections.

Neutrino absorption on free nucleons $({N \in \lbrace n, p \rbrace})$:
\begin{equation} \label{eq:sigma_abs}
\begin{split}
\sigma_{\nu N, \mathrm{abs}} = \sigma_{0} \left( \dfrac{1 + 3 g_{A}^{2}}{4} \right) \left( \dfrac{ E \pm \Delta_{n p}}{ m_{e} c^{2} } \right)^{2}
\\
\qquad \qquad \times \left[ 1 - \left( \dfrac{ m_{e} c^{2} }{ E \pm \Delta_{n p} } \right)^{2} \right]^{1/2},
\end{split}
\end{equation}
where ${\nu \in \lbrace \nu_{e}, \bar{\nu}_{e} \rbrace}$,
$\sigma_{0} = \approx 1.705 \times 10^{-44} \, \mathrm{cm}^{2},
g_{A} \approx -1.23, 
\Delta_{n p} = ( m_{n} - m_{p} ) c^{2} \approx 1.29 \, \mathrm{MeV}$, and $E$ corresponds to the neutrino energy.
The plus sign corresponds to electron neutrino absorption on free neutrons $(\sigma_{\nu_{e} n, \mathrm{abs}})$, 
while the minus sign refers to electron antineutrino absorption on free protons $(\sigma_{\bar{\nu}_{e} p, \mathrm{abs}})$.
 
For the elastic neutrino-nucleon scattering, the momentum-transfer cross section is the relevant cross sections 
for our calculations of the (transport) optical depth. It is obtained from the differential cross section
weighted by a factor $(1 - \cos{\Theta})$, and integrated over the solid angle.
Here, $\Theta$ denotes the scattering angle.
Explicitly, the momentum-transfer cross sections for neutrino-neutron scattering turns out to be
\begin{equation}
\sigma_{\nu n, \mathrm{tr}} = \dfrac{\sigma_{0}}{4} \left( \dfrac{ E }{m_{e} c^{2}} \right)^{2} \left( \dfrac{1 + 5 g_{A}^{2}}{6} \right)
\end{equation}
and for neutrino-proton scattering
\begin{equation}
\sigma_{\nu p, \mathrm{tr}} = \dfrac{\sigma_{0}}{6} \left( \dfrac{ E }{m_{e} c^{2}} \right)^{2} \left[ (C_{V}' - 1)^{2} + 5 g_{A}^{2} (C_{A}' - 1)^{2} \right],
\end{equation}
where 
\begin{align}
C_{V}' & = \dfrac{1}{2} + 2 \sin^{2}{\theta_{W}} \approx 0.96,
\\
C_{A}' & = \dfrac{1}{2}.
\end{align}

\end{section}

\begin{section}{Luminosities and neutrino fluxes} \label{app:fluxes}

\begin{figure}[tp!]
 \centering
 \includegraphics[width=0.35\textwidth]{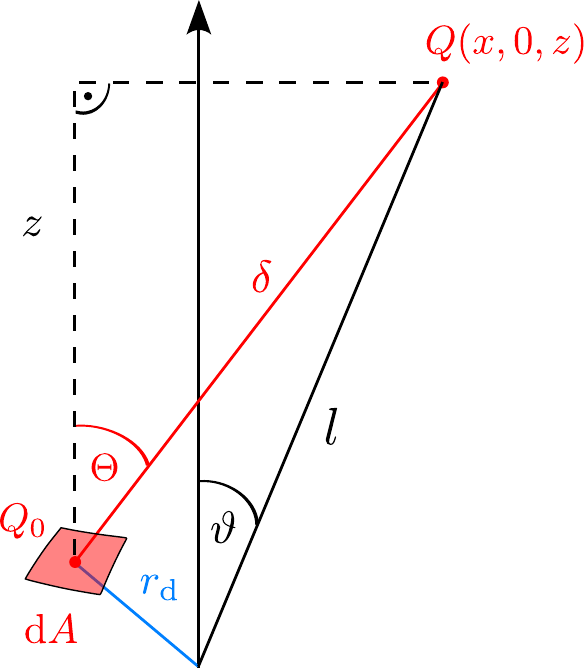}
 \caption{Geometry: $Q_{0}(r_{\mathrm{d}} \cos{\varphi}, r_{\mathrm{d}} \sin{\varphi}, 0)$, $x = l \sin{\vartheta}$, $z = l \cos{\vartheta}$, $\mathrm{d} A = \mathrm{d}r_{\mathrm{d}} \, r_{\mathrm{d}} \, \mathrm{d} \varphi$.}
 \label{disk_lum}
\end{figure}

In the following we derive an expression for the neutrino luminosities for an infinitesimal thin disk explicitly writing the constants $\hbar$ and $c$. 
For this purpose, we assume that the disk behaves like a blackbody source at a fixed temperature $T_{\nu}$ and take the neutrino number flux 
per unit energy per solid angle to be of Fermi-Dirac shape:
\begin{equation}
\mathcal{J}_{\nu}(E) = \dfrac{c}{(2 \pi \hbar c)^{3}} \dfrac{E^{2}}{\mathrm{e}^{E/ T_{\nu}} + 1}.
\end{equation}
The differential neutrino number flux per unit energy 
in a beam with differential solid angle $\mathrm{d} \Omega = \mathrm{d} A \cos(\Theta) / \delta^{2}$ at $Q(x, 0, z)$ with  a distance $l$ from the center of the disk is then given by (see 
Fig.~\ref{disk_lum}): 
\begin{equation}
\mathrm{d} \Phi_{\nu}(l, \vartheta, E) = \mathcal{J}_{\nu}(E) \dfrac{\mathrm{d} A \, \cos{\Theta}}{\delta^{2}},
\end{equation}
where
\begin{equation}
\delta \equiv \vert \mathbf{Q} - \mathbf{Q}_{0} \vert = \sqrt{ x^{2} + r_{\mathrm{d}}^{2} - 2 x r_{\mathrm{d}} \cos{\varphi} + z^{2} }
\end{equation}
corresponds to the distance from the differential area $\mathrm{d} A$ to $Q$. The $\cos(\Theta)$-factor accounts for the fact that a distant observer at $Q$ only sees an effective area which accordingly reduces the flux.
An integration over the disk yields the neutrino number flux per unit energy at $Q$:
\begin{equation}
\Phi_{\nu}(l, \vartheta, E) = \mathcal{J}_{\nu}(E) \int_{0}^{2 \pi} \, \mathrm{d} \varphi \int_{0}^{R_{\nu}} \, \mathrm{d}r_{\mathrm{d}} r_{\mathrm{d}} \, \dfrac{ \cos{\Theta} }{\delta^{2}}.
\end{equation}
Using $\cos{\Theta} = z / \delta$ and $z = l \cos{\vartheta}$, it follows:
\begin{equation}
\Phi_{\nu}(l, \vartheta, E) = l \cos{\vartheta} \, \mathcal{J}_{\nu}(E) \int_{0}^{R_{\nu}} \, \mathrm{d}r_{\mathrm{d}} r_{\mathrm{d}} \int_{0}^{2 \pi} \, \dfrac{\mathrm{d} \varphi}{\delta^{3}}.
\end{equation}
For the explicit calculation, it is convenient \cite{Malkus:2012} to introduce the quantities
\begin{equation} \label{eq:M_and_L}
L \equiv ( x - r_{\mathrm{d}} )^{2} + z^{2}, \qquad M \equiv ( x + r_{\mathrm{d}} )^{2} + z^{2}
\end{equation}
so that 
\begin{equation} \label{eq:delta}
\delta = \sqrt{ \dfrac{L + M}{2} - \dfrac{M - L}{2} \cos{\varphi} }.
\end{equation}
With the relation
\begin{equation}
\int_{0}^{2 \pi} \, \dfrac{\mathrm{d} \varphi}{\delta^{3}} = \dfrac{4 E \left(\sqrt{\frac{M - L}{M}}\right)}{L \sqrt{M}},
\end{equation}
where 
\begin{equation}
E(k) \equiv \int_{0}^{\pi / 2} \, \mathrm{d} \theta \, \sqrt{1 - k^{2} \sin^{2}{\theta}}
\end{equation}
denotes Legendre's complete elliptic integral of the second kind \cite{NIST}, we find:
\begin{align}
\Phi_{\nu}(l, \vartheta, E) = 4 l \cos{\vartheta} \, \mathcal{J}_{\nu}(E) \int_{0}^{R_{\nu}} \, \mathrm{d} r_{\mathrm{d}} \, r_{\mathrm{d}} \, \dfrac{E \left(\sqrt{\frac{M - L}{M}} \right)}{L \sqrt{M}}.
\end{align}
For an observer located at infinity, we compute the isotropized luminosity, i.e., the luminosity obtained by assuming an isotropically emitted flux:
\begin{align}
L_{\nu}(\vartheta) & = 2 \pi l^{2} \int_{0}^{\infty} \, \mathrm{d} E \, E \, \Phi_{\nu}(l, \vartheta, E) \biggr\rvert_{l \to \infty}
\\
\begin{split}
& = \dfrac{7}{120} \dfrac{\pi^{2} c}{(\hbar \, c)^{3}} \, T_{\nu}^{4} \, l^{3} \, \cos{\vartheta} \,
\\
& \qquad \times \int_{0}^{R_{\nu}} \, \mathrm{d} r_{\mathrm{d}} \, r_{\mathrm{d}} \, \dfrac{E\left(\sqrt{\frac{M - L}{M}}\right)}{L \sqrt{M}} \biggr\rvert_{l \to \infty},
\end{split}
\end{align}
where we used
\begin{align}
\int_{0}^{\infty} \, \mathrm{d} E \, \dfrac{E^{3}}{\mathrm{e}^{E / T_{\nu}} + 1} = \dfrac{7 \pi^{4}}{120} T_{\nu}^{4}.
\end{align}
The asymptotic relation
\begin{equation}
\dfrac{E(\sqrt{\frac{M - L}{M}})}{L \sqrt{M}} = \dfrac{\pi}{2} \dfrac{1}{l^{3}} + \mathcal{O}(1 / l^{5}) \qquad (l \to \infty),
\end{equation}
yields:
\begin{align}
L_{\nu}(\vartheta) & = \dfrac{7 \pi}{4} \sigma_{\mathrm{SB}} \, T_{\nu}^{4}  \, \cos{\vartheta} \, \int_{0}^{R_{\nu}} \, \mathrm{d} r_{\mathrm{d}} \, r_{\mathrm{d}}
\\
& = \dfrac{7 \pi}{8} \sigma_{\mathrm{SB}} \, T_{\nu}^{4} \, R_{\nu}^{2} \, \cos{\vartheta},
\end{align}
where $\sigma_{\mathrm{SB}} \equiv \pi^{2} k_{\mathrm{B}}^{4} / ( 60 \hbar^{3} c^{2} )$ denotes the Stefan-Boltzmann constant.

Note that in the equations above, we considered a neutrino moving in the upward direction ($\cos{\vartheta} > 0$). 
If we want to follow a neutrino moving downwards, we have to replace $\cos{\vartheta}$ by $- \cos{\vartheta}$, i.e. if $\frac{\pi}{2} < \vartheta < \pi$. 

An integration over the upper hemisphere leads to the total neutrino luminosity:
\begin{align} \label{eq:lum}
L_{\nu} & = \int_{0}^{1} \mathrm{d}(\cos{\vartheta}) \, L_{\nu}(\vartheta)
\\
& = \dfrac{7 \pi}{16} \sigma_{\mathrm{SB}} \, T_{\nu}^{4} \, R_{\nu}^{2}.
\end{align}

As an additional check, we also compute the total neutrino energy flux emitted from $\mathrm{d}A$:
\begin{align}
\mathcal{F}_{\nu} & = \int_{0}^{\infty} \mathrm{d} E \, E^{2} \, \mathcal{J}_{\nu}(E) \int \mathrm{d} \Omega \cos{\vartheta}
\\
& = \dfrac{7}{8} \dfrac{\pi}{120 \hbar^{3} c^{2}} T_{\nu}^{4} \, 2 \pi \int_{0}^{1} \mathrm{d}(\cos{\vartheta}) \cos{\vartheta}
\\
& = \dfrac{7}{16} \sigma_{\mathrm{SB}} T_{\nu}^{4}
\end{align}
and from that, we infer the total neutrino luminosity:
\begin{align}
L_{\nu} & = \pi R_{\nu}^{2} \mathcal{F}_{\nu}
\\
& = \dfrac{7 \pi}{16} \sigma_{\mathrm{SB}} \, T_{\nu}^{4} \, R_{\nu}^{2}.
\end{align}
This expression agrees with the one derived above 
(Eq.~\eqref{eq:lum}).
For a simple estimate, it is useful to express this in the convenient form:
\begin{equation} \label{lum_compact}
L_{\nu} \sim 15 \times 10^{51} \, \mathrm{erg / s} \left( \dfrac{T_{\nu}}{3.4 \, \mathrm{MeV}} \right)^{4} \left( \dfrac{R_{\nu}}{90 \, \mathrm{km}} \right)^{2}
\end{equation}
If we use the neutrino disks radii at $60 \, \mathrm{ms}$ from 
Table~\ref{table:neutrino_surface_radii} and the temperatures from Table~\ref{table:emission_parameters}, we obtain $L_{\nu_{e}} \approx 15 \times 10^{51} \, \mathrm{erg / s}$, $L_{\bar{\nu}_{e}} \approx 33 \times 10^{51} \, \mathrm{erg / s}$ and $L_{\nu_{x}} \approx 48 \times 10^{51} \, \mathrm{erg / s}$.
While the values for $L_{\nu_{e}}$ and $L_{\bar{\nu}_{e}}$ are compatible with the luminosities obtained in \cite{Perego:2014}, the value of $L_{\nu_{x}}$ is about 6 times larger.

Similarly, we compute the isotropized neutrino number luminosity:
\begin{align}
L_{\mathrm{N}, \nu}(\vartheta) & = 2 \pi l^{2} \int_{0}^{\infty} \mathrm{d} E \, \Phi_{\nu}(l, \vartheta, E) \biggr\rvert_{l \to \infty}
\\
& = \dfrac{c}{(\hbar c)^{3}} \dfrac{F_{2}(0)}{4 \pi} \cos{\vartheta} T_{\nu}^{3} R_{\nu}^{2}
\end{align}
and neutrino number luminosity:
\begin{align}
L_{\mathrm{N}, \nu} & = \int_{0}^{1} \mathrm{d} (\cos \vartheta) \, L_{\mathrm{N}, \nu}(\vartheta)
\\
& = \dfrac{c}{(\hbar c)^{3}} \dfrac{F_{2}(0)}{8 \pi} T_{\nu}^{3} R_{\nu}^{2}.
\end{align}
In local thermal equilibrium we can relate the mean energy, defined \cite{Perego:2014} via $\langle E_{\nu} \rangle = L_{\nu} / L_{\mathrm{N}, \nu}$, to the effective temperature:
\begin{equation}
\langle E_{\nu} \rangle = \dfrac{F_{3}(0)}{F_{2}(0)} T_{\nu}.
\end{equation}
However, we stress that the mechanism behind the spectra formation is rather demanding, since neutrinos decouple at different energy-dependent surfaces, 
and therefore our assumption of a thermal spectra which is determined by an effective temperature represents only a coarse approximation.
In order to account for deviations from thermal equilibrium, one could examine different shapes of the spectral function similarly to studies that have been done in the context of core-collapse supernovae (e.g., \cite{Keil:2003}).

If we multiply the neutrino number flux,
\begin{align}
F_{\nu} & =\int_{0}^{\infty} \mathrm{d} E \, \mathcal{J}_{\nu}(E) \int \mathrm{d} \Omega \, \cos{\vartheta}
\\
& = \dfrac{\pi c}{(2 \pi \hbar c)^{3}} F_{2}(0) T_{\nu}^{3},
\end{align}
by the disk area $\pi R_{\nu}^{2}$, the same expression for the 
neutrino number luminosity $L_{\mathrm{N}, \nu} = \pi R_{\nu}^{2} F_{\nu}$ as above is obtained:
\begin{equation}
L_{\mathrm{N}, \nu} = \dfrac{c}{(\hbar c)^{3}} \dfrac{F_{2}(0)}{8 \pi} T_{\nu}^{3} R_{\nu}^{2}.
\end{equation}

Finally, we express the neutrino number flux in terms of the luminosity and mean neutrino energy:
\begin{align}
F_{\nu} = \dfrac{1}{\pi R_{\nu}^{2}} \dfrac{L_{\nu}}{\langle E_{\nu} \rangle}.
\end{align}
and define the neutrino number flux per unit energy per solid angle as follows:
\begin{equation}
c \dfrac{\mathrm{d}^{2} n_{\nu}}{\mathrm{d} \Omega \mathrm{d} E} \equiv \dfrac{F_{\nu}}{\pi} f_{\nu}(E),
\end{equation}
where 
\begin{equation}
f_{\nu}(E) = \dfrac{1}{F_{2}(0)} \dfrac{1}{T_{\nu}^{3}} \dfrac{E^{2}}{\mathrm{e}^{E / T_{\nu}} + 1}
\end{equation}
denotes the normalized Fermi-Dirac distribution function with vanishing degeneracy parameter.

Consequently, the neutrino flux per unit energy corresponds to:
\begin{align}
\Phi_{\nu}(E, x, z) & = \int_{\Omega_{\nu}} \mathrm{d}\Omega \left( c \dfrac{\mathrm{d}^{2} n_{\nu}}{\mathrm{d} \Omega \mathrm{d} E}  \right)
\\
& = \dfrac{F_{\nu}}{4 \pi} f_{\nu}(E) \Omega_{\nu}, \label{eq:neutrino_flux}
\end{align}
where the solid angle is given by
\begin{equation}
\Omega_{\nu} = 4 z \int_{R_{0}}^{R_{\nu}} \mathrm{d}r_{\mathrm{d}} r_{\mathrm{d}} \, \dfrac{E(\sqrt{\frac{M - L}{M}})}{L \sqrt{M}}.
\end{equation}
We show a typical form of the solid angle in 
Fig.~\ref{Fig:solid_angle} along trajectory 2a.
\begin{figure}[tp!]
 \centering
 \includegraphics[width=0.5\textwidth]{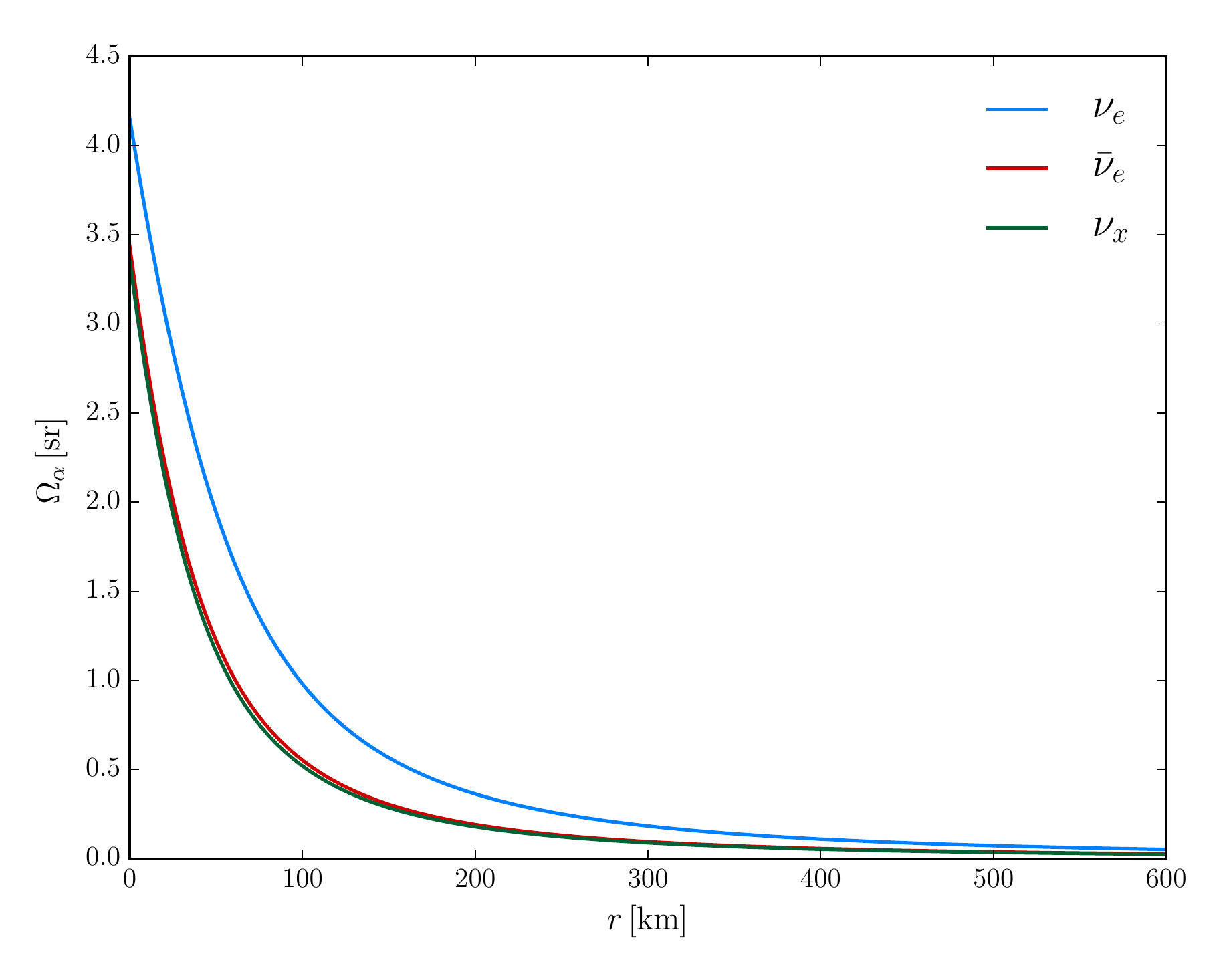} 
 \caption{Solid angle for each neutrino species computed along trajectory 2a.} 
 \label{Fig:solid_angle}
\end{figure}
Since the neutrino disk radii for $\bar{\nu}_{e}$ and $\nu_{x}$ are similar, the difference in their corresponding solid angles is minor.

\end{section}

\begin{section}{Geometric factor} \label{app:geometric_factor}

In this section we show the explicit structure of the geometric factor Eq.~\eqref{eq:geometric_factor_general}
used in the neutrino self-interaction Hamiltonian 
Eq.~\eqref{eq:Hamiltonian-nu-nu_disk}. 
An analytical calculation yields:
\begin{widetext}
\begin{align}
\int_{\Omega_{\nu_{\alpha}}} \mathrm{d} \Omega \, (1 - \cos{\Theta_{\mathbf{p} \mathbf{p}'}}) & = \int_{\cos \theta_{\mathrm{max}, \, \nu_{\alpha}}}^{+1} \mathrm{d}( \cos \theta ) \int_{0}^{2 \pi} \mathrm{d} \phi \, (1 - \cos{\Theta_{\mathbf{p} \mathbf{p}'}})
\\
& = \int_{R_{0}}^{R_{\nu_{\alpha}}} \mathrm{d} r_{\mathrm{d}} \int_{0}^{2 \pi} \mathrm{d} \varphi \, \vert \mathrm{det} \, J \vert \, (1 - \cos{\Theta_{\mathbf{p} \mathbf{p}'}})
\\
& = z \int_{R_{0}}^{R_{\nu_{\alpha}}} \mathrm{d} r_{\mathrm{d}} \, r_{\mathrm{d}} \, \Gamma(r_{\mathrm{d}}, \theta_{0}, \phi_{0}, x, z)
\\
& =: G_{\nu_{\alpha}}(\theta_{0}, \phi_{0}, x, z).
\end{align}
In the second step, we performed the previously described change of coordinates (Eqs.~\eqref{eq:coord-transformations}, \eqref{eq:Jacobi_det_2}, \eqref{eq:scattering-angle_transformed}) and introduced $\Gamma$ via:
\begin{equation} \label{eq:Gamma_factor}
\begin{split}
\Gamma(r_{\mathrm{d}}, \theta_{0}, \phi_{0}, x, z) \equiv \int_{0}^{2 \pi} \mathrm{d}\varphi \left \lbrace \dfrac{1}{\delta^{3}} \right. & \left. - \dfrac{z \cos \theta_{0} + x \sin \theta_{0} \cos \phi_{0}}{\delta^{4}} \right.
\\
& \left. + r_{\mathrm{d}} \sin \theta_{0} \cos \phi_{0} \dfrac{\cos \varphi}{\delta^{4}} + r_{\mathrm{d}} \sin \theta_{0} \sin \phi_{0} \dfrac{\sin \varphi}{\delta^{4}} \right \rbrace,
\end{split}
\end{equation}
where $\delta$ is defined as in Eq.~\eqref{eq:delta}.
In the single-trajectory approximation, the geometric factor $G_{\nu_{\alpha}}$ should be understood as an averaging over the angles.
The explicit $\varphi$-integration in Eq.~\eqref{eq:Gamma_factor} yields:
\begin{equation}
\Gamma = \dfrac{4 E \left( \sqrt{ \frac{M - L}{M} } \right)}{L \sqrt{M}} 
- \dfrac{\pi \left \lbrace (L+M)(z \cos \theta_{0} + x \sin \theta_{0} \cos \phi_{0}) - 4 x r_{\mathrm{d}}^{2} \sin \theta_{0} \cos \phi_{0} \right \rbrace}{(L M)^{3/2}},
\end{equation}
where in the second step, the relation $M - L = 4 x r_{\mathrm{d}}$, with $M$ and $L$ defined in Eq.~\eqref{eq:M_and_L}, was used and $E(k)$ denotes Legendre's complete elliptic integral of the second kind \cite{NIST}:
\begin{equation}
E(k) \equiv \int_{0}^{\pi/2} \mathrm{d} \theta \, \sqrt{1 - k^{2} \sin^{2}{\theta}}.
\end{equation}
\end{widetext}
Due to our definition of the angle $\phi$ that differs from the definition in \cite{Malkus:2012}, 
we note that $-\Gamma / 2$ with the replacement $\phi_{0} \mapsto \pi - \phi_{0}$ corresponds to the geometric factor $C$ given in \cite{Malkus:2012}. 
Here, one also has to keep in mind the different convention used to denote the elliptic integral.

\end{section}

\begin{section}{Collected data} \label{app:collected_data}
In the following table, we report a summary of published data concerning the
neutrino luminosities and mean energies of a binary neutron
star merger. We considered simulations of binary NS mergers \cite{Ruffert:1997, Rosswog:2003, Rosswog:2013, Sekiguchi:2015, Foucart:2016} or merger 
aftermaths \cite{Dessart:2009, Metzger:2014,Perego:2014}, including neutrino emission and characterized by the presence of a (possibly unstable) massive neutron star surrounded by a thick accretion disk. Since they run for very different amounts of time, for the merger simulations we choose the values when the luminosities have reached quasi-stationary values, while for the aftermath simulations the values close to the beginning of the calculation.

\begingroup
\squeezetable
\begin{table*}
\center
\caption{Table with a summary of published data. 
We considered simulations of binary NS mergers or of their aftermath. In the former case, we report the time when 
the data were taken with respect to the beginning of the merger (in ms). 
In the latter case, we report data close to the beginning of the 
simulation. For all cases, we list the neutrino luminosities (in $10^{51} \, \mathrm{erg} \,
\mathrm{s}^{-1}$), where for $\nu_{x}$, the values correspond to each single species, and, when 
available, the neutrino mean energies (in MeV, in the case of \cite{Foucart:2016} the quantities provided are the rms energies).}
\label{Table_Uncertainties}
\begin{ruledtabular}
\begin{tabular}{lcccccccl}
Source & $t - t_{\mathrm{merger}}$ & $L_{\nu_{e}}$ & $L_{\bar{\nu}_{e}}$ & $L_{\nu_{x}}$ & $\langle E_{\nu_{e}} \rangle$ & $\langle E_{\bar{\nu}_{e}} \rangle$ & $\langle E_{\nu_{x}} \rangle$ & Notes \hfill \\ \hline
\cite{Perego:2014} & $-$ & $15$ & $32$ & $8$ & $10.6$ & $15.3$ & $17.3$ & \footnotesize Grid, Newtonian, spectral leakage, \\
& & & & & & & & \footnotesize HS(TM1) EOS, MNS + disk.
\\ \hline
\cite{Ruffert:1997} & $8$ & $16$ & $43$ & $6$ & $13$ & $20$ & $28$ & \footnotesize Grid, Newtonian, gray leakage, \\
& & & & & & & & \footnotesize LS(180) EOS, $1.65 - 1.65 \, M_{\odot}$, \\
& & & & & & & & \footnotesize no spin.
\\ \hline 
\cite{Rosswog:2003} & $18$ & $45$ & $130$ & $25$ & $9$ & $15$ & $20$ & \footnotesize SPH, Newtonian, gray leakage, \\
& & & & & & & & \footnotesize Shen EOS, $1.4 - 1.4 \, M_{\odot}$, \\
& & & & & & & & \footnotesize no spin.
\\ \hline 
\cite{Dessart:2009} & $-$ & $20$ & $32$ & $6$ & $11$ & $16$ & $22$ & \footnotesize Grid, Newtonian, MGFLD, \\
& & & & & & & & \footnotesize Shen EOS, MNS + disk.
\\ \hline 
\cite{Rosswog:2013} & $16$ & $30$ & $60$ & $10$ & $8$ & $14.4$ & $26.3$ & \footnotesize SPH, Newtonian, gray leakage, \\
& & & & & & & & \footnotesize Shen EOS, $1.3 - 1.4 \, M_{\odot}$, \\
& & & & & & & & \footnotesize no spin.
\\ \hline 
\cite{Metzger:2014} & $-$ & $55$ & $45$ & $-$ & $12$ & $15$ & $-$ & \footnotesize Grid, Newtonian, gray leakage, \\
& & & & & & & & \footnotesize Timmes \& Swesty EOS, \\
& & & & & & & & \footnotesize MNS + disk.
\\ \hline 
\cite{Sekiguchi:2015} & $30$ & $120$ & $200$ & $15$ & $-$ & $-$ & $-$ & \footnotesize Grid, GR, gray GR leakage + \\ 
& & & & & & & & \footnotesize moment formalism for \\
& & & & & & & & \footnotesize free streaming $\nu$'s, \\
& & & & & & & & \footnotesize HS(TM1) EOS, $1.35 - 1.35 \, M_{\odot}$, \\
& & & & & & & & \footnotesize no spin.
\\ \hline 
\cite{Sekiguchi:2015} & $30$ & $100$ & $150$ & $10$ & $-$ & $-$ & $-$ & \footnotesize Grid, GR, $\nu$'s as above, \\
& & & & & & & & \footnotesize HS(DD2) EOS, $1.35 - 1.35 \, M_{\odot}$, \\
& & & & & & & & \footnotesize no spin.
\\ \hline 
\cite{Sekiguchi:2015} & $10$ & $175$ & $280$ & $100$ & $-$ & $-$ & $-$ & \footnotesize Grid, GR, $\nu$ as above, \\
& & & & & & & & \footnotesize SFHo EOS, $1.35 - 1.35 \, M_{\odot}$, \\
& & & & & & & & \footnotesize no spin.
\\ \hline 
\cite{Foucart:2016} & $10$ & $120$ & $210$ & $30$ & $13$ & $20$ & $26$ & \footnotesize Grid, GR, gray GR leakage, \\
& & & & & & & & \footnotesize LS(220) EOS, $1.2 - 1.2 \, M_{\odot}$, \\
& & & & & & & & \footnotesize no spin.
\\ \hline 
\cite{Foucart:2016} & $10$ & $160$ & $220$ & $22.5$ & $13$ & $20$ & $24$ & \footnotesize Grid, GR, gray GR leakage, \\
& & & & & & & & \footnotesize HS(DD2) EOS, $1.2 - 1.2 \, M_{\odot}$, \\
& & & & & & & & \footnotesize no spin.
\\ \hline
\cite{Foucart:2016} & $10$ & $190$ & $300$ & $55$ & $14$ & $21$ & $29$ & \footnotesize Grid, GR, gray GR leakage, \\
& & & & & & & & \footnotesize SFHo EOS, $1.2 - 1.2 \, M_{\odot}$, \\
& & & & & & & & \footnotesize no spin.
\\ \hline
\cite{Foucart:2016} & $10$ & $60$ & $210$ & $70$ & $13$ & $20$ & $26$ & \footnotesize Grid, GR, gray GR M1, \\
& & & & & & & & \footnotesize LS(220) EOS, $1.2 - 1.2 \, M_{\odot}$, \\
& & & & & & & & \footnotesize no spin. \\
\end{tabular}
\end{ruledtabular}
\end{table*}
\endgroup
\end{section}

\clearpage

\begin{section}{Matter-neutrino resonance locations} \label{app:MNR_locations}

\begin{figure}[tp!]
 \centering
 \includegraphics[width=0.5\textwidth]{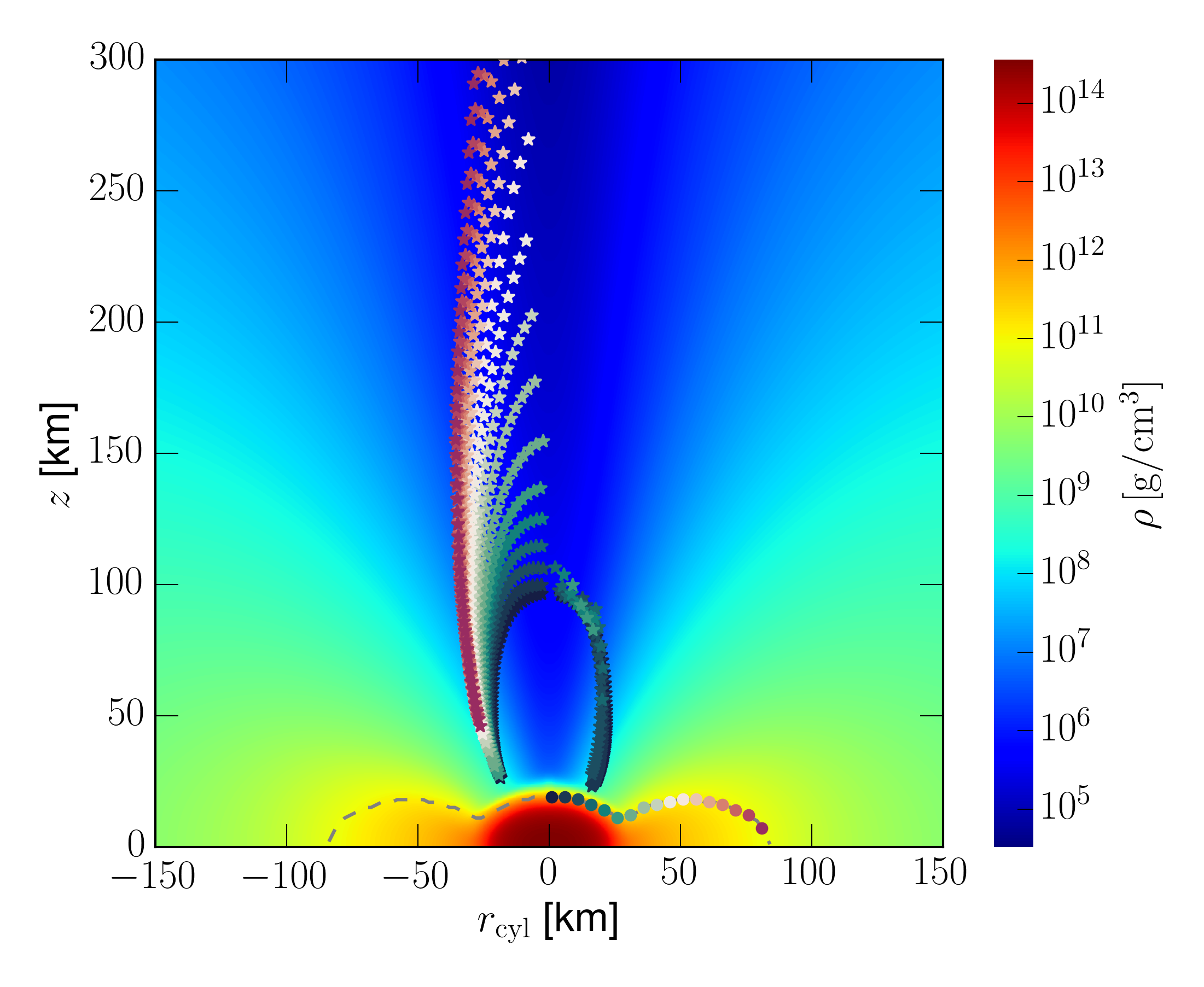} 
 \caption{Cancellation points ($\star$), i.e. points where $\lambda + \mu \approx 0$, for several trajectories above the BNS merger remnant. The emission points are represented by $\bullet$. Emission and cancellation points with the same color refer to a single trajectory. Points where the crossing starts dominated by matter are excluded. Color coded is the density of matter.}
 \label{Fig:MNR_locations}
\end{figure}

The initial conditions employed in this work are close to the ones used by \citet{Zhu:2016}, who also investigated the flavor evolution in BNS mergers. This allows a close comparison between the two different works. In particular, we compute the locations above the remnant where MNR occurs and compare with the results reported by \citet{Zhu:2016} in their figure 6. These locations can be identified as cancellation points, i.e., points where the unoscillated neutrino self-interaction potential cancels with the matter potential along a specific trajectory, $\lambda + \mu \approx 0$. In Fig.~\ref{Fig:MNR_locations}, we represent the cancellation points for several different trajectories starting from the disk. The emerging picture suggests that the resonances for neutrinos originating from one side of the disk are preferably located at the edge of the funnel. Neutrinos emerging from the MNS will all encounter MNR, probably with some flavor transformations depending on the adiabaticity. In contrast, neutrinos emitted from the disk will only encounter MNR if they propagate through the central regions of the funnel, where the matter density is relatively low.
Remarkably, this picture is qualitatively consistent with the results found in \citet{Zhu:2016}. Nevertheless, quantitative differences between the two works can be present and traced back to the different approaches used in the two works to model the radiation field outside the neutrino surfaces. On the one hand, we have assumed a thin disk model, emitting thermal radiation from a spectrally averaged, single-temperature neutrino surface. On the other hand, \citet{Zhu:2016} used an energy-dependent leakage scheme and a ray-tracing algorithm to model the emission from a finite size, thick disk. They also take explicitly into account radiation damping effects outside the neutrino surfaces.

\end{section}

\bibliographystyle{apsrev4-1}
\bibliography{mybiblio}

\end{document}